\documentclass[11pt]{article}
\usepackage{amssymb}

\usepackage{graphicx}
\usepackage{amsmath}
\usepackage{makeidx}
\usepackage{indentfirst}

\newcounter{resultnum}[section]

\newcounter{conclusionnum}[section]

\newcounter{conditionnum}[section]

\newcounter{conjecturenum}[section]

\newcounter{examplenum}[section]

\newcounter{exercisenum}[section]

\newcounter{lemmanum}[section]

\newcounter{notationnum}[section]

\newcounter{theoremnum}[section]

\newcounter{definitionnum}[section]

\newcounter{corollarynum}[section]

\newcounter{remarknum}[section]

\newcounter{propositionnum}[section]

\newcounter{acknowledgementnum}[section]

\newcounter{algorithmnum}[section]

\newcounter{axiomnum}[section]

\newcounter{casenum}[section]

\newcounter{claimnum}[section]

\newcounter{summarynum}[section]

\newcounter{problemnum}[section]

\begin{document}

\title{On Relativistic Generalization of Perelman's W--entropy and
 Thermodynamic Description of Gravitational Fields and Cosmology}
\date{February 27, 2017}
\author{${}$ \\
 Vyacheslav Ruchin\\
\small\it Heinrich-Wieland-Str. 182,\  81735 M\"{u}nchen, Germany  \\
\small\it email: v.ruchin-software@freenet.de \\
${}$ \\
 Olivia Vacaru\\
\small\it National College of Ia\c s i, 4 Arcu street,  Ia\c si, Romania, 700115  \\
\small\it email: olivia.vacaru@yahoo.com \\
${}$ \\
\vspace{.2 in} Sergiu I. Vacaru\\
{\small \textit{Quantum Gravity Research; 101 S. Topanga Canyon Blvd \#
1159. Topanga, CA 90290, USA}}  \\
{\small and \textit{University "Al. I. Cuza" Ia\c si, Project IDEI }}\footnote{Address for contact:  \textit{Flat 4 Brefney house, Fleet street, Ashton-under-Lyne, OL6 7PG, the UK  }} \\
{\small and\thanks{two DAAD fellowship visiting affiliations in Germany,  where the paper was performed} }\\
{\small \textit{Max-Planck-Institute for Physics,
Werner-Heisenberg-Institute, }}\\
{\small Foehringer Ring 6, M\"{u}nchen, Germany D-80805 ;}\\
{\small and }\\
{\small \textit{    Leibniz University of Hannover, Institute for Theoretical Physics, }}\\
{\small Appelstrasse 2, Hannover, Germany 30167}\\
{\small \textit{emails: sergiu.vacaru@uaic.ro; sergiu.vacaru@gmail.com}}
}
\maketitle

\begin{abstract}
Using double 2+2 and 3+1 nonholonomic fibrations on Lorentz manifolds, we
extend the concept of W--entropy for gravitational fields in the general
relativity, GR, theory. Such F-- and W--functionals were introduced in the
Ricci flow theory of three dimensional, 3-d, Riemannian metrics by G.
Perelman, arXiv: math.DG/0211159. Nonrelativistic 3--d Ricci flows are
characterized by associated statistical thermodynamical values determined by
W--entropy. Generalizations for geometric flows of 4--d pseudo--Riemannian
metrics are considered for models with local thermodynamical equilibrium and
separation of dissipative and non--dissipative processes in relativistic
hydrodynamics. The approach is elaborated in the framework of classical
filed theories (relativistic continuum and hydrodynamic models) without an
underlying kinetic description which will be elaborated in other works. The
3+1 splitting allows us to provide a general relativistic definition of
gravitational entropy in the Lyapunov--Perelman sense. It increases
monotonically as structure forms in the Universe. We can formulate a
thermodynamic description of exact solutions in GR depending, in general, on
all spacetime coordinates. A corresponding 2+2 splitting with nonholonomic
deformation of linear connection and frame structures is necessary for
generating in very general form various classes of exact solutions of the
Einstein and general relativistic geometric flow equations. Finally, we
speculate on physical macrostates and microstate interpretations of the
W--entropy in GR, geometric flow theories and possible connections to string
theory (a second unsolved problem also contained in Perelman's works) in the
Polyakov's approach.

\vskip0.2cm


\vskip0.2cm

PACS 2010:\ 02.40.Vh, 02.90.+p, 04.20.Cv, 04.20.Jb, 04.90.+e, 05.90.+m

\end{abstract}

\tableofcontents

\section{Introduction}

G. Perelman defined the W--entropy \cite{perelman1,perelman2,perelman3} as a
functional with non--decreasing Lyapunov--type property from which
Hamilton's equations \cite{ham1,ham2,ham3} for Ricci flows can be derived
following the variational procedure. The approach was elaborated upon for
the geometric evolution of three dimensional (3--d) Riemannian metrics.
There were obtained a number of fundamental results in geometric analysis
and topology. Such directions in modern mathematics became famous after the
elaborated methods allowed to prove the Poincar\'{e} and Thorston
conjectures. In this paper we show that using nonholonomic double 3+1 and
2+2 splitting in general relativity, GR, the geometric and statistical
thermodynamics methods considered in Perelman's works can be developed for
theories of generalized relativistic geometric flows. We consider how such
constructions can be applied in modern cosmology and astrophysics.

There are different ways for generalizing models of 3--d Ricci flow
evolution for 4--d spacetimes with pseudo--Euclidean signature. For
instance, there were formulated theories of stochastic / diffusion and
kinetic processes with local anisotropy, fractional geometric evolution etc
\cite{vkin,vfracrf,velatdif}. It is possible to construct thermo field
models of Ricci flow evolution on imaginary time $\varsigma =-it(0\leq
\varsigma \leq 1/\kappa T,$ where $\kappa $ is Boltzmann's constant and $T$
is the temperature). In such a case, the pseudo--Riemannian spacetime is
transformed into a Riemannian configuration space like in thermal and/or
finite temperature quantum field theory (see \cite{zinn,umezawa} and
references therein). Here we recall that G. Perelman treated $\tau
=\varsigma ^{-1}$ as a temperature parameters and derived his W--entropy
following analogy to formulas for the entropy in statistical mechanics.%
\footnote{%
It is interesting the Remark 5.3 in \cite{perelman1} (and next paragraph,
just before section 6 in that paper) which we reproduce here:\
\textquotedblright An entropy formula for the Ricci flow in dimension two
was found by Chow [C]; there seems to be no relation between his formula and
ours. .... The interplay of statistical physics and (pseudo)-riemannian
geometry occurs in the subject of Black Hole Thermodynamics, developed by
Hawking et al. Unfortunately, this subject is beyond my understanding at the
moment.\textquotedblright} In his works, it was not specified what type of
underlying microstates and their energy should be taken in order to explain
the geometric flows corresponding to certain thermodynamical and gravity
models.

The (nonrelativistinc) Ricci flow evolution equations postulated
heuristically by R. Hamilton can be writen in the form
\begin{equation}
\frac{\partial g_{\grave{\imath}\grave{j}}}{\partial \tau }=-2\ R_{\grave{%
\imath}\grave{j}}.  \label{heq1a}
\end{equation}%
In these formulas, $\tau $ is an evolution real parameter and the local
coordinates $u^{\grave{\imath}}$ with indices $\grave{\imath},\grave{j}%
=1,2,3 $ are defined on a real 3-d Riemannian manifold. We can consider that
equations (\ref{heq1a}) describe a nonlinear diffusion process for geometric
flow evolution of 3-d Riemannian metrics. For small deformations of a 3--d
Euclidean metric $g_{\grave{\imath}\grave{j}}\approx \delta _{\grave{\imath}%
\grave{j}}+$ $h_{\grave{\imath}\grave{j}},$ with $\delta _{\grave{\imath}%
\grave{j}}=diag[1,1,1]$ and $h_{\grave{\imath}\grave{j}}|\ll 1$, the Ricci
tensor approximates the Laplace operator $\Delta =\frac{\partial ^{2}}{%
(\partial u^{1})^{2}}+\frac{\partial ^{2}}{(\partial u^{2})^{2}}+\frac{%
\partial ^{2}}{(\partial u^{3})^{2}}$. We obtain a linear diffusion
equation, $R_{\grave{\imath}\grave{j}}\sim \Delta h_{\grave{\imath}\grave{j}%
}.$ In modified and normalized form, equations of type (\ref{heq1a}) can be
proven following a corresponding variational calculus for Perelman's W- and
F--functionals. Using the W--entropy, an analogous statistical mechanics and
thermodynamics was formulated. Respective thermodynamic values (mean energy,
entropy and fluctuation dispersion) can be considered as certain physical
characteristics of flow evolution of Riemannian metrics. Summaries of most
important mathematical results and methods can be found in \cite%
{monogrrf1,monogrrf2,monogrrf3}.

Geometric flow evolution models of pseudo--Riemannian metrics have not been
formulated and studied in modern physical mathematics. Such ideas have not
be developed and do not exist among gravitational and related relativistic
thermodynamics/ diffusion / kinetic theories. In quantum field theory there,
were considered in relativistic form some examples of low dimensional
geometric flow equations of type (\ref{heq1a}). That was even before
mathematicians formulated in rigorous form respective directions in
geometric analysis and topology which are related to the Ricci flow theory.
D. Friedan published during 1980-1985 a series of works on nonlinear sigma
models, $\sigma $--models, in two + epsilon dimensions, see \cite%
{friedan1,friedan2,friedan3}. There were studied certain topological
properties of the $\beta $ function and solutions of the fixed-point
equation (latter called the Ricci soliton equation) and further developments
on renormalization of the $\mathit{O}(N)$--invariant nonlinear $\sigma $%
--models in low--temperature regime dominated by small fluctuations around
ordered states \cite{polyakov1}.

There were studied generalized Perelman's functionals for various models of
non--Riemannian geometries and (modified) gravity theories, see \cite%
{vnoncjmp,vnoncjmprf,lin} and references therein. The corresponding
generalized Ricci soliton equations describe modified Einstein equations for
certain classes of modified gravity theories, MGTs. Such models posses an
important decoupling property of the fundamental evolution/ dynamical
equations and can be integrated in general form. We can study new classes of
exact solutions and search for application in modern gravity and cosmology.
Nevertheless, we can not argue that certain analogs of generalized/ modified
R. Hamilton's equations would describe in a self-consistent manner certain
evolution processes if additional assumptions on geometric and physical
properties of GR flows are not considered. We can not treat as an entropy
functional any formal relativistic modification of Perelman's functionals
defined only in terms of the Levi--Civita connection for pseudo-Riemannian
metrics. For the geometric flow evolution of 4--d metrics with Lorentz
signature and non--stationary solutions in GR, it is not possible to
formulate a statistical thermodynamic interpretation like in the case of 3-d
Riemannian ones.

The main goal of this paper is to study how the concept of W--entropy can be
generalized in order to characterize 3-d hypersurface gravitational
thermodynamic configurations and their GR evolution determined by exact
solutions in Einstein gravity. The approach involves two other less
established (general) relativistic theories: the relativistic statistical
thermodynamics and the nonlinear diffusion theory on curved spacetimes and
in gravity. Historically, the first relativistic generalizations of
thermodynamics due to M. Plank \cite{rtplanck} and A. Einstein \cite{rteinst}
were subjected to certain critics and modifications more than half century
latter \cite{rtcrit1,rtcrit2,rtcrit3,rtcrit4,rtcrit5,land,moe}. Various
ideas and constructions exclude each other and various debates continue even
today \cite{rttoday1,rttoday2,rttoday3,rttoday4,rttoday5,rttoday6,lopez}.
For such models, there were postulated different covariant relativistic
thermodynamical and/or statistical thermodynamical values, with respective
transformation laws under local Lorentz transforms. There is an explicit
dependence on the 4th time like coordinate and the main issue is how to
define the concept of temperature and provide a physical interpretation. We
plan to elaborate on a general thermodynamic treatment of relativistic Ricci
flows using methods of relativistic kinetics and nonlinear diffusion theory
in our further works. In this paper, we consider a generalization of G.
Perelman's W--thermodynamic model to flows of entropy and effective energy
in the framework of "most simple" relativistic fluids theory and
hydrodynamics with stability and causality.

The relativistic 4--d geometric flows depend, in general, on a time like
coordinate being described also by evolution on a temperature like
parameter. We argue that an appropriate re--definition of effective
thermodynamic variables for the Ricci flow theory allows us to compute the
entropy of gravitational fields and elaborate upon an effective statistical
mechanics and thermodynamical formalism both for cosmology and black holes.
We shall use the 3+1 decomposition formalism (see \cite{misner} and
references therein) and specify the conditions when a 3--d geometric
evolution is "driven" in relativistic form by solutions of the Einstein
equations in GR. Vacuum stationary solutions with Killing symmetries and
horizons consist a special class of gravitational configurations when the
thermodynamic models are constructed for the entropy determined by the
horizon gravity for a corresponding 2+1+1 splitting.

We compare our relativistic geometric flow approach to a recently proposed
thermodynamical theory of gravitational fields with a measure of
gravitational entropy was proposed in Ref. \cite{clifton} (the so--called
CET model). Those constructions are based on the square--root of the
Bel--Robinson tensor when the measure is non--negative in contrast to other
proposals. There were analyzed some examples, for instance, for the
Schwarzschild black hole and Friedmann--Lema\^{\i}tre--Robertson--Worker,
FLRW, cosmology. The black hole thermodynamics was derived as a particular
case for stationary spacetimes. We study how the CET model can be obtained
by corresponding nonholonomic parameterizations from the statistical
mechanics and thermodynamic models based on the concept of W--entropy (see
further developments in \cite{vnoncjmp,vnhrf,vnoncjmprf,lin} and references
therein). There is not a unique way to generalize standard black hole
thermodynamical constructions in order to include cosmological solutions. A
geometric self--consistent variant is to address such problems using
relativistic models with generalized W--entropy. In this work, we prove that
there are such nonholonomic double 3+1 and 2+2 splitting when a relativistic
generalization of statistical thermodynamics of geometric flows is possible
for general classes of cosmological and other type solutions.

A definition of gravitational entropy which would be compatible with time
depending and structure formation cosmological processes need to be valid
for very general classes of solutions with non--stationary and/or
non--vacuum spacetimes. In order to prove that our approach really provides
such a possibility, we shall apply the so-called anholonomic frame
deformation method, AFDM, (see a recent review in \cite{afdm}, and
references therein) for constructing generic off--diagonal exact solutions
in various models of gravity theories and geometric flows with commutative
and noncommutative variables \cite{vepc,vnoncjmp,vnoncjmprf}. This method
involves 2+2 nonholonomic fibrations and a geometric techniques which allows
us to integrate systems of partial differential equations, PDEs, with
functional and parametric dependencies on generating and integration
functions and constants depending, in general, on all spacetime coordinates
and with various types of Killing and non--Killing symmetries.

The article is organized as follows:\ In section \ref{s2}, we provide an
introduction into the geometry of double nonholonomic 3+1 and 2+2 fibrations
of Lorentz manifolds. The main geometric and physical objects and the
Einstein equations are written in nonholonomic variables adapted to a
general double splitting.

Section \ref{s3} is devoted to the theory of relativistic Ricci flows with
nonholonomic constraints for 3+1 splitting and auxiliary connections
completely defined by the metric and nonlinear connection structures. The
geometric evolution of the Levi--Civita configurations (with zero
nonholonomically induced torsion) is considered as a special case defined by
nonholonomic constraints. There are considered generalizations of Perelman's
functionals on Lorentz manifolds and derived the equations for the geometric
relativistic evolution.

Section \ref{s4} is a summary of the anholonomic frame deformation method,
AFDM, of constructing generic off--diagonal solutions in GR. There are
considered examples of inhomogeneous and locally anisotropic cosmological
solutions and black hole/ellipsoid solutions with self--consistent
off--diagonal deformations and solitonic interactions.

In section \ref{s5}, it is elaborated a model of statistical thermodynamics
for gravitational 3-d hypersurface configurations and with relativistic
evolution on (associated) Einstein manifolds. We show how to compute
Perelman's thermodynamical values for explicit examples of generic
off--diagonal solutions in GR. It is analyzed how the relativistic
W--entropy thermodynamics can be parameterized in order to model the CET
thermodynamics and standard black hole physics. A general relativistic
thermodynamic model for geometric flows of exact solutions in GR is
elaborated following ideas and methods of relativistic hydrodynamics. Final
remarks and conclusions are presented in section \ref{s6}.

\section{Double 2+2 and 3+1 Nonholonomic Fibrations in GR}

\label{s2} We provide an introduction into the geometry double 2+2 and 3+1
fibrations of Lorentz spacetime manifolds defined as follow in this section.
The 2+2 splitting with nonholonomic deformations of the local frame and
linear connection structures allowed to decouple the Einstein equations in
general form and to construct exact solutions with generic off--diagonal
metrics depending on all space coordinates. Such nonholonomic 2+2 methods
were applied for deformation and A--brane quantization of theories of
gravity \cite{afdm,vepc,vnoncjmp}. The 3+1 decomposition was introduced in
GR with the aim to elaborate canonical approaches, for instance, to
perturbative quantum gravity, relativistic thermodynamics etc. (for review
of results see \cite{misner}). Working with nonholonomic distributions
adapted to a conventional double 3+1 and 2+2 splitting, we can formulate an
unified geometric approach to general relativistic Ricci flow and
thermodynamical theories and apply the geometric methods for generating
exact solutions of fundamental evolution and/or gravitational and matter
fields equations.

\subsection{Nonholonomic 2+2 splitting and nonlinear connections in GR}

We consider a spacetime $\ _{1}^{3}V$ \ in GR as a 4--d real smooth, i.e. $%
\mathcal{C}^{\infty }$, Lorentz manifold $(V,\mathbf{g})$ determined by a
pseudo--Riemannian metric $\mathbf{g}$ of signature $(+,+,$ $+,-).$\footnote{%
The left labels $3$ and $1$ for $V$ will be used always when it is important
to state that it is considered a $3+1$ splitting with three space like and
one time like local coordinates.} It is assumed that it is possible to
divide continuously over $V$ each light cone of the metric $\mathbf{g}$ into
past and future paths. This means that $\ _{1}^{3}V$ is time orientable. The
tangent bundle of $V_{1}^{3}$ is defined as the union of all tangent spaces $%
T_{u}(\ _{1}^{3}V)$ is considered for all points $u\in \ _{1}^{3}V,$ i.e. $%
T(\ _{1}^{3}V):=\bigcup_{u}T_{u}(\ _{1}^{3}V)$ $.$ The typical fibers of $%
T(\ _{1}^{3}V)$ are Minkowski spaces with pseudo-Euclidean signature. The
dual bundle (equivalently, the cotangent bundle) of $T(\ _{1}^{3}V)$ $\ $is
denoted $T^{\ast }(\ _{1}^{3}V).$ We shall write $TV$ for the space of
smooth vector fields and, respectively, $T^{\ast }V$ for the space of
1--forms on a real 4-d spacetime manifold $V$ omitting left labels on
signature if that will not result in ambiguities. In rigorous mathematical
form, we can use the axiomatic approach to GR begining 1996 due to J.
Ehlers, F. A. E. Pirani and A. Schild (the so-called EPS axioms) \cite{eps},
see further developments in \cite{eps1,eps2,eps3}. In order to formulate in
standard form the gravitational field equations in GR, the Einstein
equations, it is used the unique metric compatible and torsionless
Levi--Civita (LC) connection $\nabla .$ The GR theory was reformulated in
various type of tetradic, spinor etc. variables and corresponding connection
structures introduced with various purposes, for instance, to study
solutions of the Einstein--Yang--Mills--Higgs--Dirac, EYMHD, systems \cite%
{veym1,veym2}. In so--called nonholonomic variables with and auxiliary
canonical distinguished connection structure $\widehat{\mathbf{D}}$ (see
below the definition given in formula (\ref{lcconcdcon})), the EYMHD
equations can be decoupled and integrated in very general forms with
solutions depending, in general, on all spacetime coordinates. Fortunately,
the EPS axiomatic can be generalized for spacetimes enabled with
nonholonomic distributions and fibrations (for GR and various MGTs, see \cite%
{veps1,veps2}). This provides a rigorous mathematical background for
elaborating theories of general relativistic geometric evolution with
nonholonomic variables on Lorentz manifolds which is the main subject for
study in this work.

A nonholonomic 2+2 splitting of $\ _{1}^{3}V$ is determined by a
nonholonomic distribution into local 2-d horizontal, h, and 2-d vertical, 2,
subspaces (the local subspaces are of different signature). This defines a
nonlinear connection (N--connection) structure,
\begin{equation}
\mathbf{N}:T\mathbf{V}=h\mathbf{V}\oplus v\mathbf{V,}  \label{whitney}
\end{equation}%
where $\mathbf{\oplus }$ is the Whitney sum and $h\mathbf{V}$ and $v\mathbf{V%
}$ are conventional horizontal, h, and vertical, v, subspaces. In our
approach, boldface symbols are used in order to emphasize that certain
spaces and/or geometric objects are for spaces endowed with N--connection
structure. For simplicity, we shall write $\mathbf{V}=(\ _{1}^{3}V,\mathbf{N}%
)$ for a Lorentz manifold with a $h$--$v$--decomposition (\ref{whitney}).
This is an example of a nonholonomic manifold consisting, in our case, from
a pseudo--Riemannian manifold and a nonholonomic (equivalently, anholonomic,
or non--integrable) distribution $\mathbf{N.}$\footnote{%
In this work, we use the following conventions: Local coordinates for a 2+2
splitting are denoted $u^{\mu }=(x^{i},y^{a}),$ (in brief, we shall write $%
u=(x,y)$), where indices run respectively values of type $i,j,...=1,2$ and $%
a,b,...=3,4.$ The small Greek indices run values $\alpha ,\beta ,...=1,2,3,4$
considering that $u^{4}=y^{4}=t$ is a time like coordinate. An arbitrary
local basis will be denoted by $e^{\alpha }=(e^{i},e^{a})$ and the
corresponding dual one, co-basis, is $e_{\beta }=(e_{j},e_{b}).$ There are
always nontrivial frame transforms to corresponding coordinate bases, $%
\partial _{\alpha ^{\prime }}=(\partial _{i^{\prime }},\partial _{a^{\prime
}})$ [for instance, $\partial _{i^{\prime }}=\partial /\partial x^{i^{\prime
}}],$ and cobasis $du^{\alpha ^{\prime }}=(dx^{i^{\prime }},dy^{a^{\prime
}}),$ when $e_{\beta }=A_{\beta }^{\ \beta ^{\prime }}\partial _{\beta
^{\prime }}$ and $e^{\alpha }=A_{\ \alpha ^{\prime }}^{\alpha }(u)du^{\alpha
^{\prime }}$ are arbitrary frame (vierbein) transforms. We shall use also
various types of primed, underlined indices etc. The Einstein summation rule
on repeating up--low indices will be applied if the contrary will be not
stated.} In local form, a N--connection is stated by a set of coefficients $%
N_{i}^{a}(u)$ when $\mathbf{N}=N_{i}^{a}dx^{i}\otimes \partial _{a},$ where $%
\partial _{a}=\partial /\partial y^{a}.$

There are structures of N--adapted local bases, $\mathbf{e}_{\nu }=(\mathbf{e%
}_{i},e_{a}),$ and cobases, $\mathbf{e}^{\mu }=(e^{i},\mathbf{e}^{a}),$ when
\begin{eqnarray}
\mathbf{e}_{i} &=&\partial /\partial x^{i}-\ N_{i}^{a}(u)\partial /\partial
y^{a},\ e_{a}=\partial _{a},  \label{nader} \\
e^{i} &=&dx^{i},\qquad \qquad \mathbf{e}^{a}=dy^{a}+\ N_{i}^{a}(u)dx^{i}.
\label{nadif}
\end{eqnarray}%
In general, N--adapted frames are nonholonomic because a frame basis $%
\mathbf{e}_{\nu }=(\mathbf{e}_{i},e_{a})$ satisfies the relations
\begin{equation}
\lbrack \mathbf{e}_{\alpha },\mathbf{e}_{\beta }]=\mathbf{e}_{\alpha }%
\mathbf{e}_{\beta }-\mathbf{e}_{\beta }\mathbf{e}_{\alpha }=W_{\alpha \beta
}^{\gamma }\mathbf{e}_{\gamma },  \label{anhcoef}
\end{equation}%
with nontrivial anholonomy coefficients $W_{ia}^{b}=\partial
_{a}N_{i}^{b},W_{ji}^{a}=\Omega _{ij}^{a}=\mathbf{e}_{j}\left(
N_{i}^{a}\right) -\mathbf{e}_{i}(N_{j}^{a}).$ We obtain holonomic
(integrable) configurations if and only if $W_{\alpha \beta }^{\gamma }=0.$%
\footnote{%
We can elaborate a N--adapted covariant and/or local differential and
integral calculus and a corresponding variational formalism in GR using the
N--elongated operators (\ref{nader}) and (\ref{nadif}). The geometric
constructions are performed for distinguished objects, in brief, d--objects
with coefficients determined with respect to N--adapted (co) frames and
their tensor products. A vector $Y(u)\in T\mathbf{V}$ can be parameterized
as a d--vector, $\mathbf{Y}=$ $\mathbf{Y}^{\alpha }\mathbf{e}_{\alpha }=%
\mathbf{Y}^{i}\mathbf{e}_{i}+\mathbf{Y}^{a}e_{a},$ or $\mathbf{Y}=(hY,vY),$
with $hY=\{\mathbf{Y}^{i}\}$ and $vY=\{\mathbf{Y}^{a}\}.$ Similarly, we can
determine and compute the coefficients of d--tensors, N--adapted
differential forms, d--connections, d--spinors etc. All fundamental
geometric and physical equations can be re--written equivalently in
N-adapted frames, see details in \cite{afdm}.}

On any nonholonomic spacetime $\mathbf{V,}$ we can consider covariant
derivatives determined by affine (linear) connections which are adapted to
the N--connection splitting. A \textit{distinguished connection,
d--connection, }\ is a linear connection $\mathbf{D}=(h\mathbf{D},v\mathbf{D}%
)$ which preserves under parallel transport the splitting (\ref{whitney}).
In general, a linear connection $D$ is not adapted to a prescribed $h$-$v$%
--decomposition, i.e. it is not a d--connection (we do not use a boldface
symbols for not N--adapted connections).

For any d--connection $\mathbf{D}$ and using any d--vectors $\mathbf{X,Y\in }%
T\mathbf{V,}$ we can define and compute in standard form the d--torsion, $%
\mathbf{T,}$ the nonmetricity, $\mathbf{Q},$ and the d--curvature, $\mathbf{R%
},$ tensors
\begin{eqnarray*}
&&\mathbf{T}(\mathbf{X,Y}):=\mathbf{D}_{\mathbf{X}}\mathbf{Y}-\mathbf{D}_{%
\mathbf{Y}}\mathbf{X}-[\mathbf{X,Y}],\mathbf{Q}(\mathbf{X}):=\mathbf{D}_{%
\mathbf{X}}\mathbf{g,} \\
&&\mathbf{R}(\mathbf{X,Y}):=\mathbf{D}_{\mathbf{X}}\mathbf{D}_{\mathbf{Y}}-%
\mathbf{D}_{\mathbf{Y}}\mathbf{D}_{\mathbf{X}}-\mathbf{D}_{\mathbf{[X,Y]}}.
\end{eqnarray*}
Any d--connection $\mathbf{D}$ acts as an operator of covariant derivative, $%
\mathbf{D}_{\mathbf{X}}\mathbf{Y}$, for a d--vector $\mathbf{Y}$ in the
direction of a d--vector $\mathbf{X}.$ We omit boldface symbols and consider
similar formulas for a linear connection which is not a N--connection.

We can compute in N--adapted form (with respect to (\ref{nader}) and (\ref%
{nadif})) the coefficients of any d--connection $\mathbf{D}=\{\mathbf{\Gamma
}_{\ \alpha \beta }^{\gamma
}=(L_{jk}^{i},L_{bk}^{a},C_{jc}^{i},C_{bc}^{a})\}.$ The N--adapted
coefficients of torsion, nonmetricity and curvature d--tensors are
correspondingly labeled using $h$- and $v$--indices,
\begin{eqnarray}
\mathcal{T} &=&\{\mathbf{T}_{\ \alpha \beta }^{\gamma }=\left( T_{\
jk}^{i},T_{\ ja}^{i},T_{\ ji}^{a},T_{\ bi}^{a},T_{\ bc}^{a}\right) \},%
\mathcal{Q}=\mathbf{\{Q}_{\ \alpha \beta }^{\gamma }\},  \label{rnmc} \\
\mathcal{R} &\mathbf{=}&\mathbf{\{R}_{\ \beta \gamma \delta }^{\alpha }%
\mathbf{=}\left( R_{\ hjk}^{i}\mathbf{,}R_{\ bjk}^{a}\mathbf{,}R_{\ hja}^{i}%
\mathbf{,}R_{\ bja}^{c},R_{\ hba}^{i},R_{\ bea}^{c}\right) \}.  \notag
\end{eqnarray}%
The coefficients formulas for such values can be obtained using $\mathbf{%
\Gamma }_{\ \alpha \beta }^{\gamma }$ e determined for the h--v--components
of $\mathbf{D}_{\mathbf{e}_{\alpha }}\mathbf{e}_{\beta }:=$ $\mathbf{D}%
_{\alpha }\mathbf{e}_{\beta }$ using $\mathbf{X}=\mathbf{e}_{\alpha }$ and $%
\mathbf{Y}=\mathbf{e}_{\beta }.$

Any metric tensor $\mathbf{g}$ on $\mathbf{V}$ can be written as a d--tensor
(d--metric), $\mathbf{g}=(h\mathbf{g},v\mathbf{g}),$ i.e.
\begin{equation}
\mathbf{g}=g_{\alpha }(u)\mathbf{e}^{\alpha }\otimes \mathbf{e}^{\beta
}=g_{i}(x)dx^{i}\otimes dx^{i}+g_{a}(x,y)\mathbf{e}^{a}\otimes \mathbf{e}%
^{a},  \label{dm}
\end{equation}%
for a N--adapted $\mathbf{e}^{\mu }=(e^{i},\mathbf{e}^{a})$ (\ref{nadif}).
With respect to a dual local coordinate basis $du^{\alpha },$ the same
metric field is expressed
\begin{equation}
\mathbf{g}=\underline{g}_{\alpha \beta }du^{\alpha }\otimes du^{\beta },%
\mbox{\ where \ }\underline{g}_{\alpha \beta }=\left[
\begin{array}{cc}
g_{ij}+N_{i}^{a}N_{j}^{b}g_{ab} & N_{j}^{e}g_{ae} \\
N_{i}^{e}g_{be} & g_{ab}%
\end{array}%
\right] .  \label{ofdans}
\end{equation}%
Using frame transforms (in general, not N--adapted), we can trasform any
metric into a d--metric (\ref{dm}) or in an off-diagonal form with
N--coefficients.

For any metric field $\mathbf{g}$ on a $\mathbf{V}=(\ _{1}^{3}V,\mathbf{N}),$
\ there are two 'preferred' linear connection structures. The first one is
the well--known Levi--Civita connection, $\nabla ,$ and the canonical
d--connection, $\widehat{\mathbf{D}}.$ Such geometric objects are defined
following correponding geometric conditions: {%
\begin{equation}
\mathbf{g}\rightarrow \left\{
\begin{array}{ccccc}
\nabla : &  & \nabla \mathbf{g}=0;\ ^{\nabla }\mathcal{T}=0, &  &
\mbox{ the
Levi--Civita connection;} \\
\widehat{\mathbf{D}}: &  & \widehat{\mathbf{D}}\ \mathbf{g}=0;\ h\widehat{%
\mathcal{T}}=0,\ v\widehat{\mathcal{T}}=0, &  &
\mbox{ the canonical
d--connection.}%
\end{array}%
\right.  \label{lcconcdcon}
\end{equation}%
} In these formulas, $h\widehat{\mathcal{T}}$ and $\ v\widehat{\mathcal{T}}$
are respective torsions on conventional h- and v--subspaces. We note that
there are nonzero torsion components, $hv\widehat{\mathcal{T}},$ with
nonzero mixed indices with respect to a N-adapted basis (\ref{nader}) and/or
(\ref{nadif}). Nevertheless, this torsion field $\widehat{\mathcal{T}}$ is
completely defined by metric field following a parameterization (\ref{ofdans}%
) with $(h\mathbf{g},v\mathbf{g;N}).$

All geometric constructions on $\mathbf{V,}$ can be performed equivalently
using $\nabla $ and/or $\widehat{\mathbf{D}}$ and related via the canonical
distorting relation
\begin{equation}
\widehat{\mathbf{D}}\mathbf{[g,N]}=\nabla \lbrack \mathbf{g}]+\widehat{%
\mathbf{Z}}[\widehat{\mathcal{T}}(\mathbf{g,N)}].  \label{distr}
\end{equation}%
By squared brackets $[...],$ we state a functional dependence when both
linear connections $\nabla $ and $\widehat{\mathbf{D}}$ and the distorting
tensor $\widehat{\mathbf{Z}}$ are uniquely determined by data the $(\mathbf{%
g,N)}$ and an algebraic combination of coefficients of torsion $\widehat{%
\mathcal{T}}(\mathbf{g,N).}$

The Ricci tensors of $\widehat{\mathbf{D}}$ and $\nabla $ are defined and
computed in the standard way and denoted, respectively, as $\ \widehat{%
\mathcal{R}}ic=\{\widehat{\mathbf{R}}_{\ \beta \gamma }:=\widehat{\mathbf{R}}%
_{\ \alpha \beta \gamma }^{\gamma }\}$ and $Ric=\{R_{\ \beta \gamma }:=R_{\
\alpha \beta \gamma }^{\gamma }\}.$ The N--adapted coefficients for $%
\widehat{\mathbf{D}}$ and corresponding torsion, $\widehat{\mathbf{T}}_{\
\alpha \beta }^{\gamma },$ Ricci d--tensor, $\widehat{\mathbf{R}}_{\ \beta
\gamma },$ and Einstein d--tensor, $\widehat{\mathbf{E}}_{\ \beta \gamma },$
are computed in \cite{afdm,vepc}. Any (pseudo) Riemannian geometry can be
equivalently described by both geometric data $\left( \mathbf{g,\nabla }%
\right) $ and $(\mathbf{g,N,}\widehat{\mathbf{D}}),$ when the canonical
distortion relations $\widehat{\mathcal{R}}=\ ^{\nabla }\mathcal{R+}\
^{\nabla }\mathcal{Z}$ and $\widehat{\mathcal{R}}ic=Ric+\widehat{\mathcal{Z}}%
ic,$ with respective distortion d-tensors $\ ^{\nabla }\mathcal{Z}$ and $%
\widehat{\mathcal{Z}}ic,$ are computed for the canonical distortion
relations $\widehat{\mathbf{D}}=\nabla +\widehat{\mathbf{Z}}$

By unique distortion relations (compted by introducing (\ref{distr}) into \ (%
\ref{rnmc}) and re-grouping the terms with $\nabla $ and $\widehat{\mathbf{D}%
}),$ we can relate, for instance, $\widehat{\mathbf{R}}_{\ \beta \gamma }$
to $R_{\ \beta \gamma },$%
\begin{equation*}
\widehat{\mathbf{R}}_{\ \beta \gamma }\mathbf{[g,N]}=R_{\ \beta \gamma }%
\mathbf{[g,N]}+\widehat{\mathbf{Z}}_{\ \beta \gamma }\mathbf{[g,N].}
\end{equation*}
We note that the Ricci d-tensor $\widehat{\mathcal{R}}ic$ is not symmetric, $%
\widehat{\mathbf{R}}_{\alpha \beta }\neq \widehat{\mathbf{R}}_{\beta \alpha
},$ and characterized by four subsets of $h$-$v$ N-adapted coefficients,
\begin{equation}
\widehat{\mathbf{R}}_{\alpha \beta }=\{\widehat{R}_{ij}:=\widehat{R}_{\
ijk}^{k},\ \widehat{R}_{ia}:=-\widehat{R}_{\ ika}^{k},\ \widehat{R}_{ai}:=%
\widehat{R}_{\ aib}^{b},\ \widehat{R}_{ab}:=\widehat{R}_{\ abc}^{c}\}.
\label{driccic}
\end{equation}%
It is possible to compute the scalar of canonical d--curvature, $\ \widehat{%
\mathbf{R}}:=\mathbf{g}^{\alpha \beta }\widehat{\mathbf{R}}_{\alpha \beta
}=g^{ij}\widehat{R}_{ij}+g^{ab}\widehat{R}_{ab}.$ This geometric object is
different from the LC-scalar curvature, $\ R:=\mathbf{g}^{\alpha \beta
}R_{\alpha \beta }.$

The Einstein equations in GR are written in standard form,
\begin{equation}
R_{\alpha \beta }-\frac{1}{2}g_{\alpha \beta }R=\varkappa \ ^{m}T_{\alpha
\beta },  \label{einsteq}
\end{equation}%
using the Ricci tensor $R_{\alpha \beta }$ and scalar $R$ are taken for the
Levi--Civita, LC, connection $\nabla $ of $g_{\alpha \beta }.$ In these
formulas, $\ ^{m}T_{\alpha \beta }$ is the energy--momentum tensor of matter
fields $\ ^{A}\varphi $ determined by a general Lagrangian $\ ^{m}\mathcal{L}%
(\mathbf{g,}\nabla ,\ \ ^{A}\varphi )$ and $\varkappa $ is the gravitational
coupling constant for GR.\footnote{%
We use abstract left labels $A$ and $m$ in order to distinguish the values
from similar notations, for instance, $\widehat{\mathbf{T}}_{\ \alpha \beta
}^{\gamma }.$} The gravitational field equations in GR can be rewritten
equivalently using the canonical d--connection \cite{afdm,vepc},%
\begin{eqnarray}
\widehat{\mathbf{R}}_{\alpha \beta } &=&\widehat{\mathbf{\Upsilon }}_{\alpha
\beta },  \label{deinst} \\
\widehat{\mathbf{T}}_{\ \alpha \beta }^{\gamma } &=&0.  \label{lccond}
\end{eqnarray}%
In these formulas, the effective matter fields source $\mathbf{\Upsilon }%
_{\mu \nu }$ is constructed via a N--adapted variational calculus with
respect to (\ref{nadif}) for $\ ^{m}\mathcal{L}(\mathbf{g,}\widehat{\mathbf{D%
}},\ \ ^{A}\varphi )$ in such a form that
\begin{equation*}
\widehat{\mathbf{\Upsilon }}_{\mu \nu }=\varkappa (\ ^{m}\widehat{\mathbf{T}}%
_{\mu \nu }-\frac{1}{2}\mathbf{g}_{\mu \nu }\ ^{m}\widehat{\mathbf{T}}%
)\rightarrow \varkappa (\ ^{m}T_{\mu \nu }-\frac{1}{2}\mathbf{g}_{\mu \nu }\
^{m}T)
\end{equation*}
for [coefficients of $\widehat{\mathbf{D}}]$ $\rightarrow $ [coefficients of
$\nabla $] even, in general, $\widehat{\mathbf{D}}\neq \nabla .$ In these
formulas, $\ ^{m}\widehat{\mathbf{T}}=\mathbf{g}^{\mu \nu }\ ^{m}\widehat{%
\mathbf{T}}_{\mu \nu }$ for
\begin{equation}
\ ^{m}\widehat{\mathbf{T}}_{\alpha \beta }:=-\frac{2}{\sqrt{|\mathbf{g}_{\mu
\nu }|}}\frac{\delta (\sqrt{|\mathbf{g}_{\mu \nu }|}\ \ ^{m}\mathcal{L})}{%
\delta \mathbf{g}^{\alpha \beta }}.  \label{ematter}
\end{equation}

The canonical d--connection $\widehat{\mathbf{D}}$ has a very important role
in our approach. With respect to N--adapted frames of reference, it allows
to decouple in general form the gravitational and matter field equations in
the form (\ref{deinst}) with (\ref{ematter}).\footnote{%
Such a decoupling is possible various types of geometric flow equations, see
details in sections \ref{s4} and \ref{s5}.} We can integrate nonholonomic
deformations of the Einstein equations in very general form and construct
exact solutions parameterized by generic off--diagonal metrics depending on
all spacetime coordinates via respective classes of generating functions and
integration functions and constants. Having constructed certain general
classes of solutions, we can impose at the end the LC--conditions (\ref%
{lccond}) and extract LC--configurations $\widehat{\mathbf{D}}_{\mid
\widehat{\mathcal{T}}=0}=\nabla .$ This allows, for instance, to construct
new classes of generic off--diagonal solutions of (\ref{einsteq}) in GR and
various MGTs. We note that to find nontrivial off--diagonal solutions is
important to impose the LC--conditions (\ref{lccond}) after a class of
solutions of (\ref{deinst}) for $\widehat{\mathbf{D}}$ are constructed in
general form. If we work only with $\nabla ,$ we are not able to decouple in
general form the Einstein equations.

\subsection{3+1 decompositons adapted to nonholonomic 2+2 spliting}

We foliate a 4-d Lorentzian nonholonomic manifold $\mathbf{V}=(\ _{1}^{3}V,%
\mathbf{N})$ enabled with a pseudo--Rieman\-nian metric $\mathbf{g}%
=\{g_{\alpha \beta }\}$ of signature $(+++-)$ into a family of
non-intersecting spacelike 3-d hypersurfaces $\Xi _{t}$ parameterized by a
scalar field, i.e. "time function", $t(u^{\alpha }),$ as described as
follows. \ Such a 3+1 spacetime decompositon is necessary for elaborating
various thermodynamic and flow models when a conventional splitting into
time and space like coordinates is important for definition of physical
important values (like entropy, effective energy etc.) and fundamental
geometric evolution equations. We have to generalize the well--known
geometric 3+1 formalism \cite{misner} to the case of spacetimes enables with
nontrivial N--connection structure \cite{afdm,vepc}.

A hypersurface $\Xi \subset \mathbf{V}$ is considered as an one--to--one
image of a 3--d manifold $\ _{\shortmid }\Xi .$ This image is given by an
embedding $\Xi =\zeta (\ _{\shortmid }\Xi )$ constructed as an homeomorphism
with both continuous maps $\zeta $ and $\zeta ^{-1}.$ This guarantees that $%
\Xi $ does not intersect itself. We shall use a left "up" or "low" label by
a vertical bar "$\ _{\shortmid }\ $" in order to emphasize that a manifold
is 3-d, or certain geometric objects refer to 3--d manifolds /
hypersurfaces. \ Locally, a hypersurface is considered as the set of points
for which a scalar field $t$ on $\mathbf{V}$ is constant, for instance, i.e.
$t(p)=0,\forall $ $p\in \Xi .$ We assume that $t$ spans $\mathbb{R}$ and $%
\Xi $ is a connected submanifold of $\mathbf{V}$ with topology $\mathbb{R}%
^{3}.$ The local coordinates for a 3+1 splitting are labeled $u^{\alpha
}=(x^{\grave{\imath}},t),$ where indices $\alpha ,\beta ,...=1,2,3,4$ and $%
\grave{\imath},\grave{j},...=1,2,3.$ In brief, we shall write $u=(\breve{u}%
,t).$ The mapping $\zeta $ "carries along" curves/ vectors in $\ _{\shortmid
}\Xi $ to curves / vectors in $\mathbf{V,}$ for $\zeta :(x^{\grave{\imath}%
})\longrightarrow (x^{\grave{\imath}},0).$ This defines respective local
bases $\partial _{\grave{\imath}}:=\partial /\partial x^{\grave{\imath}}\in T%
\mathbf{(\mathbf{\ }_{\shortmid }\Xi )}$ and $\partial _{\alpha }:=\partial
/\partial u^{\alpha }\in T\mathbf{V.}$ Correpsondingly, the coefficients of
3--vectors and 4--vectors are expressed $\ _{\shortmid }\mathbf{a}=a^{\grave{%
\imath}}\partial _{\grave{\imath}}$ and $\mathbf{\ a}=a^{\alpha }\partial
_{\alpha }$ (on convenience, we shall use also capital letters, for
instance, $\mathbf{\ }_{\shortmid }\mathbf{A}=A^{\grave{\imath}}\partial _{%
\grave{\imath}}$ and $\mathbf{\ A}=A^{\alpha }\partial _{\alpha }).$ For
dual forms to vectors, 1--forms, we use respective dual bases $\mathbf{d}x^{%
\grave{\imath}}\in T^{\ast }\mathbf{(\mathbf{\ }_{\shortmid }\Xi )}$ and $%
du^{\alpha }\in T^{\ast }\mathbf{V.}$ We shall write for 1--forms $\
_{\shortmid }\tilde{\mathbf{A}}=A_{\grave{\imath}}\mathbf{d}x^{\grave{\imath}%
}$ and $\mathbf{\ \tilde{A}}=A_{\alpha }du^{\alpha }$ and omit the left/up
label by a tilde $\sim $ (writing $\mathbf{\ }_{\shortmid }\mathbf{A}$ and $%
\mathbf{A)}$ if that will not result in ambiguities.

Using the push--forward mapping%
\begin{equation*}
\begin{array}{cccc}
\zeta _{\ast }: & T_{u}\mathbf{\ }_{\shortmid }\Xi \longrightarrow T_{p}%
\mathbf{V;} &  & \mathbf{\ }_{\shortmid }\mathbf{v}=(v^{\grave{\imath}%
})\longrightarrow \zeta _{\ast }\mathbf{\ }_{\shortmid }\mathbf{v}=(v^{%
\grave{\imath}},0)%
\end{array}%
;
\end{equation*}%
we can transport geometric objects from $\ _{\shortmid }\Xi $ to $\Xi ,$ and
inversely. In dual form, the pull--back mapping acts as\textbf{\ }
\begin{equation*}
\begin{array}{ccccccc}
\zeta ^{\ast }: & T_{p}^{\ast }\mathbf{V} & \longrightarrow & T_{u}^{\ast }%
\mathbf{\ }_{\shortmid }\Xi , &  &  &  \\
& \mathbf{\ }_{\shortmid }\mathbf{\tilde{A}} & \longrightarrow & \zeta
^{\ast }\ _{\shortmid }\mathbf{\tilde{A}:} & T_{u}\ _{\shortmid }\Xi
\longrightarrow \mathbb{R}; &  & \ _{\shortmid }\mathbf{A}\longrightarrow <\
_{\shortmid }\mathbf{\tilde{A}},\zeta _{\ast }\ _{\shortmid }\mathbf{A}>,%
\end{array}%
\end{equation*}%
for $<...>$ denoting the scalar product and $T_{p}^{\ast }\mathbf{V\ni }$ $%
\mathbf{\tilde{A}}=(A_{\grave{\imath}},A_{4})\longrightarrow \zeta ^{\ast }%
\mathbf{\tilde{A}=}(A_{i})\in T_{u}^{\ast }\mathbf{\ }_{\shortmid }\Xi .$ In
this work, we identify $\ _{\shortmid }\Xi $ and $\Xi =\zeta (\ _{\shortmid
}\Xi )$ and write simply a d--vector $\mathbf{\ v}$ instead of $\zeta _{\ast
}(\ _{\shortmid }\mathbf{v).}$ For holonomic configurations, the same maps
and objects are labelled in non-boldface form.

\subsubsection{Induced N--adapted 3-d hypersurface metrics}

We define the first fundamental form (the induced 3--metric) on $\Xi :$%
\begin{equation*}
\begin{array}{ccc}
\mathbf{q}:=\zeta ^{\ast }\ \mathbf{g}, &  & \mbox{ i.e. }\mathbf{q}_{\grave{%
\imath}\grave{j}}:=\mathbf{g}_{\grave{\imath}\grave{j}} \\
\forall (\ ^{1}\mathbf{a,}\ ^{2}\mathbf{a})\in T_{u}\mathbf{\ }\Xi \times
T_{u}\mathbf{\ }\Xi , &  & \ ^{1}\mathbf{a\cdot }\ ^{2}\mathbf{a=g}(\ ^{1}%
\mathbf{a,}\ ^{2}\mathbf{a})=\mathbf{q}(\ ^{1}\mathbf{a,}\ ^{2}\mathbf{a}).%
\end{array}%
\end{equation*}%
The 4--d metric $\mathbf{g}$ is constrained to be a solution of the Einstein
equations (\ref{deinst}) \ written in nonholonomic variables. The
hypersurfaces are classified following the types of induced 3--metric (for a
nontrivial N--connection, this can be represented as an induced d--metric)%
\begin{equation*}
\left\{
\begin{array}{ccc}
\mbox{ spacelike\ }\Xi , & \mathbf{q} &
\mbox{is positive definite with
signature }(+,+,+); \\
\mbox{ timelike\ }\Xi , & \mathbf{q} & \mbox{is Lorentzian with signature }%
(+,+,-); \\
\mbox{ null\ }\Xi , & \mathbf{q} & \mbox{is degenerate with signature }%
(+,+,0).%
\end{array}%
\right.
\end{equation*}%
In this article, we shall work with continuous sets of spacelike
hypersurfaces $\Xi _{t},t\in \mathbb{R},$ covering some finite, or infinite,
regions on $\mathbf{V}$. For simplicity, we use only spacelike hypersurfaces
$\Xi $ (which can be closed and compact if necessary) endowed with
Riemannian 3--metric $\mathbf{q}$ if other conditions will be not stated.

We introduce the concept of unite normal d--vector, $\mathbf{n}$, to a $\Xi $
which is constructed following such a procedure. Such d--vectors can be used
for various models of geometric flow evolution and thermodynamic models. Let
us consider a scalar field $t(u^{\alpha })$ on an open region $U\subset
\mathbf{V}$ such as the level surface is identified to $\Xi .$ We construct
in N--adapted form the gradient 1--form $\mathbf{d}t$ and it dual d-vector $%
\overrightarrow{\mathbf{e}}t=\{\mathbf{e}^{\mu }t=\mathbf{g}^{\mu \nu }%
\mathbf{e}_{\nu }t=\mathbf{g}^{\mu \nu }(\mathbf{d}t)_{\nu }\},$ see
operators (\ref{nader}) and (\ref{nadif}). For any d--vector $\mathbf{v}$
which is tangent to $\Xi ,$ the conditions $<\mathbf{d}t,\mathbf{v}>=0$ and $%
\overrightarrow{\mathbf{e}}t$ allow to define the unique direction normal to
a not null $\Xi .$ Normalizing such a d-vector, we define
\begin{equation}
\mathbf{n:=\pm }\overrightarrow{\mathbf{e}}t/\sqrt{|\overrightarrow{\mathbf{e%
}}t|},\mbox{ where }\left\{
\begin{array}{ccc}
\mathbf{n\cdot n}=-1, &  & \mbox{ for spacelike }\Xi ; \\
\mathbf{n\cdot n}=1, &  & \mbox{ for timelike }\Xi .%
\end{array}%
\right.  \label{ndv}
\end{equation}%
The unit normal vector to supersurfaces, $\mathbf{n}_{\alpha }\propto
\partial _{\alpha }t,$ when $\partial _{\alpha }:=\partial /\partial
u^{\alpha },$ can be constructed a future--directed time--like vector field.
We can use $t$ as a parameter for a congruence of curves $\chi (t)\subset \
_{1}^{3}V$ intersecting $\Sigma _{t},$ when the vector $\mathbf{t}^{\alpha
}:=du^{\alpha }/dt$ is tangent to the curves and $\mathbf{t}^{\alpha
}\partial _{\alpha }t=1.$ For any system of coordinates $u^{\alpha
}=u^{\alpha }(x^{\grave{\imath}},t),$ there are defined the vector $\mathbf{t%
}^{\alpha }:=(\partial u^{\alpha }/\partial t)_{x^{\grave{\imath}}}$ and the
(tangent) vectors $e_{~\grave{\imath}}^{\alpha }:=(\partial u^{\alpha
}/\partial x^{\grave{\imath}})$ and the Lie derivative along $\mathbf{t}%
^{\alpha }$ results in $\pounds _{t}e_{~^{\grave{\imath}}}^{\alpha }=0.$

Any 2+2 splitting $\mathbf{N}:T\mathbf{V}=h\mathbf{V}\oplus v\mathbf{V,}$ (%
\ref{whitney}) induces a N--connection structure on $\Xi ,$ $\ _{\shortmid }%
\mathbf{N}:T\ _{\shortmid }\Xi =h\ _{\shortmid }\Xi \oplus v\ _{\shortmid
}\Xi \mathbf{.}$ In result, any induced 3--metric tensor $\mathbf{q}$ can be
written in N--adapted frames as a d--tensor (d--metric) in the form
\begin{eqnarray}
\mathbf{q} &=&(h\mathbf{q},v\mathbf{q})=q_{\grave{\imath}}(u)\mathbf{e}^{%
\grave{\imath}}\otimes \mathbf{e}^{\grave{\imath}}=q_{i}(x^{k})dx^{i}\otimes
dx^{i}+q_{3}(x^{k},y^{3})\ _{\shortmid }\mathbf{e}^{3}\otimes \ _{\shortmid }%
\mathbf{e}^{3},  \label{3dm} \\
\mathbf{\ }_{\shortmid }\mathbf{e}^{3} &=&du^{3}+\mathbf{\ }_{\shortmid
}N_{i}^{3}(u)dx^{i},  \label{3nel}
\end{eqnarray}%
where $\ _{\shortmid }N_{i}^{3}(u)$ can be identified with $N_{i}^{3}(u)$
choosing common frame and coordinate systems for $\Xi \subset \mathbf{V}.$ \
We can imbed naturaly such a metric into a d--metric (\ref{dm})
reparameterized in a form adapted both to 2+2 and 3+1 nonholonomic
splitting,
\begin{eqnarray}
\mathbf{g} &=&(h\mathbf{g},v\mathbf{g})=\breve{g}_{\grave{\imath}\grave{j}}%
\mathbf{e}^{\grave{\imath}}\otimes \mathbf{e}^{\grave{j}}+g_{4}\mathbf{e}%
^{4}\otimes \mathbf{e}^{4}=q_{\grave{\imath}}(u)\mathbf{e}^{\grave{\imath}%
}\otimes \mathbf{e}^{\grave{\imath}}-\breve{N}^{2}\mathbf{e}^{4}\otimes
\mathbf{e}^{4},  \label{offd3122} \\
\mathbf{\ e}^{3} &=&\ _{\shortmid }\mathbf{e}^{3}=du^{3}+\mathbf{\ }%
_{\shortmid }N_{i}^{3}(u)dx^{i},\mathbf{\ e}^{4}=\delta t=dt+\mathbf{\ }%
N_{i}^{4}(u)dx^{i}  \notag
\end{eqnarray}
Let us explain this construction. \ The\textbf{\ lapse function} $\breve{N}%
(u)>0$ is defined as a positive scalar field which ensues that the unite
d--vector $\mathbf{n}$ is a unite one, see (\ref{ndv}). \ We use an "inverse
hat" in order to distiguish such a symbol $N$ is used traditionally in
literature on GR \cite{misner} but in another turn, the symbol $N_{i}^{a}$
is used traditionally for the N--connection. We write
\begin{equation*}
\mathbf{n:=-}\breve{N}\overrightarrow{\mathbf{e}}t\mbox{ and/or }\underline{%
\mathbf{n}}\mathbf{:=-}\breve{N}\mathbf{d}t,
\end{equation*}%
with $\breve{N}:=1/\sqrt{|\overrightarrow{\mathbf{e}}t\cdot \overrightarrow{%
\mathbf{e}}t|}=1/\sqrt{|<\mathbf{d}t\cdot \overrightarrow{\mathbf{e}}t>|}.$
For geometric constructions, it is convenient to use also the normal
evolution d--vector
\begin{equation}
\mathbf{m:=}\breve{N}\mathbf{n}  \label{nevdv}
\end{equation}%
subjected to the condition $\mathbf{m}\cdot \mathbf{m}=-\breve{N}^{2}.$ This
is justified by the property of the Lie N-adapted derivative that\ $\mathcal{%
L}_{\mathbf{m}}\mathbf{a\in }(h\Xi \oplus v\Xi )_{t},\forall \mathbf{a}\in
T\Xi _{t}.$ For a $3+1$ spacetime splitting, we consider also the \textbf{%
shift} functions (a 3--vector $\breve{N}^{\grave{\imath}}(u),$ or a
d--vector $\mathbf{\breve{N}}^{\grave{\imath}}(u)$). It is useful to define
the unit normal $\breve{n}^{\alpha }$ to the hypersurfaces when $\breve{n}%
_{\alpha }=-\breve{N}\partial _{\alpha }t$ and $\breve{n}_{\alpha }e_{~^{%
\breve{\imath}}}^{\alpha }=0.$ In N--adaptd form, we can consider that $%
\breve{n}_{\alpha }$ is a normalized version of $\mathbf{n}$ used in (\ref%
{nevdv}). This allows us to consider the decompositions
\begin{equation*}
\mathbf{t}^{\alpha }=\breve{N}^{\grave{\imath}}e_{~^{\grave{\imath}%
}}^{\alpha }+\breve{N}\breve{n}^{\alpha }
\end{equation*}%
and
\begin{equation*}
du^{\alpha }=e_{~^{\grave{\imath}}}^{\alpha }dx^{\grave{\imath}}+\mathbf{t}%
^{\alpha }dt=(dx^{\grave{\imath}}+\breve{N}^{\grave{\imath}}dt)e_{~^{\grave{%
\imath}}}^{\alpha }+(\breve{N}dt)\breve{n}^{\alpha }.
\end{equation*}%
Using the quadratic line element $ds^{2}=g_{\alpha \beta }du^{\alpha
}du^{\beta }$ of a metric tensor $\mathbf{g,}$ we can chose such fram
transforms when $\breve{g}_{\grave{\imath}\grave{j}}=q_{\grave{\imath}\grave{%
j}}=$ $g_{\alpha \beta }e_{~\grave{\imath}}^{\alpha }e_{~\grave{j}}^{\beta }$
is the induced metric on $\Xi _{t}.$ For the determinants of 4-d and 3-d
metrics parameterized in above mentioned form, we compute $\sqrt{|g|}=\breve{%
N}\sqrt{|\breve{g}|}=\breve{N}\sqrt{|q|}.$ Using coordinates $(x^{\grave{%
\imath}},t),$ the time partial derivatives are $\pounds _{t}q=\partial
_{t}q=q^{\ast }$ and the spacial derivatives are $q_{,^{\grave{\imath}%
}}:=e_{~^{^{\grave{\imath}}}}^{\alpha }$ $q_{,\alpha }.$

\subsubsection{Induced 3-d hypersurface 'preferred' linear connections}

There are two induced linear connections completely determined by an induced
3--d hypersurface metric $\mathbf{q,}$ {\
\begin{equation}
\mathbf{q}\rightarrow \left\{
\begin{array}{cccc}
\ _{\shortmid }\nabla : &  & \mathbf{\ }_{\shortmid }\nabla \mathbf{q}=0;\
\mathbf{\ }_{\shortmid }^{\nabla }\mathcal{T}=0, & \mbox{LC--connection} \\
\ _{\shortmid }\widehat{\mathbf{D}}: &  & \ _{\shortmid }\widehat{\mathbf{D}}%
\ \mathbf{q}=0;\ h\ _{\shortmid }\widehat{\mathcal{T}}=0,\ v\ _{\shortmid }%
\widehat{\mathcal{T}}=0. & \mbox{canonical d--connection.}%
\end{array}%
\right.  \label{hlccdc}
\end{equation}%
Such formulas are induced from the 4--d similar ones, see }(\ref{lcconcdcon}%
). Both linear connections, $\ _{\shortmid }\nabla $ and $\ _{\shortmid }%
\widehat{\mathbf{D}},$ are involved in a distortion relation,
\begin{equation}
\mathbf{\ }_{\shortmid }\widehat{\mathbf{D}}\mathbf{[q,\mathbf{\ }%
_{\shortmid }N]}=\mathbf{\ }_{\shortmid }\nabla \lbrack \mathbf{q}]+\mathbf{%
\ }_{\shortmid }\widehat{\mathbf{Z}}[\mathbf{\ }_{\shortmid }\widehat{%
\mathcal{T}}(q\mathbf{,\mathbf{\ }_{\shortmid }N)}],  \label{dist3}
\end{equation}%
induced by (\ref{distr}).

There are two classes of trivial or nontrivial intrinsic torsions,
nonmetricity and curvature fields for any data $(\Xi ,\mathbf{q,\mathbf{\ }%
_{\shortmid }N}),$ defined by corresponding hypersurface linear connections
when for any $\mathbf{a,b\in }T\Xi $
\begin{eqnarray*}
\ _{\shortmid }T(\mathbf{a,b}):= &&\ _{\shortmid }\nabla _{\mathbf{a}}%
\mathbf{Y}-\ _{\shortmid }\nabla _{\mathbf{Y}}\mathbf{a}-[\mathbf{a,Y}]=0,%
\mathbf{\ }_{\shortmid }Q(\mathbf{a}):=\ _{\shortmid }\nabla _{\mathbf{a}}%
\mathbf{q=0,} \\
\ _{\shortmid }R(\mathbf{a,b}):= &&\ _{\shortmid }\nabla _{\mathbf{a}}\
_{\shortmid }\nabla _{\mathbf{b}}-\ _{\shortmid }\nabla _{\mathbf{b}}\
_{\shortmid }\nabla _{\mathbf{a}}-\ _{\shortmid }\nabla _{\mathbf{[a,b]}},
\end{eqnarray*}%
and%
\begin{eqnarray*}
\mathbf{\ }_{\shortmid }\widehat{\mathbf{T}}(\mathbf{a,b}):= &&\ _{\shortmid
}\mathbf{D}_{\mathbf{a}}\mathbf{b}-\ _{\shortmid }\mathbf{D}_{\mathbf{Y}}%
\mathbf{a}-[\mathbf{a,b}]=0,\mathbf{\ }_{\shortmid }\mathbf{Q}(\mathbf{a}):=%
\mathbf{\ }_{\shortmid }\mathbf{D}_{\mathbf{a}}\mathbf{q=0,} \\
\mathbf{\ }_{\shortmid }\widehat{\mathbf{R}}(\mathbf{a,b}):= &&\ _{\shortmid
}\widehat{\mathbf{D}}_{\mathbf{a}}\ _{\shortmid }\widehat{\mathbf{D}}_{%
\mathbf{b}}-\ _{\shortmid }\widehat{\mathbf{D}}_{\mathbf{b}}\ _{\shortmid }%
\widehat{\mathbf{D}}_{\mathbf{a}}-\ _{\shortmid }\widehat{\mathbf{D}}_{%
\mathbf{[a,b]}}.
\end{eqnarray*}

We can compute the N--adapted coefficient formulas for nonholonomically
induced torsion structure $\mathbf{\mathbf{\ }_{\shortmid }}\widehat{\mathbf{%
T}}=\{\ _{\shortmid }\widehat{\mathbf{T}}_{\ \grave{j}\grave{k}}^{\grave{%
\imath}}\},$ \ determined by $\ _{\shortmid }\mathbf{D,}$ and for the
Riemannian tensors $\ \mathbf{\ }_{\shortmid }R\mathbf{=}\mathbf{\{\ \
_{\shortmid }}R_{\ \grave{j}\grave{k}\grave{l}}^{\grave{\imath}}\}$ and $%
\mathbf{\mathbf{\ }_{\shortmid }}\widehat{\mathbf{R}}=\mathbf{\{\mathbf{\ }%
_{\shortmid }}\widehat{\mathbf{R}}_{\ \grave{j}\grave{k}\grave{l}}^{\grave{%
\imath}}\},$ determined respectively by $\ _{\shortmid }\nabla $ and $\
_{\shortmid }\mathbf{D.}$ Using 3--d variants of coefficient formulas, we
can compute the N--adapted coefficients of the Ricci d--tensor, $\
_{\shortmid }\widehat{\mathbf{R}}_{\ \grave{j}\grave{k}},$ and the Einstein
d--tensor, $\ _{\shortmid }\widehat{\mathbf{E}}_{\ \grave{j}\grave{k}}.$
Contracting indices, we obtain the Gaussian curvature, $\ _{\shortmid }R=q^{%
\grave{j}\grave{k}}\ _{\shortmid }R_{\grave{j}\grave{k}}$, and the Gaussian
canonical curvature, $\ _{\shortmid }^{s}R=q^{\grave{j}\grave{k}}\
_{\shortmid }\widehat{\mathbf{R}}_{\grave{j}\grave{k}},$ of $(\Xi ,\mathbf{%
q,\ _{\shortmid }N}).$ Such geometric objects do not depend on the type of
embedding of the nonholonomic manifold $(\Xi ,\mathbf{q,\ _{\shortmid }N})$
in $(\mathbf{V},\mathbf{g,N}).$

There are another types of curvatures which describe the (non) holonomic
bending of $\ \Xi ,$ i.e. depend on embedding. Such geometric values are
considered for any type of 3+1 splitting and constructed using N--adapted
Weingarten maps (shape operator). In our approach, these are N--adapted
endomorphisms%
\begin{equation*}
\begin{array}{ccccccccc}
\chi : & T_{u}\Xi & \rightarrow & T_{u}\Xi ; & \mbox{ and } & \widehat{\chi }%
: & T\ _{\shortmid }\Xi & \rightarrow & T\ _{\shortmid }\Xi =h\ _{\shortmid
}\Xi \oplus v\ _{\shortmid }\Xi \\
& \mathbf{a} & \rightarrow & \mathbf{\ }_{\shortmid }\nabla _{\mathbf{a}}%
\mathbf{n;} &  &  & \mathbf{a} & \rightarrow & \mathbf{\ }_{\shortmid }%
\widehat{\mathbf{D}}_{\mathbf{a}}\mathbf{n}=h\mathbf{\ }_{\shortmid }%
\widehat{\mathbf{D}}_{\mathbf{a}}\mathbf{n}\oplus v\mathbf{\ }_{\shortmid }%
\widehat{\mathbf{D}}_{\mathbf{a}}\mathbf{n.}%
\end{array}%
\end{equation*}%
Such maps are self--adjoint with respect to the induced 3--metric $\mathbf{q,%
}$ i.e. for any $\mathbf{a,b\in }T_{p}\Xi \times T_{p}\Xi ,$
\begin{equation*}
\mathbf{a\cdot }\chi \mathbf{(b)=}\chi (\mathbf{a)\cdot b}\mbox{ and }%
\mathbf{a\cdot }\widehat{\chi }\mathbf{(b)=}\widehat{\chi }(\mathbf{a)\cdot
b,}
\end{equation*}%
where dot means the scalar product with respect to $\mathbf{q.}$ This
property allows to define two second fundamental forms (i.e. corresponding%
\textbf{\ extrinsic curvature d--tensors)} of hypersurface $\Xi $%
\begin{equation*}
\begin{array}{cccc}
\ _{\shortmid }K: & T_{p}\Xi \times T_{p}\Xi & \longrightarrow & \mathbb{R}%
\mbox{\ LC --configurations}, \\
& \left( \mathbf{a,b}\right) & \longrightarrow & -\mathbf{a\cdot }\chi
\mathbf{(b);} \\
\ _{\shortmid }\widehat{\mathbf{K}}: & (h\ _{\shortmid }\Xi \oplus v\
_{\shortmid }\Xi )\times (h\ _{\shortmid }\Xi \oplus v\ _{\shortmid }\Xi ) &
\longrightarrow & \mathbb{R\oplus R}\mbox{\ canonical configurations}, \\
& \left( \mathbf{a,b}\right) & \longrightarrow & -\mathbf{a\cdot }\widehat{%
\chi }\mathbf{(b).}%
\end{array}%
\end{equation*}%
In explicit operator form, $\ _{\shortmid }K\left( \mathbf{a,b}\right) =-%
\mathbf{a\cdot }\ _{\shortmid }\nabla \mathbf{_{\mathbf{b}}n}$ and $\
_{\shortmid }\widehat{\mathbf{K}}\left( \mathbf{a,b}\right) =-\mathbf{a\cdot
\mathbf{\ }_{\shortmid }\widehat{\mathbf{D}}_{\mathbf{b}}n,}$ which allows
us to compute in coefficient form $\ _{\shortmid }K_{\grave{\imath},\grave{j}%
}$ and $\ _{\shortmid }\widehat{\mathbf{K}}_{\grave{\imath},\grave{j}}.$ Any
pseudo--Riemannian geometry can be written equivalently in terms of $\left(
\mathbf{q,}\ _{\shortmid }\widehat{\mathbf{K}}\right) ,$ or $\left( \mathbf{%
q,}\ _{\shortmid }K\right) ,$ for any data $(\Xi ,\mathbf{q,\ _{\shortmid }N,%
}\breve{N}^{\grave{\imath}},\breve{N}).$ Intutively, we can work on 3--d
spacelike hypersurfaces as in Riemannian geometry. Unfortunately, such 3+1
splitting nonholonomic variables are not convenient for decoupling in
general form the gravitational field equations.

\subsection{Important formulas on spacelike N--adapted hypersurfaces}

Let us consider 3--surfaces $\Xi _{t}$ enabled with Riemannian d--metrics
when the normal d-vector $\mathbf{n}$ is timelike. We follow a 4--d point
view treating d--tensor fields defined on any $\Xi $ as they are defined for
$(\mathbf{V},\mathbf{g,N}).$ This avoids the obligation to introduce special
frame/coordinate systems and complicated notations depending on double
fibration parameterizations etc.

We consider $Vect(\mathbf{n})$ as teh 1--dimensional subspace of $T_{p}%
\mathbf{V}$ generated by the d--vector $\mathbf{n}$ use $\underline{\mathbf{n%
}}=\{\mathbf{n}_{\alpha }\}$ for it dual 1--form. In a point $p\in \Xi ,$
the spaces of all spacetime vectors can be decomposed as $\ T_{p}\mathbf{V=}%
T_{p}\Xi \oplus Vect(\mathbf{n}).$ For $\mathbf{n\cdot n}=-1$ and any $%
\mathbf{v\in }T_{p}\Xi ,$ we have $\overrightarrow{\mathbf{q}}(\mathbf{n})=0$
and $\overrightarrow{\mathbf{q}}(\mathbf{v})=\mathbf{v}.$ We construct the
orthogonal projector onto $\Xi $ as the operator $\overrightarrow{\mathbf{q}}
$ following the roule%
\begin{equation*}
\begin{array}{cccc}
\overrightarrow{\mathbf{q}}: & T_{p}\mathbf{V}\rightarrow T_{p}\Xi ; &  &
\mathbf{v}\rightarrow \mathbf{v+(n\cdot v)n}%
\end{array}%
\end{equation*}%
Similarly, we can consider a mapping $\overrightarrow{\mathbf{q}}_{\mathbf{V}%
}^{\ast }:T_{p}^{\ast }\Xi \rightarrow T_{p}^{\ast }\mathbf{V}$ setting that
for any linear form $\underline{\mathbf{z}}\in T_{p}^{\ast }\Xi $%
\begin{equation*}
\begin{array}{cccc}
\overrightarrow{\mathbf{q}}_{\mathbf{V}}^{\ast }: & T_{p}\mathbf{V}%
\rightarrow \mathbb{R}; &  & \mathbf{v}\rightarrow \underline{\mathbf{z}}(%
\overrightarrow{\mathbf{q}}(\mathbf{v})).%
\end{array}%
\end{equation*}%
The maps can be extended to bilinear forms. Taking the induced 3--metric $%
\mathbf{q}$ on $\Xi ,$ we can consider
\begin{equation}
\begin{array}{ccc}
\mathbf{q:=}\overrightarrow{\mathbf{q}}_{\mathbf{V}}^{\ast }\mathbf{g}, & %
\mbox{ i.e. } & \mathbf{q=g+}\underline{\mathbf{n}}\otimes \underline{%
\mathbf{n}}, \\
q_{\ \beta }^{\alpha }=\delta _{\ \beta }^{\alpha }+\mathbf{n}^{\alpha }%
\mathbf{n}_{\beta } & \mbox{ and } & \mathbf{q}_{\alpha \beta }=\mathbf{g}%
_{\alpha \beta }+\mathbf{n}_{\alpha }\mathbf{n}_{\beta }.%
\end{array}
\label{projop}
\end{equation}%
We get the same results as for the d--metric $\mathbf{q}$ if the two
arguments of $\mathbf{q(\cdot ,\cdot )}$ are tangent d--vectors to $\Xi .$
Such an operator gives zero if a d-vector is orthogonal to $\Xi ,$ i.e.
parallel to $\mathbf{n.}$

The nonholonomic\textbf{\ matter energy density} is defined
\begin{equation*}
\ ^{m}\widehat{E}:=\ ^{m}\widehat{\mathbf{T}}(\mathbf{n,n}),\mbox{ i.e. }\
^{m}\widehat{E}=\ ^{m}\widehat{\mathbf{T}}_{\alpha \beta }\mathbf{n}^{\alpha
}\mathbf{n}^{\beta }.
\end{equation*}%
Such values are written "without hat", ($\ ^{m}\widehat{E}\rightarrow \
^{m}E $ and $\ ^{m}\widehat{\mathbf{T}}_{\alpha \beta }\rightarrow \ ^{m}%
\mathbf{T}_{\alpha \beta }$), if $\widehat{\mathbf{D}}\rightarrow \nabla $
in $\ ^{m}\widehat{\mathbf{T}},$ see (\ref{ematter}).

Similarly, the nonholonomic\textbf{\ matter momentum density} is%
\begin{equation*}
\widehat{\mathbf{p}}:=-\ ^{m}\widehat{\mathbf{T}}(\mathbf{n,}\overrightarrow{%
\mathbf{q}}(...)),\mbox{ i.e. }<\widehat{\mathbf{p}},\mathbf{a}>=-\ ^{m}%
\widehat{\mathbf{T}}(\mathbf{n,}\overrightarrow{\mathbf{q}}(\mathbf{a})),
\end{equation*}%
or $\widehat{\mathbf{p}}_{\alpha }=-\ ^{m}\widehat{\mathbf{T}}_{\mu \nu }%
\mathbf{n}^{\mu }q_{\ \alpha }^{\nu },$ for any $\mathbf{a}\in (h\Xi \oplus
v\Xi ).$

The nonholonomic\textbf{\ stress d--tensor} is the bilinear form%
\begin{equation*}
\widehat{\mathbf{S}}:=\overrightarrow{\mathbf{q}}^{\ast }\ ^{m}\widehat{%
\mathbf{T}}\mbox{ and in N--adapted components }\widehat{\mathbf{S}}_{\alpha
\beta }=\ ^{m}\widehat{\mathbf{T}}_{\mu \nu }q_{\ \alpha }^{\mu }q_{\ \beta
}^{\nu }.
\end{equation*}%
We can consider also the trace of this field, $\widehat{S}:=\mathbf{q}^{%
\grave{\imath}\grave{j}}\widehat{\mathbf{S}}_{\grave{\imath}\grave{j}}=%
\mathbf{g}^{\alpha \beta }\widehat{\mathbf{S}}_{\alpha \beta }.$

Both $\widehat{\mathbf{p}}$ and $\widehat{\mathbf{S}}$ are d--tensor fields
tangent to $\Xi _{t}.$ The data $(\widehat{E},\widehat{\mathbf{p}},\widehat{%
\mathbf{S}})$ allow to reconstruct
\begin{equation*}
\ ^{m}\widehat{\mathbf{T}}=\widehat{\mathbf{S}}+\underline{\mathbf{n}}%
\otimes \widehat{\mathbf{p}}+\widehat{\mathbf{p}}\otimes \underline{\mathbf{n%
}}+\ ^{m}E\underline{\mathbf{n}}\otimes \underline{\mathbf{n}},
\end{equation*}%
when $\widehat{T}=\widehat{S}-\widehat{E}.$

\subsubsection{Einstein equations in nonholonomic variables for double 2+2
and 3+1 splitting}

Summarizing above formulas, the nonholonomic version of Einstein equations (%
\ref{deinst}) \ with matter sources of type (\ref{ematter}) can be written
in the form {\footnotesize
\begin{equation}
\begin{array}{ccc}
\begin{array}{c}
\mbox{4-d indices} \\
\\
\mbox{3-d indices } \\
\end{array}
&  &
\begin{array}{c}
\mathcal{L}_{\mathbf{m}}\widehat{\mathbf{K}}_{\alpha \beta }:=-\ _{\shortmid
}\widehat{\mathbf{D}}_{\alpha }\ _{\shortmid }\widehat{\mathbf{D}}_{\beta
}N+N\{\ _{\shortmid }\widehat{\mathbf{R}}_{\alpha \beta }+\ _{\shortmid }%
\widehat{K}\widehat{\mathbf{K}}_{\alpha \beta }-2\widehat{\mathbf{K}}%
_{\alpha \mu }\widehat{\mathbf{K}}_{\ \beta }^{\mu }+4\pi \lbrack (\widehat{S%
}-\widehat{E})\mathbf{q}_{\alpha \beta }-2\widehat{\mathbf{S}}_{\alpha \beta
}]\}, \\
\\
\mathcal{L}_{\mathbf{m}}\widehat{\mathbf{K}}_{\grave{\imath}\grave{j}}:=-\
_{\shortmid }\widehat{\mathbf{D}}_{\grave{\imath}}\ _{\shortmid }\widehat{%
\mathbf{D}}_{\grave{j}}N+N\{\ _{\shortmid }\widehat{\mathbf{R}}_{\grave{%
\imath}\grave{j}}+\ _{\shortmid }\widehat{K}\widehat{\mathbf{K}}_{\grave{%
\imath}\grave{j}}-2\widehat{\mathbf{K}}_{\grave{\imath}\grave{k}}\widehat{%
\mathbf{K}}_{\ \grave{j}}^{\grave{k}}+4\pi \lbrack (\widehat{S}-\widehat{E})%
\mathbf{q}_{\grave{\imath}\grave{j}}-2\widehat{\mathbf{S}}_{\grave{\imath}%
\grave{j}}]\}; \\
\end{array}
\\
\mbox{Hamilt. constr.} &  & \ _{\shortmid }\widehat{R}+(\ _{\shortmid }%
\widehat{K})^{2}-\ _{\shortmid }\widehat{\mathbf{K}}_{\grave{\imath}\grave{j}%
}\ _{\shortmid }\widehat{\mathbf{K}}^{\grave{\imath}\grave{j}}=16\pi
\widehat{E}, \\
\mbox{momentum constr.} &  & \ _{\shortmid }\widehat{\mathbf{D}}_{\grave{j}%
}\ _{\shortmid }\widehat{\mathbf{K}}_{\ \grave{\imath}}^{\grave{j}}-\
_{\shortmid }\widehat{\mathbf{D}}_{\grave{\imath}}\ _{\shortmid }\widehat{K}%
=8\pi \widehat{\mathbf{p}}_{\grave{\imath}}.%
\end{array}
\label{projeq2}
\end{equation}%
} Such systems of nonlinear partial differential equations, PDEs, are useful
for stating the Cauchy problem, defining energy and momentum type values and
elaborating methods of canonical and/or loop quantization. To decouple and
solve in general analytic forms such systems of equations it is a very
difficult technical task even such constructions are used in numeric
analysis. Nevertheless, we can compute all values in (\ref{projeq2}) using
any class of solutions found for (\ref{deinst}), see section \ref{s4}.

\subsubsection{Weyl's tensor 3+1 projections adapted to nonholonomic 2+2
splitting}

Taking any timelike unit vector $\mathbf{v}^{\alpha }$, we can define a
projection tensor, $\mathbf{h}_{\alpha \beta }=\mathbf{g}_{\alpha \beta }+%
\mathbf{v}_{\alpha }\mathbf{v}_{\beta }.$ A general d--vector $\mathbf{v}%
_{\alpha }$ is necessary for study relativistic thermodynamic and
hydrodynamic models, see Refs. \cite%
{moe,rtcrit1,rttoday2,rttoday3,rttoday4,rttoday6,misner,velatdif,vkin}. For
constructions in this work, we can consider $\mathbf{v}_{\alpha }=\mathbf{n}%
_{\alpha }$ and use a 3--d hypersurface metric $\mathbf{q}_{\alpha \beta }=%
\mathbf{g}_{\alpha \beta }+\mathbf{n}_{\alpha }\mathbf{n}_{\beta }.$

A (3+1)+(2+2) covariant description of gravitational filed is possible by
splitting correspondingly into irreducible parts such that%
\begin{equation*}
\widehat{\mathbf{D}}_{\beta }\mathbf{v}_{\alpha }=-\mathbf{v}_{\beta }%
\mathbf{v}^{\gamma }\widehat{\mathbf{D}}_{\gamma }\mathbf{v}_{\alpha }+\frac{%
1}{3}\widehat{\mathbf{\Theta }}\mathbf{h}_{\alpha \beta }+\widehat{\sigma }%
_{\alpha \beta }+\widehat{\omega }_{\alpha \beta },
\end{equation*}%
where $\mathbf{v}^{\gamma }\widehat{\mathbf{D}}_{\gamma }\mathbf{v}^{\alpha
} $ is the acceleration vector, $\widehat{\mathbf{\Theta }}:=\mathbf{h}%
^{\alpha \beta }\widehat{\mathbf{D}}_{\beta }\mathbf{v}_{\alpha }$ is the
expansion scalar, $\widehat{\sigma }_{\alpha \beta }=[\mathbf{h}_{~(\alpha
}^{\gamma }\mathbf{h}_{\beta )}^{~\delta }-\frac{1}{3}\mathbf{h}_{\alpha
\beta }\mathbf{h}^{\gamma \delta }]\widehat{\mathbf{D}}_{\gamma }\mathbf{v}%
_{\delta }$ is the shear tensor and $\widehat{\omega }_{\alpha \beta }=%
\mathbf{h}_{~[\alpha }^{\gamma }\mathbf{h}_{\beta ]}^{~\delta }\widehat{%
\mathbf{D}}_{\gamma }\mathbf{v}_{\delta }$ is the vorticity tensor (in the
last two formulas $(...)$ and $[...]$ mean respectively symmetrization and
anti--symmetrization of indices). We shall use the following decompositions
of the Weyl tensor%
\begin{equation}
\widehat{\mathbf{C}}_{\ \alpha \beta }^{\tau \quad \gamma }:=\widehat{%
\mathbf{R}}_{\ \alpha \beta }^{\tau \quad \gamma }+2\widehat{\mathbf{R}}%
_{[\alpha }^{\quad \lbrack \tau }\delta _{\beta ]}^{\gamma ]}+\frac{1}{3}%
\widehat{\mathbf{R}}\delta _{\lbrack \alpha }^{\gamma }\delta _{\beta
]}^{\tau }  \label{wtens}
\end{equation}%
into, respectively, electric and magnetic like parts (similarly to the
Maxwell theory),%
\begin{equation}
\widehat{\mathbf{E}}_{\alpha \beta }=\widehat{\mathbf{C}}_{\alpha \beta
\gamma \delta }\mathbf{v}^{\gamma }\mathbf{v}^{\delta }\mbox{ and }\widehat{%
\mathbf{H}}_{\alpha \beta }=\frac{1}{2}\eta _{\alpha \gamma \delta }\widehat{%
\mathbf{C}}_{\ \quad \alpha \beta }^{\gamma \delta }\mathbf{v}^{\gamma }%
\mathbf{v}^{\varepsilon }.  \label{emgrv}
\end{equation}%
In these formulas, $\eta _{\alpha \beta \gamma }=\eta _{\alpha \beta \gamma
\delta }\mathbf{v}^{\delta },$ are for the spatial alternating tensor $\eta
_{\alpha \beta \gamma \delta }=\eta _{\lbrack \alpha \beta \gamma \delta ]}$
with $\eta _{1234}=\sqrt{|g_{\alpha \beta }|}.$ For any spacelike unit
vectors $\mathbf{x}^{\alpha },\mathbf{y}^{\alpha },\mathbf{z}^{\alpha }$
that together with $\mathbf{v}^{\alpha }$ form an orthonormal basis, we can
introduce null tetrads as%
\begin{equation*}
\mathbf{m}^{\alpha }:=\frac{1}{\sqrt{2}}(\mathbf{x}^{\alpha }-i\mathbf{y}%
^{\alpha }),\mathbf{l}^{\alpha }:=\frac{1}{\sqrt{2}}(\mathbf{v}^{\alpha }-%
\mathbf{z}^{\alpha })\mbox{ and }\mathbf{k}^{\alpha }:=\frac{1}{\sqrt{2}}(%
\mathbf{v}^{\alpha }+\mathbf{z}^{\alpha }),
\end{equation*}%
for $i^{2}=-1,$ and express the metric $\mathbf{g}_{\alpha \beta }=2\mathbf{m%
}_{(\alpha }\mathbf{m}_{\beta )}-2\mathbf{k}_{(\alpha }\mathbf{l}_{\beta )}.$
We can transform a double fibration, for instance, into a 3+1 fibration if
we substitute $\widehat{\mathbf{D}}\rightarrow \nabla ,$ omit 'hats' and
change 'boldface' symbols into similar 'non-boldface' ones and work with
arbitrary with arbitrary bases instead of N--adapted ones.

The formulas (\ref{wtens}) and (\ref{emgrv}) are important for constructing
thermodynamical values for gravitational fields following the CET model \cite%
{clifton}.

\section{Geometric Evolution of Einstein Gravitational Fields}

\label{s3}Nonholonomic Ricci flows for theories with distinguished
connections were considered in \cite{vnhrf,vnoncjmprf}. To study general
relativistic geometric flows and elaborate on thermodynamical models we use
a N--adapted 3+1 decomposition for the canonical d--connection, $\widehat{%
\mathbf{D}}=(\ _{\shortmid }\widehat{\mathbf{D}},\ ^{t}\widehat{D})$ and
d--metric $\mathbf{g}:=(\mathbf{q,}\breve{N})$ of a 4--d spacetime $\mathbf{V%
}$. In this section we formulate the general relativistic geometric flow
theory in nonholonomic variables with double splitting. The fundamental
functionals and Ricci flow evolution equations are constructed on 4--d
Lorentz manifolds determined by exact solutions in GR.

\subsection{Distortion relations on induced linear connections}

On closed 3-d spacelike hypersurfaces, geometric flow and gravitational
field theories can be formulated in two equivalent forms using the
connections $\ _{\shortmid }\nabla $ and/or $\ _{\shortmid }\widehat{\mathbf{%
D}}.$ The evolution of such connections and N--adapted frames are determined
by the evolution of the hypersurface metric $\mathbf{q}$. In N--adapted
variables, we can introduce the canonical Laplacian d-operator, $\
_{\shortmid }\widehat{\Delta }:=$ $\ _{\shortmid }\widehat{\mathbf{D}}$ $\
_{\shortmid }\widehat{\mathbf{D}}$ and fine the canonical distortion tensor $%
\ _{\shortmid }\widehat{\mathbf{Z}}$. The distortion of the Ricci d--tensor
and corresponding Ricci scalar are computed by introducing $\ _{\shortmid
}\nabla =\ _{\shortmid }\widehat{\mathbf{D}}-\mathbf{\ }_{\shortmid }%
\widehat{\mathbf{Z}}$ (\ref{dist3}) in
\begin{equation*}
\mathbf{\ }_{\shortmid }\widehat{\Delta }=\mathbf{\ }_{\shortmid }\widehat{%
\mathbf{D}}_{\alpha }\ \mathbf{\ }_{\shortmid }\widehat{\mathbf{D}}^{\alpha
}=\mathbf{\ }_{\shortmid }\Delta +\ _{\shortmid }^{Z}\widehat{\Delta }.
\end{equation*}%
We obtain
\begin{eqnarray}
\ _{\shortmid }\Delta &=&\mathbf{\ }_{\shortmid }\nabla _{\grave{\imath}}\
\mathbf{\ }_{\shortmid }\nabla ^{\grave{\imath}}=\mathbf{\ }_{\shortmid
}\nabla _{\alpha }\ \mathbf{\ }_{\shortmid }\nabla ^{\alpha },
\label{distrel2} \\
\mathbf{\ }_{\shortmid }^{Z}\widehat{\Delta } &=&\mathbf{\ }_{\shortmid }%
\widehat{\mathbf{Z}}_{\grave{\imath}}\ \mathbf{\ }_{\shortmid }\widehat{%
\mathbf{Z}}^{\grave{\imath}}-[\ _{\shortmid }\widehat{\mathbf{D}}_{\grave{%
\imath}}(\mathbf{\ }_{\shortmid }\widehat{\mathbf{Z}}^{\grave{\imath}})+%
\mathbf{\ }_{\shortmid }\widehat{\mathbf{Z}}_{\grave{\imath}}(\ _{\shortmid }%
\widehat{\mathbf{D}}^{\grave{\imath}})]=\mathbf{\ }_{\shortmid }\widehat{%
\mathbf{Z}}_{\alpha }\mathbf{\ }_{\shortmid }\widehat{\mathbf{Z}}^{\alpha
}-[\ _{\shortmid }\widehat{\mathbf{D}}_{\alpha }(\mathbf{\ }_{\shortmid }%
\widehat{\mathbf{Z}}^{\alpha })+\mathbf{\ }_{\shortmid }\widehat{\mathbf{Z}}%
_{\alpha }(\ _{\shortmid }\widehat{\mathbf{D}}^{\alpha })];  \notag \\
\mathbf{\ }_{\shortmid }\widehat{\mathbf{R}}_{\grave{\imath}\grave{j}} &=&%
\mathbf{\ }_{\shortmid }R_{\grave{\imath}\grave{j}}-\mathbf{\ }_{\shortmid }%
\widehat{\mathbf{Z}}ic_{\grave{\imath}\grave{j}},\mathbf{\ }_{\shortmid }%
\widehat{\mathbf{R}}_{\ \beta \gamma }=\mathbf{\ }_{\shortmid }R_{\ \beta
\gamma }-\mathbf{\ }_{\shortmid }\widehat{\mathbf{Z}}ic_{\beta \gamma },\
\notag \\
\ _{\shortmid }\widehat{R} &=&\mathbf{\ }_{\shortmid }R-\mathbf{g}^{\beta
\gamma }\ _{\shortmid }\widehat{\mathbf{Z}}ic_{\beta \gamma }=\mathbf{\ }%
_{\shortmid }R-\mathbf{q}^{\grave{\imath}\grave{j}}\mathbf{\ }_{\shortmid }%
\widehat{\mathbf{Z}}ic_{\grave{\imath}\grave{j}}=\mathbf{\ }_{\shortmid }R-%
\mathbf{\ }_{\shortmid }\widehat{\mathbf{Z}},  \notag \\
\mathbf{\ }_{\shortmid }\widehat{\mathbf{Z}} &=&\mathbf{g}^{\beta \gamma }%
\mathbf{\ }_{\shortmid }\widehat{\mathbf{Z}}ic_{\beta \gamma }=\mathbf{q}^{%
\grave{\imath}\grave{j}}\mathbf{\ }_{\shortmid }\widehat{\mathbf{Z}}ic_{%
\grave{\imath}\grave{j}}=\ _{h}\widehat{Z}+\ _{v}\widehat{Z},\ _{h}\widehat{Z%
}=g^{ij}\ \widehat{\mathbf{Z}}ic_{ij},\ _{v}\widehat{Z}=h^{ab}\ \widehat{%
\mathbf{Z}}ic_{ab};  \notag \\
R &=&\ _{h}R+\ _{v}R,\ \ _{h}R:=g^{ij}\ R_{ij},\ _{v}R=h^{ab}\ R_{ab}.
\notag
\end{eqnarray}%
Such values can be computed in explicit form for any class of exact
solutions of the nonholonomic Einstein equations (\ref{deinst}), when a
double 2+2 and 3+1 splitting is prescribed and the LC--conditions (\ref%
{lccond}) can be imposed additionally.

\subsection{Nonholonomc Perelman's functionals on 3--d hypersurfaces}

For standard Ricci flows on a normalized 3-d spacelike closed hypersurface $%
\ ^{c}\widehat{\Xi }\subset \mathbf{V},$ the normalized Hamilton equations
written in a coordinate basis are%
\begin{eqnarray}
\partial _{\tau }q_{\grave{\imath}\grave{j}} &=&-2\ _{\shortmid }R_{\grave{%
\imath}\grave{j}}+\frac{2\grave{r}}{5}q_{\grave{\imath}\grave{j}},
\label{heq1b} \\
q_{\grave{\imath}\grave{j}\mid \tau =0} &=&q_{\grave{\imath}\grave{j}%
}^{[0]}[x^{\grave{\imath}}].  \notag
\end{eqnarray}%
We use the left label "c" for the conditions "compact and closed" and do not
emphasize dependence on space coordinates (writing in brief $q_{\grave{\imath%
}\grave{j}}(x^{\grave{\imath}},\tau )=q_{\grave{\imath}\grave{j}}(\tau ))$
if this do not result in ambiguities. In above formulas, $\ _{\shortmid }R_{%
\grave{\imath}\grave{j}}$ is computed for the Levi--Civita connection $\
_{\shortmid }\nabla $ of $q_{\grave{\imath}\grave{j}}(\tau )$ parameterized
by a real variable $\tau ,$ $0\leq \tau <\tau _{0},$ for a differentiable
function $\tau (t).$ The boundary conditions are stated for $\tau =0$ and
the normalizing factor
\begin{equation*}
\grave{r}=\int_{\ ^{c}\widehat{\Xi }}\ _{\shortmid }R\sqrt{|q_{\grave{\imath}%
\grave{j}}|}d\grave{x}^{3}/\int_{\ ^{c}\widehat{\Xi }}\sqrt{|q_{\grave{\imath%
}\grave{j}}|}d\grave{x}^{3}
\end{equation*}%
is introduced in the form to preserve the volume of $\ \ ^{c}\widehat{\Xi },$
i.e. $\int_{\ ^{c}\Xi }\sqrt{|q_{\grave{\imath}\grave{j}}|}d\grave{x}^{3}.$
For simplicity, we can find solutions of (\ref{heq1a}) with $\grave{r}=0.$

In order to find explicit solutions of (\ref{heq1b}) for $q_{\grave{\imath}%
\grave{j}}\subset g_{\alpha \beta }$ with $g_{\alpha \beta }$ defined also
as a solution of a 4-d Einstein equations (\ref{einsteq}), we have to
consider a nontrivial $\grave{r}.$ We can re--write (\ref{heq1a}) in any
nonholonomic basis using the geometric evolution of frame fields, $\
\partial _{\chi }e_{\grave{\imath}}^{\ \underline{\grave{\imath}}}=q^{%
\underline{\grave{\imath}}\underline{\grave{j}}}\ _{\shortmid }R_{\underline{%
\grave{j}}\underline{\grave{k}}}e_{\grave{\imath}}^{\ \underline{\grave{k}}%
}, $ when $q_{\grave{\imath}\grave{j}}(\tau )=q_{\underline{\grave{\imath}}%
\underline{\grave{j}}}(\tau )e_{\grave{\imath}}^{\ \underline{\grave{\imath}}%
}(\tau )e_{\grave{j}}^{\ \underline{\grave{j}}}(\tau )$ for $e_{\grave{\imath%
}}(\tau )=e_{\grave{\imath}}^{\ \underline{\grave{\imath}}}(\tau )\partial _{%
\underline{\grave{\imath}}}$ and $e^{j}(\tau )=e_{\ \underline{\grave{j}}%
}^{j}(\tau )dx^{\underline{\grave{j}}}.$ There is a unique solution for such
systems of linear ODEs for any $\tau \in \lbrack 0,\tau _{0}).$

In nonholonomic variables and for the linear d--connection $\ _{\shortmid }%
\widehat{\mathbf{D}},$ the Perelman's functionals parameterized in
N--adapted from are written%
\begin{eqnarray}
\ _{\shortmid }\widehat{\mathcal{F}} &=&\int_{\widehat{\Xi }_{t}}e^{-f}\sqrt{%
|q_{\grave{\imath}\grave{j}}|}d\grave{x}^{3}(\ _{\shortmid }\widehat{R}+|\
_{\shortmid }\widehat{\mathbf{D}}f|^{2}),  \label{perelm3f} \\
&&\mbox{ and }  \notag \\
\ _{\shortmid }\widehat{\mathcal{W}} &=&\int_{\widehat{\Xi }_{t}}M\sqrt{|q_{%
\grave{\imath}\grave{j}}|}d\grave{x}^{3}[\tau (\ _{\shortmid }\widehat{R}+|\
\ _{\shortmid }^{h}\widehat{\mathbf{D}}f|+|\ \ _{\shortmid }^{v}\widehat{%
\mathbf{D}}f|)^{2}+f-6],  \label{perelm3w}
\end{eqnarray}%
where the scaling function $f$ satisfies $\int_{\widehat{\Xi }_{t}}M\sqrt{%
|q_{\grave{\imath}\grave{j}}|}d\grave{x}^{3}=1$ for $M=\left( 4\pi \tau
\right) ^{-3}e^{-f}$ $.$ The functionals $\ _{\shortmid }\widehat{\mathcal{F}%
}$ and $\ _{\shortmid }\widehat{\mathcal{W}}$ \ transform into standard
Perelman functionals on $\widehat{\Xi }_{t}$ if $\ _{\shortmid }\widehat{%
\mathbf{D}}\rightarrow \ _{\shortmid }\nabla .$ The W--entropy $\
_{\shortmid }\widehat{\mathcal{W}}$ is a Lyapunov type non--decreasing
functional.

\subsection{Nonholonomic Ricci flow evolution equations for 3--d
hypersurface metrics}

Considering dependencies of formulas (\ref{perelm3f}) and (\ref{perelm3w})
on a smooth parameter $\chi (\tau )$ for which $\partial \chi /\partial \tau
=-1$ (when, for simplicity, the normalization terms are not included) can
prove the geometric evolution equations for any induced 3--d metric $\mathbf{%
q}$ and canonical d--connection $\ _{\shortmid }\widehat{\mathbf{D}}.$ A
subclass of geometric flow models with general relativistic extension from
3-d to 4-d can be elaborated if one of the parameters $\chi$ or $\tau$ is
taken to be proportional to the time like 4th coordinate, $t$.

Applying the variational procedure to $\ _{\shortmid }\widehat{\mathcal{F}}$
(\ref{perelm3f}) in N--adapted form (see details and references in \cite%
{vnhrf,vnoncjmprf}; in this work for a double nonholonomic splitting), we
obtain
\begin{eqnarray}
\partial _{\chi }\mathbf{q}_{\grave{\imath}\grave{j}} &=&-2(\mathbf{\ }%
_{\shortmid }\widehat{\mathbf{R}}_{\grave{\imath}\grave{j}}+\mathbf{\ }%
_{\shortmid }\widehat{\mathbf{Z}}ic_{\grave{\imath}\grave{j}}),
\label{heq1c} \\
\mathbf{\ }_{\shortmid }\widehat{\mathbf{R}}_{i\grave{a}} &=&-\mathbf{\ }%
_{\shortmid }\widehat{\mathbf{Z}}ic_{i\grave{a}},  \label{heq1d} \\
\partial _{\chi }f &=&-(\mathbf{\ }_{\shortmid }\widehat{\Delta }-\mathbf{\ }%
_{\shortmid }^{Z}\widehat{\Delta })f+|(\ _{\shortmid }\widehat{\mathbf{D}}-%
\mathbf{\ }_{\shortmid }\widehat{\mathbf{Z}})f|^{2}-\mathbf{\ }_{\shortmid }%
\widehat{R}-\mathbf{\ }_{\shortmid }\widehat{\mathbf{Z}}.  \notag
\end{eqnarray}%
The distortions in such nonholonomic evolution equations are completely
determined by $\mathbf{q}_{\grave{\imath}\grave{j}}$ following formulas (\ref%
{distrel2}). There is also another important property that
\begin{equation*}
\partial _{\chi }\ _{\shortmid }\widehat{\mathcal{F}}(q,\ _{\shortmid }%
\widehat{\mathbf{D}},f)=2\int_{\ \ ^{c}\widehat{\Xi }_{t}}e^{-f}\sqrt{|q_{%
\grave{\imath}\grave{j}}|}d\grave{x}^{3}[|\ _{\shortmid }\widehat{\mathbf{R}}%
_{\grave{\imath}\grave{j}}+\ _{\shortmid }\widehat{\mathbf{Z}}ic_{\grave{%
\imath}\grave{j}}+(\ _{\shortmid }\widehat{\mathbf{D}}_{\grave{\imath}}-\
_{\shortmid }\widehat{\mathbf{Z}}_{\grave{\imath}})(\ _{\shortmid }\widehat{%
\mathbf{D}}_{\grave{j}}-\ _{\shortmid }\widehat{\mathbf{Z}}_{\grave{j}%
})f|^{2}],
\end{equation*}%
when $\int_{\ \ ^{c}\widehat{\Xi }_{t}}e^{-f}\sqrt{|q_{\grave{\imath}\grave{j%
}}|}d\grave{x}^{3}$ is constant for a fixed $\tau $ and $f(\chi (\tau
))=f(\tau ).$

The system of equations (\ref{heq1c}) and (\ref{heq1d}) is equivalent to (%
\ref{heq1b}) up to certain re--definition of noholonomic frames and
variables. We have to consider (\ref{heq1d}) as additional constraints
because in nonholonomic variables the Ricci d--tensor is (in general)
nonsymmetric. If we do not impose such constraints, the geometric evolution
is with nonsymmetric metrics.

\subsection{Geometric evolution to 4--d Lorentz configurations as exact
solutions in GR}

The geometric evolution of metrics of type (\ref{offd3122}), when $\mathbf{q}%
(\tau )\mathbf{\rightarrow }$\ $\mathbf{g}(\tau ):=\left( \mathbf{q}(\tau )%
\mathbf{,}\breve{N}(\tau )\right)$, is described by respective
generalizations of functionals (\ref{perelm3f}) and (\ref{perelm3w}).
Considering N--connection adapted foliations $\widehat{\Xi }_{t}$
parameterized by a spacetime coordinate $t,$ we introduce such 4--d
functionals
\begin{eqnarray}
&&\widehat{\mathcal{F}}(\mathbf{q},\ _{\shortmid }\widehat{\mathbf{D}},f)
=\int_{t_{1}}^{t_{2}}dt\ \breve{N}(\tau )\ _{\shortmid }\widehat{\mathcal{F}}%
(\mathbf{q},\ _{\shortmid }\widehat{\mathbf{D}},f),  \label{ffperel} \\
\mbox{ and } &&  \notag \\
&& \widehat{\mathcal{W}}(\mathbf{q},\ _{\shortmid }\widehat{\mathbf{D}},f)
=\int_{t_{1}}^{t_{2}}dt\ \breve{N}(\tau )\ \ _{\shortmid }\widehat{\mathcal{W%
}}(\mathbf{q},\ _{\shortmid }\widehat{\mathbf{D}},f(t)).\   \label{wfperel}
\end{eqnarray}%
For arbitrary frame transforms on 4-d nonholonomic Lorentz manifolds, such
values can be re--defined respectively in terms of data $(\mathbf{g}(\tau ),%
\widehat{\mathbf{D}}(\tau )).$ We obtain
\begin{eqnarray}
\widehat{\mathcal{F}} &=&\int_{t_{1}}^{t_{2}}\int_{\widehat{\Xi }_{t}}e^{-%
\widehat{f}}\sqrt{|\mathbf{g}_{\alpha \beta }|}d^{4}u(\widehat{R}+|\widehat{%
\mathbf{D}}\widehat{f}|^{2}),  \label{ffperelm4} \\
&&\mbox{ and }  \notag \\
\widehat{\mathcal{W}} &=&\int_{t_{1}}^{t_{2}}\int_{\widehat{\Xi }_{t}}%
\widehat{M}\sqrt{|\mathbf{g}_{\alpha \beta }|}d^{4}u[\tau (\widehat{R}+|\
^{h}\widehat{D}\widehat{f}|+|\ ^{v}\widehat{D}\widehat{f}|)^{2}+\widehat{f}%
-8],  \label{wfperelm4}
\end{eqnarray}%
where the scaling function $\widehat{f}$ satisfies $\int_{t_{1}}^{t_{2}}%
\int_{\widehat{\Xi }_{t}}\widehat{M}\sqrt{|\mathbf{g}_{\alpha \beta }|}%
d^{4}u=1$ for $\widehat{M}=\left( 4\pi \tau \right) ^{-3}e^{-\widehat{f}}$ $%
. $

We emphasize that the functionals $\widehat{\mathcal{F}}$ and $\widehat{%
\mathcal{W}}$ for 4-d pseudo-Riemannian metrics are not of entropy type like
in the 3--d Riemannian case. Nevertheless, they describe nonlinear general
relativistic diffusion type processes if $\mathbf{q}\subset \mathbf{g}\ $\
and $\ _{\shortmid }\widehat{\mathbf{D}}$ $\subset \widehat{\mathbf{D}}$ are
determined by certain lapse and shift functions as certain solutions of 4--d
gravitational equations. This is motivation to construct such functionals
using nonholonomic variables.

The formulas (\ref{heq1c}) and (\ref{heq1d}) can be considered for 4--d
configurations even the coefficients determined by the Ricci d--tensors and
distortions will remain with left label "$\shortmid ".$ Using the formulas $%
\mathbf{q}_{\alpha \beta }=\mathbf{g}_{\alpha \beta }+\mathbf{n}_{\alpha }%
\mathbf{n}_{\beta }$ (see (\ref{projop})), we get%
\begin{eqnarray}
\partial _{\tau }\mathbf{g}_{\alpha \beta } &=&-2(\mathbf{\ }_{\shortmid }%
\widehat{\mathbf{R}}_{\alpha \beta }+\mathbf{\ }_{\shortmid }\widehat{%
\mathbf{Z}}ic_{\alpha \beta })-\partial _{\tau }(\mathbf{n}_{\alpha }\mathbf{%
n}_{\beta }),  \label{heq1cc} \\
\mathbf{\ }_{\shortmid }\widehat{\mathbf{R}}_{ia} &=&-\mathbf{\ }_{\shortmid
}\widehat{\mathbf{Z}}ic_{ia},\mbox{ for }\mathbf{\ }_{\shortmid }\widehat{%
\mathbf{R}}_{\alpha \beta }\mbox{ with }\alpha \neq \beta .  \notag
\end{eqnarray}%
The term $\partial _{\tau }(\mathbf{n}_{\alpha }\mathbf{n}_{\beta })$ can be
computed in explicit form using formulas for the geometric evolution of
N--adapted frames, see below formulas (\ref{aeq5}). The formulas for $%
\partial _{\tau }f,\partial _{\tau }(_{\shortmid }\widehat{\mathcal{F}})$
and $\partial _{\tau }\widehat{\mathcal{W}}$ can be re--written for values
with 4--d indices (we omit such formulas in this work).

For 4-d configurations with a corresponding re--definition of the scaling
function, $f\rightarrow \widehat{f},$ and for necessary type N--adapted
distributions, we can construct models of geometric evolution with $h$-- and
$v$--splitting for $\mathbf{\widehat{\mathbf{D}},}$
\begin{eqnarray}
\partial _{\tau }g_{ij} &=&-2\widehat{R}_{ij},\ \partial _{\tau }g_{ab}=-2%
\widehat{R}_{ab},  \label{rfcandc} \\
\partial _{\tau }\widehat{f} &=&-\widehat{\square }\widehat{f}+\left\vert
\widehat{\mathbf{D}}\widehat{f}\right\vert ^{2}-\ ^{h}\widehat{R}-\ ^{v}%
\widehat{R}.  \notag
\end{eqnarray}%
These formulas can be derived from the functional (\ref{ffperel}), following
a similar calculus to that presented in the proof of Proposition 1.5.3 of
\cite{monogrrf1} but in N--adapted form as in Ref. \cite{vnhrf,vnoncjmprf}.
\ We have to impose the conditions $\widehat{R}_{ia}=0$ and $\widehat{R}%
_{ai}=0 $ if we wont to keep the total metric to be symmetric under Ricci
flow evolution. The general relativistic character of 4-d geometric flow
evolution is encoded in operators like $\widehat{\square }=\widehat{\mathbf{D%
}}^{\alpha }\widehat{\mathbf{D}}_{\alpha },$ d-tensor components $\widehat{R}%
_{ij}$ and $\widehat{R}_{ab},$ theirs scalars $\ ^{h}\widehat{R}=g^{ij}%
\widehat{R}_{ij}$ and $\ ^{v}\widehat{R}=g^{ab}\widehat{R}_{ab}$ with data $%
(g_{ij},g_{ab},\widehat{\mathbf{D}}_{\alpha })$ constrained to the condition
to define solutions of certain 4-d Einstein equations.

The evolution on a parameter $\chi \in \lbrack 0,\chi _{0})$ of N--adapted
frames in a 4--d nonholonomic Lorentz manifold can be computed as
\begin{equation*}
\ \mathbf{e}_{\alpha }(\chi )=\ \mathbf{e}_{\alpha }^{\ \underline{\alpha }%
}(\chi ,u)\partial _{\underline{\alpha }}.
\end{equation*}%
Up to frame/coordinate transforms by the frame coefficients are
\begin{equation*}
\ \mathbf{e}_{\alpha }^{\ \underline{\alpha }}(\chi ,u)=\left[
\begin{array}{cc}
\ e_{i}^{\ \underline{i}}(\chi ,u) & -~N_{i}^{b}(\chi ,u)\ e_{b}^{\
\underline{a}}(\tau ,u) \\
0 & \ e_{a}^{\ \underline{a}}(\chi ,u)%
\end{array}%
\right] ,\ \mathbf{e}_{\ \underline{\alpha }}^{\alpha }(\chi ,u)\ =\left[
\begin{array}{cc}
e_{\ \underline{i}}^{i}=\delta _{\underline{i}}^{i} & e_{\ \underline{i}%
}^{b}=N_{k}^{b}(\chi ,u)\ \ \delta _{\underline{i}}^{k} \\
e_{\ \underline{a}}^{i}=0 & e_{\ \underline{a}}^{a}=\delta _{\underline{a}%
}^{a}%
\end{array}%
\right] ,
\end{equation*}%
with $\tilde{g}_{ij}(\chi )=\ e_{i}^{\ \underline{i}}(\chi ,u)\ e_{j}^{\
\underline{j}}(\chi ,u)\eta _{\underline{i}\underline{j}}$ and $\tilde{g}%
_{ab}(\chi )=\ e_{a}^{\ \underline{a}}(\chi ,u)\ e_{b}^{\ \underline{b}%
}(\chi ,u)\eta _{\underline{a}\underline{b}}.$ For $\eta _{\underline{i}%
\underline{j}}=diag[+,+]$ and $\eta _{\underline{a}\underline{b}}=diag[1,-1]$
corresponding to the chosen signature of $\ \mathbf{\tilde{g}}_{\alpha \beta
}^{[0]}(u),$ we have the evolution equations
\begin{equation}
\frac{\partial }{\partial \chi }\mathbf{e}_{\ \underline{\alpha }}^{\alpha
}\ =\ \mathbf{g}^{\alpha \beta }~\widehat{\mathbf{R}}_{\beta \gamma }~\
\mathbf{e}_{\ \underline{\alpha }}^{\gamma }.  \label{aeq5}
\end{equation}%
Such equations are prescribed for models with geometric evolution determined
by the canonical d--connection $\widehat{\mathbf{D}}.$ The equations (\ref%
{aeq5}) can be written in terms of the Levi--Civita connection if
distortions of type (\ref{distrel2}) are considered for 4--d values
determined by exact solutions in GR.

\section{Generation of Off-Diagonal Solutions}

\label{s4}

Let us summarize the anholonomic frame deformation method, AFDM, of
constructing generic off--diagonal exact solutions in GR with possible
dependencies on all spacetime coordinates (see details and various examples
in \cite{vepc,veym1,veps2} and references therein). Using N--adapted 2+2
frame and coordinate transforms,
\begin{equation*}
\mathbf{g}_{\alpha \beta }(x^{i},t)=e_{\ \alpha }^{\alpha ^{\prime
}}(x^{i},y^{a})e_{\ \beta }^{\beta ^{\prime }}(x^{i},y^{a})\widehat{\mathbf{g%
}}_{\alpha ^{\prime }\beta ^{\prime }}(x^{i},y^{a})\mbox{ and }\Upsilon
_{\alpha \beta }(x^{i},t)=e_{\ \alpha }^{\alpha ^{\prime }}(x^{i},y^{a})e_{\
\beta }^{\beta ^{\prime }}(x^{i},y^{a})\widehat{\Upsilon }_{\alpha ^{\prime
}\beta ^{\prime }}(x^{i},y^{a}),
\end{equation*}%
for a time like coordinate $y^{4}=t$ ($i^{\prime },i,k,k^{\prime },...=1,2$
and $a,a^{\prime },b,b^{\prime },...=3,4),$ we can parameterize the metric
and effective source in certain adapted forms. We consider
\begin{eqnarray}
\mathbf{g} &=&\mathbf{g}_{\alpha ^{\prime }\beta ^{\prime }}\mathbf{e}%
^{\alpha ^{\prime }}\otimes \mathbf{e}^{\beta ^{\prime
}}=g_{i}(x^{k})dx^{i}\otimes dx^{j}+\omega ^{2}(x^{k},y^{3},t)h_{a}(x^{k},t)%
\mathbf{e}^{a}\otimes \mathbf{e}^{a}  \notag \\
&=&q_{i}(x)dx^{i}\otimes dx^{i}+q_{3}(x,y)\mathbf{e}^{3}\otimes \mathbf{e}%
^{3}-\breve{N}^{2}(x^{k},y^{3},t)\mathbf{e}^{4}\otimes \mathbf{e}^{4},
\label{anstrsolit} \\
\mathbf{e}^{3} &=&dy^{3}+n_{i}(x^{k},t)dx^{i},\mathbf{e}%
^{4}=dt+w_{i}(x^{k},t)dx^{i}.  \notag
\end{eqnarray}%
This ansatz is a general one for 4--d metric which can be written in the
form (\ref{offd3122}) with
\begin{equation}
\breve{N}^{2}(u)=-\omega ^{2}h_{4}=-g_{4}.  \label{shift1}
\end{equation}
It allows a straightforward extension of 3--d ansatz to 4-d configurations
by introducing the values $\breve{N}^{2}(x^{k},t)$ and $w_{i}(x^{k},t)$ in
order to generate exact solutions of the Einstein equations. The nontrivial
respective N--connection, d-metric and matter source coefficients are
denoted
\begin{eqnarray}
\ N_{i}^{3} &=&\mathbf{\ }_{\shortmid
}N_{i}^{3}=n_{i}(x^{k},t);N_{i}^{4}=w_{i}(x^{k},t);  \notag \\
\{g_{i^{\prime }j^{\prime }}\}
&=&diag[g_{i}],g_{1}=g_{2}=q_{1}=q_{2}=e^{\psi (x^{k})};\ \{g_{a^{\prime
}b^{\prime }}\}=diag[\omega ^{2}h_{a}],h_{a}=h_{a}(x^{k},t),q_{3}=\omega
^{2}h_{3};  \notag \\
&&\mbox{ and }\Upsilon _{\alpha \beta }=diag[\Upsilon _{i};\Upsilon _{a}],%
\mbox{ for }\Upsilon _{1}=\Upsilon _{2}=\widetilde{\Upsilon }%
(x^{k}),\Upsilon _{3}=\Upsilon _{4}=\Upsilon (x^{k},t).  \label{2sourc}
\end{eqnarray}%
The ansatz (\ref{anstrsolit}) determines d-metrics of type (\ref{dm}) and (%
\ref{offd3122}) as 4--d generalizations of the 3-d hypersurface metric (\ref%
{3dm}) for a nontrivial lapse function (\ref{shift1}). The N--adapted
coefficients (\ref{2sourc}) can be very general ones but for an assumption
that there are N--adapted frames with respect to which the exact solutions
for $\omega =1$ are with Killing symmetry on $\partial /\partial y^{3}$. For
such configurations, there are N--adapted bases when the geometric and
physical values do not depend on coordinate $y^{3}.$ For simplicity, we
shall consider solutions with one Killing symmetry. \footnote{%
We note that it is possible to construct very general classes of generic
off--diagonal solutions depending on all spacetime variables, see details
and examples in Refs. \cite{vepc,veym1,veps2} for "non--Killing"
configurations.}

\subsection{Decoupling of Einstein equations for nonholonomic 2+2 splitting}

Let us introduce the values $\alpha _{i}=h_{3}^{\ast }\partial _{i}\varpi
,\beta =h_{3}^{\ast }\ \varpi ^{\ast },\gamma =\left( \ln
|h_{3}|^{3/2}/|h_{4}|\right) ^{\ast }$ determined by a generating function
\begin{equation}
\varpi :=\ln |h_{3}^{\ast }/\sqrt{|h_{3}h_{4}|}|,%
\mbox{ we shall also use
the value  }\Psi :=e^{\varpi }.  \label{genf}
\end{equation}%
We shall use brief denotations for partial derivatives: $a^{\bullet
}=\partial _{1}a,b^{\prime }=\partial _{2}b,h^{\ast }=\partial
_{4}h=\partial _{t}h.$

For ansatz (\ref{anstrsolit}) written in terms of date for the generating
function (\ref{genf}) and above coefficients, the nonholonomic Einstein
equations (\ref{deinst}) transform into a system of nonlinear PDEs with
\textbf{decoupling property,}\footnote{%
see details of such computation in \cite{vepc,veym1,veps2}}
\begin{equation}
\psi ^{\bullet \bullet }+\psi ^{\prime \prime }=2~\widetilde{\Upsilon },\
\varpi ^{\ast }h_{3}^{\ast }=2h_{3}h_{4}\Upsilon ,\ n_{i}^{\ast \ast
}+\gamma n_{i}^{\ast }=0,\ \beta w_{i}-\alpha _{i}=0.\   \label{eq4}
\end{equation}%
The un--known functions for this system are $\psi
(x^{i}),h_{a}(x^{k},t),w_{i}(x^{k},t)$ and $n_{i}(x^{k},t).$

We can simplify the system (\ref{genf})--(\ref{eq4}) using an important
property which allows us to re--define the generating function, $\Psi =\exp
\varpi \longleftrightarrow \widetilde{\Psi }=\exp \widetilde{\varpi },$ and
the effective source, $\Upsilon \longleftrightarrow \Lambda =const,\Lambda
\neq 0.$ Such nonlinear transforms are given by formulas
\begin{equation}
\Lambda (\Psi ^{2})^{\ast }=|\Upsilon |(\widetilde{\Psi }^{2})^{\ast }%
\mbox{
and }\Lambda \Psi ^{2}=\widetilde{\Psi }^{2}|\Upsilon |+\int dt\widetilde{%
\Psi }^{2}|\Upsilon |^{\ast }.  \label{nonltr}
\end{equation}%
For generating off--diagonal inhomogeneous and locally anisotropic
cosmological solutions depending on $t,$ we have to consider generating
functions for which $\Psi ^{\ast }\neq 0.$ We obtain a system of nonlinear
PDEs with effective cosmological constant $\Lambda ,$
\begin{eqnarray}
\psi ^{\bullet \bullet }+\psi ^{\prime } &=&2~\widetilde{\Upsilon },\
\label{eq4a} \\
\widetilde{\varpi }^{\ast }h_{3}^{\ast } &=&2h_{3}h_{4}\Lambda ,
\label{eq4b} \\
\ n_{i}^{\ast \ast }+\gamma n_{i}^{\ast } &=&0,\ \   \label{eq4c} \\
\varpi ^{\ast }\ w_{i}-\partial _{i}\varpi &=&0  \label{eq4d} \\
\partial _{k}\omega +n_{k}\partial _{3}\omega +w_{k}\omega ^{\ast } &=&%
\mathbf{e}_{k}\omega =0.  \label{eq4e}
\end{eqnarray}
This system can be solved in very general forms by prescribing $~\widetilde{%
\Upsilon },$ $\Lambda $ and $\Psi ,$ or $\widetilde{\Psi },$ by integrating
the equations "step by step". We obtain%
\begin{eqnarray}
g_{i} &=&g_{i}[\psi ,~\widetilde{\Upsilon }]\simeq e^{\psi (x^{k})}%
\mbox{ as
a solution of 2-d Poisson/ Laplace equations (\ref{eq4a})};\text{ }
\label{solut1t} \\
h_{a} &=&h_{a}[\widetilde{\Psi },\Lambda ]=h_{a}[\Psi ,\Upsilon ],%
\mbox{
where }h_{3}=\frac{\widetilde{\Psi }^{2}}{4\Lambda }\mbox{ and
}h_{4}=\frac{(\widetilde{\Psi }^{\ast })^{2}}{\Theta };  \notag \\
n_{k} &=&\ _{1}n_{k}+\ _{2}n_{k}\int dt\ h_{4}/(\sqrt{|h_{3}|})^{3}=\
_{1}n_{k}+\ _{2}\widetilde{n}_{k}\int dt\ (\widetilde{\Psi }^{\ast })^{2}/%
\widetilde{\Psi }^{3}\Theta ;  \notag \\
w_{i} &=&\partial _{i}\varpi /\varpi ^{\ast }=\partial _{i}\Psi /\Psi ^{\ast
}=\partial _{i}(\Psi )^{2}/(\Psi ^{2})^{\ast }=\partial _{i}\Theta /\Theta
^{\ast };  \notag \\
\omega &=&\omega \lbrack \widetilde{\Psi },\Lambda ]=\omega \lbrack \Psi
,\Upsilon ]\mbox{ is any solution of 1st order system (\ref{eq4e})}.  \notag
\end{eqnarray}%
By $h_{a}[\Psi ,\Upsilon ],$ we denote that the coefficients $h_{a}$ depend
functionally on two functions, $[\Psi ,\Upsilon ].$

The solutions (\ref{solut1t}) contain also integration functions$\
^{0}h_{3}(x^{k}),$ $\ _{1}n_{k}(x^{i})$ and $\ _{2}n_{k}(x^{i}),$ or $_{2}%
\widetilde{n}_{k}(x^{i})=8\ _{2}n_{k}(x^{i})|\Lambda |^{3/2},$ and the
generating source
\begin{equation}
\Theta =\int dt\ \Upsilon (\widetilde{\Psi }^{2})^{\ast }.  \label{effsourc2}
\end{equation}%
We can satisfy the conditions for $\omega $ in the second line in (\ref{eq4d}%
) if we keep, for simplicity, the Killing symmetry on $\partial _{i}$ and
take, for example, $\omega ^{2}=|h_{4}|^{-1}.$ Such solutions are
constructed in explicit form by solving the equations (\ref{eq4}) and/or (%
\ref{eq4a})-(\ref{eq4e}) for certain prescribed values of $\Psi $ and $%
\Upsilon $ and following certain assumptions on initial/boundary/asymptotic
conditions, physical arguments on symmetries of solutions, compatibility
with observational data etc. It should be emphasized that re-definitions of
generating functions of type (\ref{nonltr}) allow to construct exact
solutions for general sources $\Upsilon $ using certain classes of solutions
with nontrivial cosmological constant $\Lambda$.

\subsection{The Levi--Civita conditions}

The solutions (\ref{solut1t}) are defined for the canonical d--connection $%
\widehat{\mathbf{D}}$ and with respect to N--adapted frames. There are
nontrivial coefficients of nonholonomically induced torsion. We have to
subject the d--metric and N--connection coefficients to additional
nonholonomic constraints (\ref{lccond}) in order to satisfy the torsionless
conditions and extract Levi--Civita, LC, configurations. For the ansatz (\ref%
{anstrsolit}), such conditions can be written (see details in \cite%
{vepc,veym1,veps2})
\begin{equation}
w_{i}^{\ast }=(\partial _{i}-w_{i}\partial _{4})\ln \sqrt{|h_{4}|},(\partial
_{i}-w_{i}\partial _{4})\ln \sqrt{|h_{3}|}=0,\partial _{i}w_{j}=\partial
_{j}w_{i},n_{i}^{\ast }=0,\partial _{i}n_{j}=\partial _{j}n_{i}.
\label{lccond1}
\end{equation}%
We must consider additional constraints on data $(\Psi ,\Upsilon ),$ or $(%
\tilde{\Psi},\ \Lambda ),$ and nonzero integration functions $\
_{1}n_{j}(x^{k})$ but $\ _{2}n_{k}(x^{i})=0.$

To generate explicit solutions, we can consider any functional dependence $H=%
\widetilde{\Psi }[\Psi ],$ for which
\begin{equation*}
\mathbf{e}_{i}H=(\partial _{i}-w_{i}\partial _{4})H=\frac{\partial H}{%
\partial \Psi }(\partial _{i}-w_{i}\partial _{4})\Psi \equiv 0
\end{equation*}%
, see (\ref{eq4d}). For instance, $H=\widetilde{\Psi }=\ln \sqrt{|\ h_{3}|}$
results in $\mathbf{e}_{i}\ln \sqrt{|\ h_{3}|}=0.$ If we work with classes
of generating functions $\Psi =\check{\Psi}$ for which
\begin{equation}
(\partial _{i}\check{\Psi})^{\ast }=\partial _{i}(\check{\Psi}^{\ast }),
\label{cond1}
\end{equation}%
we obtain $w_{i}^{\ast }=\mathbf{e}_{i}\ln |\check{\Psi}^{\ast }|.$ For a
given functional dependence $h_{4}[\Psi ,\Upsilon ],$ we can express
\begin{equation*}
\mathbf{e}_{i}\ln \sqrt{|\ h_{4}|}=\mathbf{e}_{i}[\ln |\check{\Psi}^{\ast
}|-\ln \sqrt{|\ \Upsilon |}]
\end{equation*}
(we used the property $\mathbf{e}_{i}\check{\Psi}=0)$. In result,
\begin{equation*}
w_{i}^{\ast }=\mathbf{e}_{i}\ln \sqrt{|\ h_{4}|}
\end{equation*}
if $\mathbf{e}_{i}\ln \sqrt{|\ \Upsilon |}=0.$ This is possible for any $%
\Upsilon =const, $ or any effective source expressed as a functional $\
\Upsilon (x^{i},t)=\ \Upsilon \lbrack \check{\Psi}].$

The conditions $\partial _{i}w_{j}=\partial _{j}w_{i}$ can be solved by any
function $\check{A}=\check{A}(x^{k},t)$ for which
\begin{equation}
w_{i}=\check{w}_{i}=\partial _{i}\check{\Psi}/\check{\Psi}^{\ast }=\partial
_{i}\check{A}.  \label{cond2}
\end{equation}%
This is a system of first order PDEs which allows to find a function $\check{%
A}[\check{\Psi}]$ if a functional $\check{\Psi}$ is prescribed. For the
second set of N--coefficients, we chose $\ _{1}n_{j}(x^{k})=\partial
_{j}n(x^{k})$ for a function $n(x^{k}).$

Finally, we conclude that we can generate off--diagonal torsionless
solutions of the Einstein equations (\ref{einsteq}) by choosing certain
subclasses of generating functions and effective sources in (\ref{solut1t}),
when
\begin{equation}
\check{\Upsilon}=\Upsilon (x^{i},t)=\ \Upsilon \lbrack \check{\Psi}%
],w_{i}=\partial _{i}\check{A}[\check{\Psi}],n_{i}=\partial _{i}n,
\label{cond3}
\end{equation}%
and the generating function $\Psi =\check{\Psi}$ and "associated" $\check{A}$
are subjected to the conditions (\ref{cond1}) and (\ref{cond2}).

\subsection{General solutions for (non) holonomic Einstein manifolds}

Summarizing the results obtained in previous subsections, we can construct
the quadratic linear elements for generic off--diagonal metrics defining 4-d
Einstein spaces with effective cosmological constant $\Lambda$ and
nonholonomic deformations to general sources of type (\ref{2sourc}).

\subsubsection{Solutions with nonholonomically induced torsion}

The quadratic line elements determined by coefficients (\ref{solut1t}), with
label $torsRs,$ are written
\begin{eqnarray}
ds_{torsRs}^{2} &=&g_{\alpha \beta }(x^{k},t)du^{\alpha }du^{\beta }=e^{\psi
}[(dx^{1})^{2}+(dx^{2})^{2}]  \label{qnk4d} \\
&&+\omega ^{2}\frac{\tilde{\Psi}^{2}}{4\Lambda }\left[ dy^{3}+\left( \
_{1}n_{k}+_{2}\widetilde{n}_{k}\int dt\frac{(\tilde{\Psi}^{\ast })^{2}}{%
\tilde{\Psi}^{3}\ \Theta }\right) dx^{k}\right] ^{2}+\omega ^{2}\frac{(%
\tilde{\Psi}^{\ast })^{2}}{\Theta }\ \left[ dt+\frac{\partial _{i}\Theta }{%
\Theta ^{\ast }}dx^{i}\right] ^{2}.  \notag
\end{eqnarray}%
Fixing in (\ref{qnk4d}) an effective cosmological constant $\Lambda \neq 0,$
an "associated" generating function \ $\tilde{\Psi}(x^{k},t),$ for which\ $%
\tilde{\Psi}^{\ast }\neq 0,$ and a generating source $\Theta (x^{k},t),$ for
which $\Theta ^{\ast }\neq 0,$ we generate exact solutions of the
nonholonomic Einstein equations (\ref{deinst}). Using formula (\ref%
{effsourc2}), we compute the effective source $\Upsilon =$ $\Theta ^{\ast }/(%
\widetilde{\Psi }^{2})^{\ast }.$ Such effective values and formula (\ref%
{nonltr}) determine a generating function $\Psi :=e^{\varpi }$ (\ref{genf})
computed from $\Psi ^{2}=\Lambda ^{-1}(\widetilde{\Psi }^{2}|\Upsilon |+\int
dt\widetilde{\Psi }^{2}|\Upsilon |^{\ast }).$ We can express equivalently
the N--adapted coefficients $h_{a},n_{i}$ and $w_{i}$ as functionals of $%
(\Psi ,\Upsilon )$ following formulas (\ref{solut1t}). The $h$--metric $%
e^{\psi }$ is any solution of the 2-d Poisson equation (\ref{eq4a}) with
source $\widetilde{\Upsilon }.$ The vertical conformal factor $\omega
(x^{i},y^{3},t)$ can be determined as a solution of (\ref{eq4e}). The
subclass of solutions (\ref{qnk4d}) with Killing symmetry on $\partial _{3}$
can be extracted for $\omega ^{2}=1.$

It should be noted that the effective source $\Upsilon $ determines an
effective source (\ref{effsourc2}) up to a class of frame transforms $e_{\
\alpha }^{\alpha ^{\prime }}(x^{i},y^{a}).$ Such coefficients must defined
from a system of quadratic algebraic equations for any prescribed $\Upsilon
_{\beta }^{\alpha }=\left[ \ \tilde{\Upsilon}\delta _{j}^{i},\ \Upsilon
\delta _{b}^{a}\right] .$ The coefficients $e_{\ \alpha }^{\alpha ^{\prime
}} $ and the integration functions $\ _{1}n_{k}(x^{i})$ and $_{2}\widetilde{n%
}_{k}(x^{i})$ can be defined in explicit form if we chose respective
boundary/asymptotic conditions, or solve (for another type of solutions) the
Cauchy problem.

The solutions (\ref{qnk4d}) can be parameterized in certain forms generating
nonholonomically deformed black ellipsoid, black hole, \ wormhole,
solitonic, inhomogeneous solutions in various MGTs and in GR, see discussion
and examples in \cite{afdm,vepc,veym1,veym2}.

\subsubsection{LC varieties for effective Einstein manifolds}

If the generating/effective functions and sources are subjected to the
LC--conditions (\ref{cond1})--(\ref{cond3}), we obtain quadratic linear
elements
\begin{eqnarray}
ds_{LCRs}^{2} &=&g_{\alpha \beta }(x^{k},t)du^{\alpha }du^{\beta }=e^{\psi
}[(dx^{1})^{2}+(dx^{2})^{2}]+  \label{4dlcrs} \\
&&\omega ^{2}[\check{\Psi}]\frac{(\tilde{\Psi}[\check{\Psi}])^{2}}{4\Lambda }%
\left[ dy^{3}+(\partial _{k}n)\ dx^{k}\right] ^{2}+\omega ^{2}[\check{\Psi}]%
\frac{(\tilde{\Psi}^{\ast }[\check{\Psi}])^{2}}{\Theta \lbrack \check{\Psi}]}%
\ \left[ dt+(\partial _{i}\ \check{A}[\check{\Psi}])dx^{i}\right] ^{2}.
\notag
\end{eqnarray}%
The coefficients of these generic off--diagonal metrics also generate exact
solutions of (\ref{einsteq}) with effective source (\ref{effsourc2}) but
with zero torsion. Such solutions can be modelled equivalently in GR using
the LC--connection $\nabla .$

There are such generating functions and sources when, for instance, black
ellipsoid/ hole de Sitter configurations can be extracted from (\ref{4dlcrs}%
) in the limit of certain small off--diagonal deformations, see details in
\cite{afdm}. We note that the metrics are generic off--diagonal if the
anholonomic coefficients $\ W_{\alpha \beta }^{\gamma }$ (\ref{anhcoef}) are
not zero.

\subsection{Off--diagonal deformations of physically important solutions in
GR}

In this section, we present four classes of exact solutions generated by the
AFDM. The are constructed explicit examples of generic off--diagonal metrics
(\ref{4dlcrs}) with parameterizations of the generating and integration
functions and constants, and corresponding sources, when certain "prime"
metrics in GR are transformed into "target" metrics with modified physical
constants (running type, or effective polarizations) and/or deformed
horizons and/or self--consistent interactions, for instance, with
gravitational solitonic waves.

\subsubsection{Ellipsoid Kerr -- de Sitter configurations}

Let us consider a parameterization of 4--coodrinated like in the Kerr
geometry (we cite \cite{misner}, for a review of physically important
solutons, and \cite{vepc}, for nonholonomic deformations of such solutons).
The coordinates $x^{k^{\prime }}=x^{k^{\prime }}(r,\vartheta ),$ $%
y^{3}=t,y^{4}=\varphi ,$ when $u^{\alpha }=(x^{i^{\prime }},t,\varphi ).$
The prime metric is taken to be the Kerr solution written in the so--called
Boyer--Linquist coordinates $(r,\vartheta ,\varphi ,t),$ for $r=m_{0}(1+p%
\widehat{x}_{1}),\widehat{x}_{2}=\cos \vartheta .$ The parameters $p,q$ are
related to the total black hole mass, $m_{0}$ and the total angular
momentum, $am_{0},$ for the asymptotically flat, stationary and axisymmetric
Kerr spacetime. The formulas $m_{0}=Mp^{-1}$ and $a=Mqp^{-1}$ when $%
p^{2}+q^{2}=1$ implies $m_{0}^{2}-a^{2}=M^{2}.$ In such variables, the
vacuum Kerr solution can be written%
\begin{eqnarray}
ds_{[0]}^{2} &=&(dx^{1^{\prime }})^{2}+(dx^{2^{\prime }})^{2}+\overline{A}(%
\mathbf{e}^{3^{\prime }})^{2}+(\overline{C}-\overline{B}^{2}/\overline{A})(%
\mathbf{e}^{4^{\prime }})^{2},  \label{kerrbl} \\
\mathbf{e}^{3^{\prime }} &=&dt+d\varphi \overline{B}/\overline{A}%
=dy^{3^{\prime }}-\partial _{i^{\prime }}(\widehat{y}^{3^{\prime }}+\varphi
\overline{B}/\overline{A})dx^{i^{\prime }},\mathbf{e}^{4^{\prime
}}=dy^{4^{\prime }}=d\varphi .  \notag
\end{eqnarray}%
We can consider any coordinate functions
\begin{equation*}
x^{1^{\prime }}(r,\vartheta ),\ x^{2^{\prime }}(r,\vartheta ),\ y^{3^{\prime
}}=t+\widehat{y}^{3^{\prime }}(r,\vartheta ,\varphi )+\varphi \overline{B}/%
\overline{A},y^{4^{\prime }}=\varphi ,\ \partial _{\varphi }\widehat{y}%
^{3^{\prime }}=-\overline{B}/\overline{A},
\end{equation*}%
for which $(dx^{1^{\prime }})^{2}+(dx^{2^{\prime }})^{2}=\Xi \left( \Delta
^{-1}dr^{2}+d\vartheta ^{2}\right) $, and the coefficients are%
\begin{eqnarray*}
\overline{A} &=&-\overline{\Xi }^{-1}(\overline{\Delta }-a^{2}\sin
^{2}\vartheta ),\overline{B}=\overline{\Xi }^{-1}a\sin ^{2}\vartheta \left[
\overline{\Delta }-(r^{2}+a^{2})\right] , \\
\overline{C} &=&\overline{\Xi }^{-1}\sin ^{2}\vartheta \left[
(r^{2}+a^{2})^{2}-\overline{\Delta }a^{2}\sin ^{2}\vartheta \right] ,%
\mbox{
and } \\
\overline{\Delta } &=&r^{2}-2m_{0}+a^{2},\ \overline{\Xi }=r^{2}+a^{2}\cos
^{2}\vartheta .
\end{eqnarray*}

The quadratic linear element (\ref{kerrbl}) with prime data
\begin{eqnarray*}
\mathring{g}_{1} &=&1,\mathring{g}_{2}=1,\mathring{h}_{3}=-\rho ^{2}Y^{-1},%
\mathring{h}_{4}=Y,\mathring{N}_{i}^{a}=\partial _{i}\widehat{y}^{a}, \\
(\mbox{ or \ }\mathring{g}_{1^{\prime }} &=&1,\mathring{g}_{2^{\prime }}=1,%
\mathring{h}_{3^{\prime }}=\overline{A},\mathring{h}_{4^{\prime }}=\overline{%
C}-\overline{B}^{2}/\overline{A}, \\
\mathring{N}_{i^{\prime }}^{3} &=&\mathring{n}_{i^{\prime }}=-\partial
_{i^{\prime }}(\widehat{y}^{3^{\prime }}+\varphi \overline{B}/\overline{A}),%
\mathring{N}_{i^{\prime }}^{4}=\mathring{w}_{i^{\prime }}=0)
\end{eqnarray*}%
define solutions of the vacuum Einstein equations parameterized in the form (%
\ref{deinst}) and (\ref{lccond}) with zero sources.

We construct a subclass of solutions with rotoid configurations for
generating functions $\ \tilde{\Psi}=e^{\varpi }[\mathbf{\ }\widetilde{%
\Lambda }/2\ ^{\mu }\tilde{\Lambda}+\underline{\zeta }\sin (\omega
_{0}\varphi +\varphi _{0})],$ for $\tilde{\Psi}=e^{\varpi },$
\begin{eqnarray}
&&ds^{2}=e^{\psi (x^{k^{\prime }})}(1+\varepsilon \chi (x^{k^{\prime
}}))[(dx^{1^{\prime }})^{2}+(dx^{2^{\prime }})^{2}]  \notag \\
&& -\frac{e^{2\varpi }}{4|\Lambda |}\overline{A}[1+2\varepsilon \underline{%
\zeta }\sin (\omega _{0}\varphi +\varphi _{0})][dy^{3^{\prime }}+\left(
\partial _{k^{\prime }}\ ^{\eta }n(x^{i^{\prime }})-\partial _{k^{\prime }}(%
\widehat{y}^{3^{\prime }}+\varphi \frac{\overline{B}}{\overline{A}})\right)
dx^{k^{\prime }}]^{2}  \notag \\
&&+\frac{(\varpi ^{\ast })^{2}}{\ \Lambda }(\overline{C}-\frac{\overline{B}%
^{2}}{\overline{A}})[1+\varepsilon (\frac{\Lambda }{\widetilde{\lambda }}%
\partial _{\varphi }\varpi \mathbf{\ }+2\partial _{4}\varpi \underline{\zeta
}\sin (\omega _{0}\varphi +\varphi _{0})+2\omega _{0}\ \underline{\zeta }%
\cos (\omega _{0}\varphi +\varphi _{0}))]  \notag \\
&& [d\varphi +(\partial _{i^{\prime }}\ \widetilde{A}+\varepsilon \partial
_{i^{\prime }}\ \ ^{1}\check{A})dx^{i^{\prime }}]^{2}.  \label{lkerr}
\end{eqnarray}

Such metrics have a Killing symmetry on $\partial /\partial t$ and are
completely defined by a generating function $\varpi (x^{k^{\prime }},\varphi
)$ and the source and $\ \Lambda .$ They describe $\varepsilon $%
--deformations of Kerr -- de Sitter black holes into ellipsoid
configurations with effective (polarized) cosmological constants determined
by possible geometric flows and generic off--diagonal interactions. If the
zero torsion conditions are satisfied (like in (\ref{4dlcrs})), such metrics
can be modelled in GR.

\subsubsection{Nonholonomically deformed wormhole configurations}

Let us consider a different class of stationary configuration which defines
a wormhole solutions and their off--diagonal deformations. We begin with a
diagonal prime wormhole metric
\begin{eqnarray*}
\mathbf{\mathring{g}} &=&\mathring{g}_{i}(x^{k})dx^{i}\otimes dx^{i}+%
\mathring{h}_{a}(x^{k})dy^{a}\otimes dy^{a} \\
&=&[1-b(r)/r]^{-1}dr\otimes dr+r^{2}d\theta \otimes d\theta
-e^{2B(r)}dt\otimes dt+r^{2}\sin ^{2}\theta d\varphi \otimes d\varphi ,
\end{eqnarray*}%
where $B(r)$ and $b(r)$ are called respectively the red--shift and form
functions. On holonomic wormhole solutions, we refer readers to \cite%
{morris,visser,lobo,boehm}. There are used the local coordinates $u^{\alpha
}=(r,\theta ,t,\varphi ).$ The radial coordinate has a range $r_{0}\leq r<a.$
The minimum value $r_{0}$ is for the wormhole throat. The constant $a$ is
the distance at which the interior spacetime joins to an exterior vacuum
solution ($a\rightarrow \infty $ for specific asymptotically flat wormhole
geometries). The coefficients of the diagonal stress--energy tensor
\begin{equation*}
\mathring{T}_{~\nu }^{\mu }=diag[~^{r}p=\tau (r),~^{\theta
}p=p(r),~^{\varphi }p=p(r),~^{t}p=\rho (r)]
\end{equation*}%
are subjected to certain conditions in order to generate wormhole solutions
of the Einstein equations in GR.

Wormhole metrics are constructed to possess conformal symmetry determined \
by a vector $\mathbf{X}=\{X^{\alpha }(u)\},$ when the Lie derivative
\begin{equation*}
X^{\alpha }\partial _{\alpha }\mathring{g}_{\mu \nu }+\mathring{g}_{\alpha
\nu }\partial _{\mu }X^{\alpha }+\mathring{g}_{\alpha \mu }\partial _{\nu
}X^{\alpha }=\sigma \mathring{g}_{\mu \nu },
\end{equation*}
where $\sigma =\sigma (u)$ is the conformal factor. A class of such
solutions are parameterized by
\begin{eqnarray*}
B(r) &=&\frac{1}{2}\ln (C^{2}r^{2})-\kappa \int r^{-1}\left( 1-b(r)/r\right)
^{-1/2}dr,~b(r)=r[1-\sigma ^{2}(r)], \\
\tau (r) &=&\frac{1}{\kappa ^{2}r^{2}}(3\sigma ^{2}-2\kappa \sigma -1),~p(r)=%
\frac{1}{\kappa ^{2}r^{2}}(\sigma ^{2}-2\kappa \sigma +\kappa ^{2}+2r\sigma
\sigma ^{\bullet }), \\
\rho (r) &=&\frac{1}{\kappa ^{2}r^{2}}(1-\sigma ^{2}-2r\sigma \sigma
^{\bullet }).
\end{eqnarray*}
These data generate \textquotedblright diagonal\textquotedblright\ wormhole
configurations determined by \textquotedblright exotic\textquotedblright\
matter because the null energy condition (NEC), $\mathring{T}_{\mu \nu
}k^{\mu }k^{\nu }\geq 0$, ($k^{\nu }$ is any null vector), is violated.

In this section, we analyze wormholoe configurations which match the
interior geometries to an exterior de Sitter one which (in general) can be
also determined by an off--diagonal metric. The exotic matter and effective
matter configurations are considered to be restricted to spacial
distributions in the throat neighborhood which limit the dimension of
locally isotropic and/or anisotropic wormhole to be not arbitrarily large.
The Schwarzschild -- de Sitter (SdS) metric%
\begin{equation*}
ds^{2}=q^{-1}(r)(dr^{2}+r^{2}\ d\theta ^{2})+r^{2}\sin ^{2}\theta \ d\varphi
^{2}-q(r)\ dt^{2},
\end{equation*}%
can be re--parameterized for any $(x^{1}(r,\theta ),x^{2}(r,\theta
),y^{3}=\varphi ,y^{4}=t)$ in order to express $q^{-1}(r)(dr^{2}+r^{2}\
d\theta ^{2})=e^{\mathring{\psi}{(x^{k})}}[(dx^{1})^{2}+(dx^{2})^{2}].$ Such
a metric defines two real static solutions of the Einstein equations with
cosmological constant $\Lambda $ if $M<1/3\sqrt{|\Lambda |},$ for $q(r)=1-2%
\overline{M}(r)/r,\overline{M}(r)=M+\Lambda r^{3}/6,$ where $M$ is a
constant mass parameter. For diagonal configurations, we can identify $%
\Lambda $ with the effective cosmological constant.

The next step is to consider conformal, ellipsoid and/or solitonic/toroidal
deformations related in certain limits to the SdS metric written in the form
\begin{equation*}
~_{\Lambda }\mathbf{g}=d\xi \otimes d\xi +r^{2}(\xi )\ d\theta \otimes
d\theta +r^{2}(\xi )\sin ^{2}\theta \ d\varphi \otimes d\varphi -q(\xi )\
dt\otimes \ dt.
\end{equation*}%
There are used local coordinates $x^{1}=\xi =\int dr/\sqrt{\left\vert
q(r)\right\vert },x^{2}=\vartheta ,y^{3}=\varphi ,y^{4}=t,$ for a system of $%
h$--coordinates when $(r,\theta )\rightarrow (\xi ,\vartheta )$ with $\xi $
and $\vartheta $ of length dimension. The data for this primary metric are
written
\begin{equation*}
\mathring{g}_{i}=\mathring{g}_{i}(x^{k})=e^{\mathring{\psi}{(x^{k})}},\
\mathring{h}_{3}=r^{2}(x^{k})\sin ^{2}\theta (x^{k}),\mathring{h}_{4}=-q(r{%
(x^{k})}),\mathring{w}_{i}=0,\ \mathring{n}_{i}=0.
\end{equation*}

Off--diagonal deformations on a small parameter $\varepsilon $ of the
wormhole metrics are described by quadratic elements of type (\ref{4dlcrs}),
when the generating and integration functions are chosen
\begin{eqnarray}
\mathbf{ds}^{2} &=&e^{\tilde{\psi}(\widetilde{\xi },\theta )}(d\widetilde{%
\xi }^{2}+\ d\vartheta ^{2})-\frac{e^{2\widetilde{\varpi }}}{4\Lambda }%
[1+\varepsilon \overline{\chi }_{3}(\widetilde{\xi },\varphi )]e^{2B(%
\widetilde{\xi })}[dt+\partial _{\widetilde{\xi }}(\ ^{\eta }n+\varepsilon
\overline{n})~d\widetilde{\xi }+\partial _{\vartheta }(\ ^{\eta
}n+\varepsilon \overline{n})~d\vartheta ]^{2}  \notag \\
&&+\frac{[\partial _{\varphi }\widetilde{\varpi }]^{2}}{\Lambda }%
(1+\varepsilon \frac{\partial _{\varphi }[\overline{\chi }_{3}\widetilde{%
\varpi }]}{\partial _{\varphi }\widetilde{\varpi }})\ ^{0}\overline{h}%
_{4}[d\varphi +\partial _{\widetilde{\xi }}(\ ^{\eta }\widetilde{A}%
+\varepsilon \overline{A})d\widetilde{\xi }+\partial _{\vartheta }(\ ^{\eta }%
\widetilde{A}+\varepsilon \overline{A})d\vartheta ]^{2}.  \label{ellipswh}
\end{eqnarray}%
We prescribe such generating functions $\widetilde{\varpi }$ and effective
source $\Lambda $ that the effective polarization functions can be
approximated $\widetilde{\eta }_{a}\simeq 1$ and $^{\eta }\widetilde{A}$ and
$\ ^{\eta }n$ are "almost constant", with respect to certain systems of
radial coordinates. For such conditions, the metric (\ref{ellipswh}) mimic
small rotoid wormhole like configurations with off--diagonal terms and $\ $%
possible geometric flow modifications of the diagonal coefficients, see
details in \cite{sv2014}. It is possible to chose such integration functions
and constants that this class of stationary solutions define wormhole like
metrics depending generically on three space coordinates with
self--consistent "imbedding" in an offidagonal GR background, for $\ ^{0}%
\overline{h}_{3}=r^{2}(\widetilde{\xi })\sin ^{2}\theta (\widetilde{\xi }%
,\vartheta ),$ where $\ \widetilde{\xi }=\int dr/\sqrt{|1-b(r)/r|}$ and $B(%
\widetilde{\xi })$ are determined by a wormhole metric.

\subsubsection{Solitonic waves for inhomogeneous cosmological solutions}

For simplicity, we can consider solutions of type (\ref{4dlcrs}) generated
by a nonlinear radial (solitonic, with left $s$-label) generating function $%
\tilde{\Psi}=\ ^{s}\tilde{\Psi}(r,t)=4\arctan e^{q\sigma (r-vt)+q_{0}}$ and
construct a metric
\begin{equation}
\mathbf{ds}^{2}=e^{\psi (r,\theta )}(dr^{2}+\ d\theta ^{2})+~\frac{\ ^{s}%
\tilde{\Psi}^{2}}{4\ \Upsilon }\mathring{h}_{3}(r,\theta )d\varphi ^{2}-%
\frac{(\partial _{t}\ ^{s}\tilde{\Psi})^{2}}{\Upsilon \ ^{s}\tilde{\Psi}^{2}}%
\mathring{h}_{4}(r,t)[dt+(\partial _{r}\ \widetilde{A})dr]^{2},
\label{excosms}
\end{equation}%
In this metric, for simplicity, we fixed $n(r,\theta )=0$\ and consider that
$\widetilde{A}(r,t)$ is defined as a solution of $\ ^{s}\tilde{\Psi}%
^{\bullet }/\ ^{s}\tilde{\Psi}^{\ast }=\partial _{r}\ \widetilde{A}$ and $%
\mathring{h}_{a}$ are given by a homogeneous cosmology model data. The
generating function is just a 1--soliton solution of the sine--Gordon
equation $\ ^{s}\tilde{\Psi}^{\ast \ast }-\ ^{s}\tilde{\Psi}^{\bullet
\bullet }+\sin \ ^{s}\tilde{\Psi}=0$. For any class of small polarizations
with $\eta _{a}\sim 1),$ we can consider that the source $(\ ^{m}\Upsilon +\
^{\alpha }\Upsilon )$ is polarized by $\ ^{s}\tilde{\Psi}^{-2}$ when $%
h_{3}\sim \mathring{h}_{3}$ and $h_{4}\sim \mathring{h}_{4}(\ ^{s}\tilde{\Psi%
}^{\ast })^{2}/\ ^{s}\tilde{\Psi}^{4}$ with an off--diagonal term $\partial
_{r}\ \widetilde{A}$ resulting in a stationary solitonic universe. If we
consider that $(\partial _{\widehat{R}}\widehat{f})^{-1}=\ ^{s}\tilde{\Psi}%
^{-2}$, we can model $\widehat{f}$--interactions via off--diagonal
interactions and "gravitational polarizations".

In absence of matter, the off--diagonal cosmology is completely determined
by $\ \Upsilon $ related to an effective cosmological constant induced by
geometric flows. Such configurations can be determined alternatively using
distribution of matter fields when contributions from massive gravity are
with small anisotropic polarization, see details in \cite{sv,sv2015,esv,sv2016}.

\subsubsection{Off--diagonal deformations of FLRW metrics and gravitational
solitonic waves}

Using a re--defined time like coordinate $\widehat{t},$ when $t=t(x^{i},%
\widehat{t})$, $\sqrt{|h_{4}|}\partial t/\partial \widehat{t}$, for a scale
factor ${\widehat{a}}(x^{i},\widehat{t}),$ the d--metric (\ref{excosms}) can
be represented in the form%
\begin{equation}
ds^{2}=\widehat{a}^{2}(x^{i},\widehat{t})[\eta _{i}(x^{k},\widehat{t}%
)(dx^{i})^{2}+\widehat{h}_{3}(x^{k},\widehat{t})(\mathbf{e}^{3})^{2}-(%
\widehat{\mathbf{e}}^{4})^{2}],  \label{scaledm}
\end{equation}%
where $\eta _{i}=\widehat{a}^{-2}e^{\psi },\widehat{a}^{2}\widehat{h}%
_{3}=h_{3},\mathbf{e}^{3}=dy^{3}+\partial _{k}n~dx^{k},\widehat{\mathbf{e}}%
^{4}=d\widehat{t}+\sqrt{|h_{4}|}(\partial _{i}t+w_{i}).$ This is a
non--stational solution defining inhomogeneous cosmological spacetimes.

We can model small off--diagonal deformations of the Friedmann--Lema\^{\i}%
tre--Roberstson--Worker, FLRW, metric parameterized by a small parameter $%
\varepsilon ,$ with $0\leq \varepsilon <1,$ when
\begin{equation*}
\eta _{i}\simeq 1+\varepsilon \chi _{i}(x^{k},\widehat{t}),\partial
_{k}n\simeq \varepsilon \widehat{n}_{i}(x^{k}),\sqrt{|h_{4}|}(\partial
_{i}t+w_{i})\simeq \varepsilon \widehat{w}_{i}(x^{k},\widehat{t}).
\end{equation*}%
It is possible to choose such generating functions and sources when $%
\widehat{a}(x^{i},\widehat{t})\rightarrow $ $\widehat{a}(t),\widehat{h}%
_{3}(x^{i},\widehat{t})\rightarrow \widehat{h}_{3}(\widehat{t})$ etc. \ This
results in new classes of solutions even in diagonal limits because of
generic nonlinear and the nonholonomic character of off--diagonal systems in
GR and geometric flow evolution theories. For $\varepsilon \rightarrow 0$
and $\widehat{a}(x^{i},\widehat{t})\rightarrow $ $\widehat{a}(t),$ we obtain
scaling factors which are very different from those in FLRW cosmology. They
mimic such cosmological models with re--defined interaction parameters and
possible small off--diagonal deformations of cosmological evolution both in
GR, MGTs and Ricci flow evolution, see \cite{esv,sv2016}.

A metric (\ref{excosms}) defines solitonic waves along coordinate $x^{1}$ if
we take
\begin{equation*}
\check{\Psi}=\ ^{s}\Psi (x^{1},t)=4\arctan e^{q_{1}q_{2}(x^{1}-\xi t)+q_{0}},
\end{equation*}%
for certain constants $q_{0},q_{1}$ and $q_{2}=1/\sqrt{|1-\xi ^{2}|},$ and
put, for simplicity, $n=0.$ The function $\widetilde{A}$ is a solution of $\
\ ^{s}\Psi ^{\ast }\widetilde{A}^{\bullet }=\ ^{s}\Psi ^{\bullet }.$ For
such conditions, the generating function induces 1-soliton waves as a
solutions of the sine-Gordon equation
\begin{equation*}
\ ^{s}\Psi ^{\ast \ast }-\ ^{s}\Psi ^{\bullet \bullet }+\sin (\ ^{s}\Psi )=0
\end{equation*}
in the $v$--subspace together with self--consistent off--diagonal
propagation of such waves via $w_{i}=\widetilde{A}^{\bullet }.$ There are
possible more general 3--d solitonic gravitational waves which can propagate
self--consistently in Minkowski spacetime or in a FLRW background. To
generate such solutions we need to take a generating function $\ ^{s}\Psi
(x^{1},x^{2},t)$ which is a solution of the Kadomtsev--Petviashvili, KdP,
equation \cite{kdp,tgovsv1,svmpa}.

If we consider that in the ansatz of type (\ref{4dlcrs}) $y^{3}=t$ (time
like coordinate) and $y^{4}=\varphi $ (an angular space coordinate), we
construct stationary solutions with Killing symmetry on $\partial _{3},$ see
details in \cite{afdm,vepc}. Such solutions in GR and geometric flow theory
include as particular cases black ellipsoid configurations, Taub NUT
configurations and other type of Coulomb--like gravitational fields. In
particular, for small $\varepsilon $--deformations they include black hole
and black ellipsoid solutions in GR and certain classes of modified
theories. Such gravitational configurations are characterized by different
thermodynamic values in a W--thermodynamic model of geometric flows and
their stationary Ricci soliton configurations.

\section{Nonholonomic Thermodynamics of Gravitational Fields}

\label{s5} There is a standard theory of thermodynamics of black hole, BH,
solutions originally elaborated by Bekenstein--Hawking constructions for
stationary solutions in gravity theories (for a review, see \cite{misner}
and references therein). Following the formalism of double fibrations, the
BH thermodynamics can be derived for a very special example of Lyapunov type
functional on 3-d hypersurfaces with further 2+1 splitting (or 2+1+1
holonomic fibrations with horizon configurations and corresponding
asymptotic conditions). In general, we can not apply the BH thermodynamics
to study general gravitational configurations, for instance, models of
inhomogeneous cosmology and/or other spacetime configurations in GR and
MGTs. There were proposed certain ideas to formulate more general
thermodynamic descriptions of gravitational fields, see \cite{clifton} \
(the so--called CET model)\ and references therein.

In this section, we develop a W--functional thermodynamic model on a 3--d
closed spacelike hypersuface in GR, following the standard theory of Ricci
flows. In general, such an approach is very different from that considered
in standard BH thermodynamics. As a matter of principle, it is possible to
establish certain equivalence conditions (with very special holonomic vacuum
gravity configurations or diagonal de Sitter BH solutions) when the
W--functional approach will give same result as in the Bekenshein--Hawking
BH thermodynamics. The goal of this section is to elaborate on models on
nonholonomic thermodynamics of gravitational fields using 4-d
generalizations of Perelman's W--functionals. We shall also speculate how
such constructions can be related to the CET approach if certain additional
conditions are imposed. In this work, we restrict our approach to general
relativistic flows of 3-d to 4-d configuration to a special class of models
when the effective temperature $\beta ^{-1}(t)$ depends, in general, on time
like coordinate $t. $

\subsection{General relativistic models of W--entropy and geometric flow
evolution}

We emphasize that Perelman's functional $\ _{\shortmid }\mathcal{W}$ (\ref%
{perelm3w}) is in a\ sense analogous to minus entropy. Similar values can be
considered for various metric compatible nonholonomic Ricci flows and (non)
commutative, fractional derivative and other type modifications \cite%
{vfracrf,velatdif,vnhrf,vnoncjmprf}. This allows us to associate certain
analogous thermodynamical values characterizing (non) holonomic modified
Ricci flow evolution of metrics with local Euclidean structure and
generalized connections. Using $\ _{\shortmid }\mathcal{W}$ $\ $for $(%
\mathbf{q},\ _{\shortmid }\widehat{\mathbf{D}})$ and its 4--d generalization
$\widehat{\mathcal{W}}$ \ in a form (\ref{wfperel}), or (\ref{wfperelm4})
for $(\ \mathbf{g},\widehat{\mathbf{D}}),$ we can construct relativistic
geometric evolution models with generalized Hamilton equations of type (\ref%
{rfcandc}).

For 4-d general relativistic configurations, we can not provide a standard
statistical thermodynamic interpretation. We shall have to elaborate on
relativistic hydrodynamical type generalizations (see below the section \ref%
{ssrhth}). Here we emphasize that we can always characterise the geometric
flows by analogous thermodynamic systems on corresponding families of 3-d
closed hypersurfaces $\ \widehat{\Xi }_{t}$ using a nonholonomically
deformed entropy $\ _{\shortmid }\widehat{\mathcal{W}}.$ \ Let us consider a
standard partition function
\begin{equation*}
\breve{Z}=\exp \left\{ \int_{\widehat{\Xi }_{t}}M\sqrt{|q_{\grave{\imath}%
\grave{j}}|}d\grave{x}^{3}[-\breve{f}+n]~\right\}
\end{equation*}%
for the conditions stated for definition of (\ref{perelm3f}) \ and (\ref%
{perelm3w}). This allows us to compute main thermodynamical values for the
Levi--Civita connection $\ _{\shortmid }\nabla $ \cite{perelman1,monogrrf1}
and $n=3.$ A statistical model can be elaborated for any prescribed
partition function $Z=\int \exp (-\beta E)d\omega (E)$ for a corresponding
canonical ensemble at temperature $\beta ^{-1}$ being defined by the measure
taken to be the density of states $\omega (E).$ The standard thermodynamical
values are computed for the average energy, $\mathcal{E}=\ \left\langle
E\right\rangle :=-\partial \log Z/\partial \beta ,$ the entropy $S:=\beta
\left\langle E\right\rangle +\log Z$ and the fluctuation $\sigma
:=\left\langle \left( E-\left\langle E\right\rangle \right)
^{2}\right\rangle =\partial ^{2}\log Z/\partial \beta ^{2}.$

To elaborate the constructions in N--adapted form we consider a family of $%
q_{\grave{\imath}\grave{j}}(\tau (\chi ))$, with $\partial \tau /\partial
\chi =-1,$ being a real re-parametrization of $\chi .$ For 4--d, it can be
considered as a timelike parameter, $\chi \sim t).$ We change $\ _{\shortmid
}\nabla \rightarrow \ _{\shortmid }\widehat{\mathbf{D}}$ as it is determined
by distortions (\ref{distrel2}), see similar constructions in \cite{vnhrf}.
By a corresponding re-scaling $\breve{f}\rightarrow \tilde{f}$ and $\tau
\rightarrow \tilde{\tau}(t)$ (such a re-scaling is useful if we wont to
compare thermodynamical values for different linear connections in GR), we
compute
\begin{eqnarray}
\ _{\shortmid }\widehat{\mathcal{E}}\ &=&-\tilde{\tau}^{2}\int_{\widehat{\Xi
}_{t}}M\sqrt{|q_{\grave{\imath}\grave{j}}|}d\grave{x}^{3}\left( \mathbf{\ }%
_{\shortmid }\widehat{R}+|\ _{\shortmid }\widehat{\mathbf{D}}\tilde{f}|^{2}-%
\frac{3}{\tilde{\tau}}\right) ,  \label{3dthv} \\
\ _{\shortmid }\widehat{S} &=&-\int_{\widehat{\Xi }_{t}}M\sqrt{|q_{\grave{%
\imath}\grave{j}}|}d\grave{x}^{3}\left[ \tilde{\tau}\left( \ \mathbf{\ }%
_{\shortmid }\widehat{R}+|\ _{\shortmid }\widehat{\mathbf{D}}\tilde{f}%
|^{2}\right) +\tilde{f}-6\right] ,  \notag \\
\ _{\shortmid }\widehat{\sigma } &=&2\ \tilde{\tau}^{4}\int_{\widehat{\Xi }%
_{t}}M\sqrt{|q_{\grave{\imath}\grave{j}}|}d\grave{x}^{3}[|\ _{\shortmid }%
\widehat{\mathbf{R}}_{\grave{\imath}\grave{j}}+\ _{\shortmid }\widehat{%
\mathbf{D}}_{\grave{\imath}}\ _{\shortmid }\widehat{\mathbf{D}}_{\grave{j}}%
\tilde{f}-\frac{1}{2\tilde{\tau}}q_{\grave{\imath}\grave{j}}|^{2}].  \notag
\end{eqnarray}%
Such formulas can be considered for 4--d configurations considering that the
laps function $\breve{N}=1$ for N-adapted Gaussian coordinates but in such
cases it will be more difficult to computing in explicit form using standard
forms of solutions of 4--d physically important equations.

Using thermodynamical (\ref{3dthv}), we can compute the corresponding
average energy, entropy and fluctuations for evolution both on redefined
parameter $\tau $ and on a time like parameter $t$ of any family of closed \
hypersurfaces all determined by $\ _{\shortmid }\widehat{\mathbf{D}}$,
\begin{equation}
\widehat{\mathcal{E}}(\tau )=\int_{t_{1}}^{t_{2}}dt\breve{N}(\tau )\
_{\shortmid }\widehat{\mathcal{E}}(\tau ),\ \widehat{\mathcal{S}}(\tau
)=\int_{t_{1}}^{t_{2}}dt\breve{N}(\tau )\ _{\shortmid }\widehat{S}(\tau ),\
\widehat{\Sigma }(\tau )=\int_{t_{1}}^{t_{2}}dt\breve{N}(\tau )\ _{\shortmid
}\widehat{\sigma }(\tau ).  \label{thermodv}
\end{equation}%
These formulas are related by distortion formulas (\ref{distrel2}) with
corresponding values determined by $\ _{\shortmid }\nabla $,
\begin{equation*}
\ ^{\nabla }\mathcal{E}(\tau )=\int_{t_{1}}^{t_{2}}dt\breve{N}(\tau )\
_{\shortmid }^{\nabla }\mathcal{E}(\tau ),\ ^{\nabla }\mathcal{S}(\tau
)=\int_{t_{1}}^{t_{2}}dt\breve{N}(\tau )\ _{\shortmid }^{\nabla }S(\tau ),\
^{\nabla }{\Sigma }(\tau )=\int_{t_{1}}^{t_{2}}dt\breve{N}(\tau )\
_{\shortmid }^{\nabla }\sigma (\tau ).
\end{equation*}

Such values with "hat" or left label $\nabla $ are different even $\
_{\shortmid }\widehat{\mathbf{D}}\rightarrow \nabla $ for certain
topologically nontrivial configurations.

\subsection{W--thermodynamic values for exact solutions in GR}

For nonholonomic 4--d Lorentz--Ricci solitonic equations defined by systems
of type (\ref{qnk4d}), or (\ref{4dlcrs}), the generating function $\Psi
(x^{k},t)$ (or $\widetilde{\Psi }(x^{k},t),$ or $\check{\Psi}(x^{k},t)$ for
torsionless configurations) and the effective source $\Upsilon (x^{k},t)$
can be prescribed in some forms not depending one on another. The vertical
conformal factor $\omega $ can be an arbitrary function on $(x^{k},y^{3},t)$
subjected to the conditions (\ref{eq4e}). Such values should be subjected to
an additional constraint if we consider that the 4--d solutions for a
d--metric $\mathbf{g}$ encode also 3-d d--metrics $\mathbf{q}$. The 3-d
metric is parameterized in the form (\ref{anstrsolit}) with a Ricci flow
evolution determined by the Hamilton equations (\ref{heq1b}) on a
relativistic parameter. To construct explicit solutions, we consider the
parameter $\chi =t$ for 3-d evolution equations written with respect to
N--adapted frames.

\subsubsection{ Modified 3-d Ricci flows with induced nonholonomic torsion}

On a 4-d Lorentz--Ricci solitonic space determined by a quadratic line
element (\ref{qnk4d}), the 3--d Ricci flow evolution equation (\ref{heq1b})
is written
\begin{equation}
\mathbf{q}_{\grave{\imath}\grave{j}}^{\ast }=-2\ _{\shortmid }\widehat{%
\mathbf{R}}_{\grave{\imath}\grave{j}}+\frac{2\grave{r}}{5}\mathbf{q}_{\grave{%
\imath}\grave{j}},  \notag
\end{equation}%
for $\mathbf{q}_{\grave{\imath}\grave{j}}^{\ast }=\partial _{t}\mathbf{q}_{%
\grave{\imath}\grave{j}},$ where $\ _{\shortmid }\widehat{\mathbf{R}}_{%
\grave{\imath}\grave{j}}=\{\widehat{\mathbf{R}}_{1}^{1}\mathbf{q}_{ij},%
\widehat{\mathbf{R}}_{3}^{3}\mathbf{q}_{3}\}.$ In non-explicit form, $%
\partial _{t}$ is related to the partial derivation on temperature $\beta
^{-1}(t)$, which in this work can be related to the time like variable. For
components with $i,j,k...=1,2,$ in a 3+1 nonholonomic distribution, these
evolution equations with $q_{ij}^{\ast }=0$ and $q_{3}^{\ast }\neq 0,$
decouple in the form
\begin{eqnarray}
_{\shortmid }\widehat{\mathbf{R}}_{j}^{i} &=&\delta _{j}^{i}\frac{\grave{r}}{%
5}=\delta _{j}^{i}\widetilde{\Upsilon }(x^{k}),  \label{3rfleq} \\
\partial _{t}\ln |\omega ^{2}h_{3}| &=&-2\widehat{\mathbf{R}}_{3}^{3}+\frac{2%
\grave{r}}{5},  \notag
\end{eqnarray}%
for $\widehat{\mathbf{R}}_{3}^{3}=\Upsilon (x^{k},t).$ The h--source $%
\widetilde{\Upsilon }(x^{k})$ can be considered as a normalizing factor for
the 3--d Ricci flows when $\grave{r}=\widetilde{\Upsilon }.$ Taking $h_{3}=%
\frac{\tilde{\Psi}^{2}}{4\Lambda }$ and introducing the first equation in (%
\ref{3rfleq}) into the second one, we obtain
\begin{equation}
\partial _{t}\ln |\widetilde{\Psi }|=\widetilde{\Upsilon }-\Upsilon
-\partial _{t}\ln |\omega |.  \label{relat1}
\end{equation}%
We conclude that 4-d solitons of the Einstein equations describe also 3-d
solutions of the normalized N--adapted Hamilton equations if the associated
generating function $\tilde{\Psi},$ the effective sources $\widetilde{%
\Upsilon }-\Upsilon $ and v--conformal factor $\omega $ are subjected to the
conditions (\ref{relat1}). We can use this for definition of a
self--consistent parametrization of the effective source, $\Upsilon =$ $%
\widetilde{\Upsilon }-(\ln |\omega \widetilde{\Psi }|)^{\ast }.$ Introducing
this value in (\ref{nonltr}), we can find this nonlinear symmetry for
re--defining generating functions:
\begin{equation*}
\Psi ^{2}=\Lambda ^{-1}\left[ \widetilde{\Psi }^{2}|\widetilde{\Upsilon }%
-(\ln |\omega \widetilde{\Psi }|)^{\ast }|+\int dt\widetilde{\Psi }^{2}|%
\widetilde{\Upsilon }-(\ln |\omega \widetilde{\Psi }|)^{\ast }|^{\ast }%
\right] .
\end{equation*}%
As particular cases, we can consider $\widetilde{\Upsilon }=const$ and $%
\omega =1$ in order to generate a self-consistent 3--d Ricci flow evolution
on a 4--d effective Einstein spaces. In such cases, all geometric and
physical objects are with Killing symmetry on $\partial _{3}.$

By definition of the generating effective source (\ref{effsourc2}), $\Theta
^{\ast }=\ \Upsilon (\widetilde{\Psi }^{2})^{\ast }.$ We can write the
equation (\ref{relat1}) using values $(\omega ,\widetilde{\Psi },\widetilde{%
\Upsilon },\Theta ^{\ast }),$%
\begin{equation*}
\Theta ^{\ast }=(\widetilde{\Psi }^{2})^{\ast }\ [\widetilde{\Upsilon }-%
\frac{\omega ^{\ast }}{\omega }]-(\widetilde{\Psi }^{\ast })^{2}.
\end{equation*}%
We conclude that we can model Ricci flows of 3-d metrics $q_{\grave{\imath}%
\grave{j}}$ in N--adapted form on effective 4-d Einstein configurations by
imposing additional constraints on the generating functions, effective
sources, or effective generating sources. Such geometric evolutions are
characterized by nontrivial nonholonomic torsion completely defined by $%
\mathbf{g.}$

\subsubsection{ Modified 3-d Ricci flows of LC--configurations}

Torsionless 4-d Einstein configurations (\ref{4dlcrs}) are determined by
generating functions $\tilde{\Psi}[\check{\Psi}]$ and sources $\check{%
\Upsilon}=\Upsilon (x^{i},t)=\ \Upsilon \lbrack \check{\Psi}]$ subjected to
the conditions (\ref{cond1})--(\ref{cond3}). The 3-d Ricci flow evolution in
N--adapted form is described by (\ref{relat1}) re-defined for generating
functions and sources subjected to the condition
\begin{equation*}
\partial _{t}\ln |\widetilde{\Psi }[\check{\Psi}]|=\widetilde{\Upsilon }-%
\check{\Upsilon}-\partial _{t}\ln |\omega |.
\end{equation*}%
The LC geometric flow configurations are characterized by effective source $%
\check{\Upsilon}=$ $\widetilde{\Upsilon }-(\ln |\omega \widetilde{\Psi }[%
\check{\Psi}]|)^{\ast }$ and effective generating source $\check{\Theta}%
^{\ast }=\ \check{\Upsilon}(\widetilde{\Psi }^{2}[\check{\Psi}])^{\ast }.$

\subsubsection{ N--adapted 3-d Ricci flows on exact solutions in GR}

Exact solutions of 3-d Hamilton like equations (\ref{3rfleq}) are considered
for 4-d solutions (\ref{4dlcrs}). The evolution equation (\ref{relat1}) is
modified
\begin{equation*}
\partial _{t}\ln |\widetilde{\Psi }|=\widetilde{\Upsilon }-\Upsilon
-\partial _{t}\ln |\omega |.
\end{equation*}%
The generalization for additional geometric flow effective source and
effective generating source is given, respectively, by formulas%
\begin{equation*}
\ \Upsilon =\ \widetilde{\Upsilon }-(\ln |\omega \widetilde{\Psi }|)^{\ast }%
\mbox{ and }\ \Theta ^{\ast }=\ \Upsilon (\widetilde{\Psi }^{2})^{\ast }.
\end{equation*}%
The generic off--diagonal contributions to non--Riemannian evolution models
are with nontrivial nonholonomic torsion. LC--configurations can be
extracted by respective functional dependencies $\widetilde{\Psi }^{2}[%
\check{\Psi}]$ and effective sources and effective generating sources,
respectively, $\ \check{\Upsilon}=\ $ $\widetilde{\Upsilon }-(\ln |\omega
\widetilde{\Psi }[\check{\Psi}]|)^{\ast }$ and $\check{\Theta}^{\ast }=\
\check{\Upsilon}(\widetilde{\Psi }^{2}[\check{\Psi}])^{\ast }.$ One of the
fundamental consequences of such nonholonomic evolution theories is that
various massive, modified and GR effects can be modelled by nonholonomic
constraints even any value of effective sources $\Upsilon $ can be zero for
certain configurations on a 3-d hypersurface $\Xi _{0}.$

\subsubsection{Relativistic thermodynamic values for N--adapted 3--d
modified Ricci flows}

One of motivations to find 3-d Ricci flow solutions of Hamilton equations (%
\ref{3rfleq}) embedded in 4--d spacetimes in GR is that such solutions are
characterized by analogous thermodynamical models which can be generalized
for associated 4-d modified Lorentz-Ricci solitons, i.e. effecive Einstein
spaces. We note that the lapse function $N(u)=-\omega ^{2}h_{4}=-g_{4}$ (\ref%
{shift1}) is contained in explicit form in the integration measure for
formulas (\ref{3dthv}) if we do not work in N--adapted Gaussian coordinates.
For 3--d thermodynamical values, we obtain formulas
\begin{eqnarray}
\ _{\shortmid }\widehat{\mathcal{E}}\ \ &=&\tilde{\tau}^{2}\int_{\widehat{%
\Xi }_{t}}\tilde{M}\omega ^{2}h_{4}\sqrt{|q_{1}q_{2}q_{3}|}d\grave{x}%
^{3}\left( \mathbf{\ }_{\shortmid }\widehat{R}+|\ _{\shortmid }\widehat{%
\mathbf{D}}\tilde{f}|^{2}-\frac{3}{\tilde{\tau}}\right) ,  \label{thval3d} \\
\ _{\shortmid }\widehat{S} &=&\int_{\widehat{\Xi }_{t}}\tilde{M}\omega
^{2}h_{4}\sqrt{|q_{1}q_{2}q_{3}|}d\grave{x}^{3}\left[ \tilde{\tau}\left( \
\mathbf{\ }_{\shortmid }\widehat{R}+|\ _{\shortmid }\widehat{\mathbf{D}}%
\tilde{f}|^{2}\right) +\tilde{f}-6\right] ,  \notag \\
\ _{\shortmid }\widehat{\sigma } &=&-2\ \tilde{\tau}^{4}\int_{\widehat{\Xi }%
_{t}}\tilde{M}\omega ^{2}h_{4}\sqrt{|q_{1}q_{2}q_{3}|}d\grave{x}^{3}[|\
_{\shortmid }\widehat{\mathbf{R}}_{\grave{\imath}\grave{j}}+\ _{\shortmid }%
\widehat{\mathbf{D}}_{\grave{\imath}}\ _{\shortmid }\widehat{\mathbf{D}}_{%
\grave{j}}\tilde{f}-\frac{1}{2\tilde{\tau}}q_{\grave{\imath}\grave{j}}|^{2}],
\notag
\end{eqnarray}%
up to any parametric function $\tilde{\tau}(t)$ in $\tilde{M}=\left( 4\pi
\tilde{\tau}\right) ^{-3}e^{-\tilde{f}}$ with any $\tilde{\tau}(t)$ for $%
\partial \tilde{\tau}/\partial t=-1$ and $\chi >0$ Taking respective 3-d
coefficients of a solution (\ref{qnk4d}), or (\ref{4dlcrs}) [or any solution
of type (\ref{lkerr}), (\ref{ellipswh}), (\ref{excosms}), (\ref{scaledm})],
and prescribing a closed $\widehat{\Xi }_{0}$ we can compute such values for
any closed $\widehat{\Xi }_{t}.$ We have to fix an explicit N--adapted
system of reference and scaling function $\tilde{f}$ in order to find
certain explicit values for corresponding average energy, entropy and
fluctuations for evolution on a time like parameter $t$ of any family of
closed hypersurfaces. We can decide if certain solutions with effective
Lorentz-Ricci soliton source and/or with contributions from additional MGT
sources may be more convenient thermodynamically than other configurations.

Let us compute the values (\ref{thval3d}) for systems (\ref{eq4a})-(\ref%
{eq4e}) with solutions (\ref{qnk4d}) with $q_{1}=q_{2}=e^{\psi },q_{3}=%
\tilde{\Psi}^{2}/4\Lambda ,h_{4}=(\tilde{\Psi}^{\ast })^{2}/\ ^{F}\Theta .$
Using the nonlinear symmetry (\ref{nonltr}) with effective $\Lambda ,$ for $%
\ _{\shortmid }\widehat{\mathbf{R}}_{\grave{\imath}\grave{j}}=\Lambda
\mathbf{q}_{\grave{\imath}\grave{j}}$ and $\ \mathbf{\ }_{\shortmid }%
\widehat{R}=3\Lambda ;$ taking, for simplicty, $\ \tilde{f}=0,$ and
re--defining $\frac{3}{2(4\pi )^{3}}\omega ^{2}\rightarrow \omega ^{2},$ we
compute
\begin{eqnarray}
\ _{\shortmid }\widehat{\mathcal{E}}\ \ &=&\tilde{\tau}^{2}(t)\int_{\widehat{%
\Xi }_{t}}\frac{\tilde{\Psi}(\tilde{\Psi}^{\ast })^{2}}{\ \Theta \tilde{\tau}%
^{3}(t)}e^{\psi }\omega ^{2}d\grave{x}^{3}\left[ \sqrt{|\Lambda |}-\frac{1}{%
\tilde{\tau}(t)\sqrt{|\Lambda |}}\right] ,  \label{thval3dsol} \\
\ _{\shortmid }\widehat{S} &=&\frac{1}{2}\int_{\widehat{\Xi }_{t}}\frac{%
\tilde{\Psi}(\tilde{\Psi}^{\ast })^{2}}{\Theta \tilde{\tau}^{3}(t)}e^{\psi
}\omega ^{2}d\grave{x}^{3}\left[ \tilde{\tau}(t)\sqrt{|\Lambda |}-\frac{2}{%
\sqrt{|\Lambda |}}\right] ,  \notag \\
\ _{\shortmid }\widehat{\sigma } &=&-\ \tilde{\tau}^{4}(t)\int_{\widehat{\Xi
}_{t}}\frac{\tilde{\Psi}(\tilde{\Psi}^{\ast })^{2}}{\Theta \tilde{\tau}%
^{3}(t)}e^{\psi }\omega ^{2}d\grave{x}^{3}\left[ \sqrt{|\Lambda |}-\frac{1}{2%
\tilde{\tau}(t)\sqrt{|\Lambda |}}\right] ^{2},  \notag
\end{eqnarray}%
where $\ \Theta =\int dt\ \ \Upsilon (\widetilde{\Psi }^{2})^{\ast }$.

It is possible to define and compute thermodynamic like values (\ref%
{thval3dsol}) for generic off--diagonal solutions in GR as we explained for
solutions (\ref{4dlcrs}) and corresponding functionals $\tilde{\Psi}[\check{%
\Psi}],\Theta \lbrack \check{\Psi}]$ and $\omega \lbrack \check{\Psi}].$ We
emphasize that in both cases with zero, or nonzero nonholonomic torsion,
above formulas for thermodynamical values are defined for a nonzero
effective $\widetilde{\Lambda }$ and non--zero source $\Upsilon .$ For
(effective) vacuum configurations, such formulas have to computed using
corresponding classes of off--diagonal solutions.

The values (\ref{thval3d}) and/or (\ref{thval3dsol}) can be computed for 4-d
configurations determined by modified Lorentz--Rieman solitons, $\widehat{%
\mathcal{E}}=\int_{t_{1}}^{t_{2}}dt\ _{\shortmid }\widehat{\mathcal{E}},$ $%
\widehat{\mathcal{S}}=\int_{t_{1}}^{t_{2}}dt\ _{\shortmid }\widehat{S},$ $%
\widehat{{\Sigma }}=\int_{t_{1}}^{t_{2}}dt\ _{\shortmid }\widehat{\sigma }$
determined by $_{\shortmid }\widehat{\mathbf{D}}.$ In a similar form, we can
use distortion formulas (\ref{distrel2}) and compute $\ ^{\nabla }\mathcal{E}%
=\int_{t_{1}}^{t_{2}}dt\ _{\shortmid }^{\nabla }\mathcal{E},\ ^{\nabla }%
\mathcal{S}=\int_{t_{1}}^{t_{2}}dt\ _{\shortmid }^{\nabla }S,\ ^{\nabla }{%
\Sigma }=\int_{t_{1}}^{t_{2}}dt\ _{\shortmid }^{\nabla }\sigma $ for $\
_{\shortmid }\nabla .$ Such values are positively different even $\
_{\shortmid }\widehat{\mathbf{D}}\rightarrow \ _{\shortmid }\nabla $ for
certain topologically nontrivial configurations. In result, we can analyze
if a nonholonomic configuration with N--adapted $_{\shortmid }\widehat{%
\mathbf{D}}$ may be more, or less, convenient thermodynamically than a
similar holonomic one determined by $\ _{\shortmid }\nabla .$

Finally, we note that geometric thermodynamics values (\ref{thval3d}) are
defined both for (modified) black hole solutions and inhomogeneous
cosmological solutions. The physical meaning of such a thermodynamical
approach is very different from that of standard black hole thermodynamics
in GR. Nevertheless, certain criteria for equivalent modelling can be
analyzed in various MGTs, see \cite{vepc,sv2014,sv,sv2015,esv,sv2016,tgovsv1}.

\subsection{General relativistic hydrodynamics and thermodynamics\newline
for geometric flows of gravitational fields}

\label{ssrhth}Thermodynamic values (\ref{3dthv}) and (\ref{thermodv}) are
formulate in variables with coefficients computed with respect to N--adapted
frame of reference which allows extensions of 3-d thermodynamic values to
4-d ones. This is a hidden aspect of models of nonrelativistic hydrodynamics
and an apparent property of relativistic theories, see modern approaches in
\cite{rttoday1,rttoday2,rttoday3,rttoday4,rttoday5,rttoday6}. In this
section, we speculate on \ a model of geometric flows with local
thermodynamic equilibrium for which exists a single distinguished velocity
field $\mathbf{v}_{\alpha }$, (if it is convenient, we can take $\mathbf{v}%
_{\alpha }=\mathbf{n}_{\alpha }),$ and a corresponding N--adapted rest
frame, where all flow evolutions of related physical quantities are
described by simple formulas.

For a fluid model with particle production, we characterize general
relativistic Ricci flows by the particle number density d--vector $\widehat{%
\mathcal{N}}^{\alpha },$ the entropy density d--vector $\ \widehat{\mathcal{S%
}}^{\alpha }$ and the effective energy momentum density tensor \ $\widehat{%
\mathcal{T}}^{\alpha \beta }.$ Using the LC--connection $\nabla ,$ we
postulate a covariant forms for conservation of the particle number and the
effective energy momentum,%
\begin{equation}
\nabla _{\alpha }\widehat{\mathcal{N}}^{\alpha }(u^{\gamma })=0\mbox{ and }%
\nabla _{\alpha }\ \widehat{\mathcal{T}}^{\alpha \beta }(u^{\gamma })=0.
\label{balanceeq}
\end{equation}%
Using distortion relations $\widehat{\mathbf{D}}=\nabla +\widehat{\mathbf{Z}}
$ (\ref{distr}), we can compute nonholonomic deformations of such values
when $\widehat{\mathbf{Z}}$ determine respective sources induced by
nonholonomic torsion. This is typical for models of nonholonomic continuous
mechanics and generalized hydrodynamic/fluid models. We note that the
effective entropy is not conserved. It is supposed that the (general)
relativistic effective entropy is zero only in the thermodynamic
equilibrium:
\begin{equation*}
\nabla _{\alpha }\widehat{\mathcal{S}}^{\alpha }(u^{\gamma })\geq 0.
\end{equation*}

With the help of N--adapted velocity field $\mathbf{v}^{\alpha }(u^{\gamma
}) $ and velocity orthogonal projection operator $\pi ^{\alpha \beta
}=\delta ^{\alpha \beta }-\mathbf{v}^{\alpha }\mathbf{v}^{\beta },$ we
introduce necessary values for an effective hydrodynamical model:%
\begin{equation*}
\begin{array}{cc}
\widehat{n}:=\widehat{\mathcal{N}}^{\alpha }\mathbf{v}_{\alpha } &
\mbox{
is the particle number density }; \\
\widehat{j}^{\alpha }:=\pi ^{\alpha \beta }\widehat{\mathcal{N}}_{\beta } & %
\mbox{ is the diffusion current }; \\
\ \widehat{S}:=\widehat{\mathcal{S}}^{\alpha }\mathbf{v}_{\alpha } &
\mbox{
is the entropy density }; \\
\widehat{\mathcal{J}}^{\alpha }:=\pi ^{\alpha \beta }\widehat{\mathcal{S}}%
_{\beta } & \mbox{ is the entropy current }; \\
\widehat{\mathcal{E}}:=\mathbf{v}_{\alpha }\ \widehat{\mathcal{T}}^{\alpha
\beta }\mathbf{v}_{\beta } & \mbox{ is the energy density }; \\
\widehat{p}^{\alpha }:=\pi _{\ \gamma }^{\alpha }\ \widehat{\mathcal{T}}%
^{\gamma \beta }\mathbf{v}_{\beta } &
\mbox{ is the momentum density
(i.e. the energy current) }; \\
\widehat{\mathcal{P}}^{\alpha \beta }:=\pi _{\ \alpha ^{\prime }}^{\alpha
}\pi _{\ \beta ^{\prime }}^{\beta }\ \widehat{\mathcal{T}}^{\alpha ^{\prime
}\beta ^{\prime }} & \mbox{ is the pressure tensor }.%
\end{array}%
\end{equation*}%
"Hats" are used in order to emphasize that necessary coefficients are
defined in N--adapted form. As a matter of principle, all constructions can
be re--defined in arbitrary frames of reference (when symbols "lose" their
hats). In addition to these local N-adapted rest frame quantities, we
consider such values:
\begin{equation*}
\begin{array}{cc}
\mbox{ the effective energy-momentum vector } & \widehat{\mathcal{E}}%
^{\alpha }:=\mathbf{v}_{\beta }\ \widehat{\mathcal{T}}^{\alpha \beta }; \\
\mbox{ the energy-momentum current density } & \widehat{\mathcal{B}}^{\alpha
\beta }:=\pi _{\ \gamma }^{\alpha }\ \widehat{\mathcal{T}}^{\gamma \beta };
\\
\mbox{ momentum density } & \widehat{p}^{\alpha }:=\pi _{\ \gamma }^{\alpha
}\ \widehat{\mathcal{E}}^{\gamma }; \\
\mbox{ the energy current } & \widehat{p}^{\alpha }:=\mathbf{v}_{\beta }%
\widehat{\mathcal{B}}^{\alpha \beta }.%
\end{array}%
\end{equation*}%
Using above formulas, we compute the local rest N--adapted frame densities
determined by general relativistic Ricci flows as%
\begin{eqnarray}
\widehat{\mathcal{N}}^{\alpha } &=&\widehat{n}\mathbf{v}^{\alpha }+\widehat{j%
}^{\alpha },  \label{balance2} \\
\widehat{\mathcal{S}}^{\alpha } &=&\widehat{S}\mathbf{v}^{\alpha }+\widehat{%
\mathcal{J}}^{\alpha },  \notag \\
\ \widehat{\mathcal{T}}^{\alpha \beta } &=&\widehat{\mathcal{E}}^{\alpha }%
\mathbf{v}^{\beta }+\widehat{\mathcal{B}}^{\alpha \beta }=\widehat{\mathcal{E%
}}\mathbf{v}^{\alpha }\mathbf{v}^{\beta }+\widehat{p}^{\alpha }\mathbf{v}%
^{\beta }+\mathbf{v}^{\alpha }\widehat{p}^{\beta }+\widehat{\mathcal{P}}%
^{\alpha \beta }.  \notag
\end{eqnarray}%
These formulas express conveniently the particle number density for 4-d
d--vector and the energy-momentum density d--tensor with the help of local
rest frame quantities relative to a velocity field. Such formulas can be
related to a 3--d hypersurface Perelman's thermodynamic model (\ref{3dthv})
if $\widehat{S}=\ _{\shortmid }\widehat{S}$ and $\widehat{\mathcal{E}}=\
_{\shortmid }\widehat{\mathcal{E}}.$ For $\mathbf{v}^{\beta }=\mathbf{n}%
^{\beta }$ determined by an exact solution in GR, the densities (\ref%
{balance2}) can be normalized to respective values in (\ref{thermodv}).

Using the canonical d--connection $\widehat{\mathbf{D}}$ and above formulas,
the balance equations (\ref{balanceeq}) can be written in the form
\begin{equation*}
(\widehat{\mathbf{D}}_{\alpha }-\widehat{\mathbf{Z}}_{\alpha })\widehat{%
\mathcal{N}}^{\alpha }=\mathbf{v}^{\alpha }(\widehat{\mathbf{D}}_{\alpha }-%
\widehat{\mathbf{Z}}_{\alpha })\widehat{n}+\widehat{n}(\widehat{\mathbf{D}}%
_{\alpha }-\widehat{\mathbf{Z}}_{\alpha })\mathbf{v}^{\alpha }+(\widehat{%
\mathbf{D}}_{\alpha }-\widehat{\mathbf{Z}}_{\alpha })\widehat{j}^{\alpha }
\end{equation*}%
and
\begin{eqnarray*}
(\widehat{\mathbf{D}}_{\beta }-\widehat{\mathbf{Z}}_{\beta })\widehat{%
\mathcal{T}}^{\alpha \beta } &=&\mathbf{v}^{\gamma }(\widehat{\mathbf{D}}%
_{\gamma }-\widehat{\mathbf{Z}}_{\gamma })\widehat{\mathcal{E}}^{\alpha }+%
\widehat{\mathcal{E}}^{\alpha }(\widehat{\mathbf{D}}_{\gamma }-\widehat{%
\mathbf{Z}}_{\gamma })\mathbf{v}^{\gamma }+(\widehat{\mathbf{D}}_{\beta }-%
\widehat{\mathbf{Z}}_{\beta })\widehat{\mathcal{B}}^{\alpha \beta } \\
&=&\mathbf{v}^{\alpha }\mathbf{v}^{\gamma }(\widehat{\mathbf{D}}_{\gamma }-%
\widehat{\mathbf{Z}}_{\gamma })\widehat{\mathcal{E}}+\widehat{\mathcal{E}}%
\mathbf{v}^{\alpha }(\widehat{\mathbf{D}}_{\gamma }-\widehat{\mathbf{Z}}%
_{\gamma })\mathbf{v}^{\gamma }+\mathbf{v}^{\gamma }(\widehat{\mathbf{D}}%
_{\gamma }-\widehat{\mathbf{Z}}_{\gamma })\widehat{p}^{\alpha }+\widehat{p}%
^{\alpha }(\widehat{\mathbf{D}}_{\gamma }-\widehat{\mathbf{Z}}_{\gamma })%
\mathbf{v}^{\gamma } \\
&&+\widehat{\mathcal{E}}\mathbf{v}^{\gamma }(\widehat{\mathbf{D}}_{\gamma }-%
\widehat{\mathbf{Z}}_{\gamma })\mathbf{v}^{\alpha }+\mathbf{v}^{\alpha }(%
\widehat{\mathbf{D}}_{\gamma }-\widehat{\mathbf{Z}}_{\gamma })\widehat{p}%
^{\gamma }+\widehat{p}^{\gamma }(\widehat{\mathbf{D}}_{\gamma }-\widehat{%
\mathbf{Z}}_{\gamma })\mathbf{v}^{\alpha }+(\widehat{\mathbf{D}}_{\beta }-%
\widehat{\mathbf{Z}}_{\beta })\widehat{\mathcal{P}}^{\alpha \beta }.
\end{eqnarray*}%
These formulas are written in N--adapted form. In general, they contain
non--dissipative and dissipative components of general relativistic Ricci
flows. In terms of LC--connection $\nabla =\widehat{\mathbf{D}}-\widehat{%
\mathbf{Z}},$ we can eliminate certain sources of nonholonomic torsion
dissipation but in LC-variables the formulas can not be decoupled in general
form.

Let us discuss the fact and the conditions when the particle number density
d--vector and the energy--momentum d--tensor split into a non--dissipative
and dissipative parts. Such parts are easily distinguished in a local rest
N--adapted frame when the non--dissipative particle number density d--vector
$\widehat{\mathcal{N}}_{0}^{\alpha }$ is parallel to the d--velocity and the
non-dissipative energy momentum d--tensor $\ \widehat{\mathcal{T}}^{\gamma
\beta }$ is diagonal. We have a particular case of balance equations (\ref%
{balance2}) when%
\begin{equation}
\widehat{\mathcal{N}}_{0}^{\alpha }=\widehat{n}_{0}\mathbf{v}^{\alpha }%
\mbox{ and }\ \widehat{\mathcal{T}}_{0}^{\alpha \beta }=\widehat{\mathcal{E}}%
_{0}\mathbf{v}^{\alpha }\mathbf{v}^{\beta }-p_{0}\mathcal{\pi }^{\alpha
\beta },  \notag
\end{equation}%
with an effective static (scalar) pressure determined by the state equation
of state of the effective fluid. Such approximations are possible for
gravitational configurations without singularities. For instance, black
holes may have nonzero entropy (even determined in a different form than in
an approach with $W$--entropy) together with a nonzero particle production.
In such a case, we must introduce a term like $\widehat{\mathcal{S}}%
_{0}^{\alpha }=\widehat{S}_{0}\mathbf{v}^{\alpha }+\widehat{\mathcal{J}}%
_{0}^{\alpha }$ with an entropy current $\widehat{\mathcal{J}}_{0}^{\alpha }$
determined by geometric flows and $\widehat{S}_{0}$ identified with the
standard black hole entropy. Therefore the diffusion current density $%
\widehat{j}^{\alpha }$ consists the dissipative part of the particle number
(and the momentum density/energy) currents. This defines the difference of
the total and the equilibrium pressure $\Pi ^{\alpha \beta }=\widehat{%
\mathcal{P}}^{\alpha \beta }+p_{0}\mathcal{\pi }^{\alpha \beta }.$ The
effective viscous pressure determines the dissipative parts of the energy
momentum.

One should be mentioned that a splitting into some dissipative and
non--dissipative parts of certain physical quantities is related and depends
on the local rest frame (how it is chosen and N--adapted). Additionally to
the velocity field of the geometric flow continuum, this requires a
particular thermostatics to determine the static pressure. \ However, the
dissipation of thermodynamical systems (in our approach, by general
relativistic Ricci flows) is principally defined by the entropy production.
This is implicitly related to the background thermostatics, which is a
concept for systems in local equilibrium. On the other hand, the rest frame
is not determined in the case of dissipation, neither by any special form of
the physical quantities. For geometric flows related to 4-d exact solutions
in GR, one may consider to extend the Perelman's thermodynamic approach by
additional construction with the velocity field. We have to address to
nonholonomic kinetic theory and possible Finsler like modifications of GR,
relativistic thermodynamics and diffusion as in Refs. \cite%
{vkin,vfracrf,velatdif}. Such models are planned to be elaborated in our
further works.

The general relativistic Ricci flow evolution is related to problems of
stability and causality similarly to those in relativistic hydrodynamics.
This involves a detailed analysis of the Second Law as in Refs. \cite%
{rttoday4,rttoday5,rttoday6}. We can assume the possibility of acceleration
independent entropy production when the local rest frame entropy density is
a function of type $\widehat{S}(\widehat{\mathcal{E}},\widehat{n})=\widehat{S%
}(\sqrt{\widehat{\mathcal{E}}^{\alpha }\widehat{\mathcal{E}}_{\alpha }},%
\widehat{n})$ for values defined above. Then, it is possible to write a
Gibbs like relation%
\begin{equation*}
d\widehat{\mathcal{E}}+\widehat{\mathcal{E}}^{-1}\widehat{p}_{\gamma }d%
\widehat{p}^{\gamma }=\beta ^{-1}d\widehat{S}+\widehat{\mu }d\widehat{n}
\end{equation*}%
for an effective temperature $\beta $ and effective particle production
potential $\widehat{\mu }.$ Such an approach to relativistic thermodynamics
of Ricci flows eliminates the generic instability which exists in the
original Eckart theory and the stability conditions are independent of any
flow frames, see \cite{rtcrit4,rtcrit5}. Such conditions are the
non-negativity of the geometric flow transport coefficients and the
effective thermodynamic stability and the concavity of the entropy density.
This is in contrast to the former variants of the Eckart theory (which is
unstable), or to the Israel--Stewart theory. In the last case, there are
several complicated conditions, see details in \cite%
{rttoday2,rttoday4,rttoday5}.

Finally, the main conclusion is that we can characterize gravitational field
equations by Perelman's W--entropy on any closed 3-d hypersurface. In such a
case, a natural statistical thermodynamic model can be associated. To extend
the approach in a 4-d general relativistic form is necessary to elaborate on
modified relativistic thermodynamic and hydrodynamic theories which is the
main purpose of this work. The constructions can be performed in explicit
form and related to general classes of exact solutions in GR if we formulate
all gravitational field, geometric evolution and effective relativistic
thermodynamic theories in certain N-adapted nonholonomic variables. In next
section, we show how this can be connected to exact solutions.

\subsection{Parameterizations for the CET model}

A thermodynamic model is relativistic if it is derived for the energy
momentum conservation equations,%
\begin{equation}
\widehat{\mathbf{D}}_{\beta }(\mathbf{v}_{\alpha }\mathbf{T}^{\alpha \beta
})=\widehat{\mathbf{D}}_{\beta }(\mathbf{v}_{\alpha })\mathbf{T}^{\alpha
\beta }-\mathbf{v}_{\alpha }\widehat{\mathbf{J}}^{\alpha },  \label{conseq}
\end{equation}%
considering the heat flow $\mathbf{v}_{\alpha }\widehat{\mathbf{J}}^{\alpha
} $ into an effective fluid with $\widehat{\mathbf{J}}^{\alpha }=-\widehat{%
\mathbf{D}}_{\beta }\mathbf{T}^{\alpha \beta }.$\footnote{%
Such formulas are similar to those for $\nabla $ if we consider distorting
relations $\widehat{\mathbf{D}}=\nabla +\widehat{\mathbf{Z}}$ uniquely
determined by data $(\mathbf{g,N})$ and when $\widehat{\mathbf{D}}_{\mid
\widehat{\mathbb{T}}=0}\rightarrow \nabla .$} For perfect (pressureless)
matter, $\mathbf{T}_{\alpha \beta }=p\mathbf{g}_{\alpha \beta }+(\rho +p)%
\mathbf{v}_{\alpha }\mathbf{v}_{\beta },$ where $\mathbf{v}_{\alpha }\mathbf{%
v}^{\alpha }=-1$ , $\rho $ and $p$ are respectively the density and
pressure. The 4--velocity of the fluid can be taken as $v^{\alpha
}=(0,0,0,1) $ in a certain co-moving N-adapted frames when $\mathbf{T}%
_{~\beta }^{\alpha }=diag[0,0,0,-\rho ].$ We can consider also fluids with
nontrivial momentum density $\mathbf{q}^{\alpha }$ and anisotropic
(tracefree) pressure $\pi _{\alpha \beta },$ when
\begin{equation}
\mathbf{T}_{\alpha \beta }=p\mathbf{h}_{\alpha \beta }+\rho \mathbf{v}%
_{\alpha }\mathbf{v}_{\beta }+\mathbf{q}_{\alpha }\mathbf{v}_{\beta }+%
\mathbf{v}_{\alpha }\mathbf{q}_{\beta }+\pi _{\alpha \beta }.
\label{anisotrfl}
\end{equation}%
Defining $\widehat{\mathbf{\Theta }}:=(\mathbf{v}^{\gamma }\widehat{\mathbf{D%
}}_{\gamma }v)/v$ for a spatial domain $\upsilon $ with volume $%
V=\int\nolimits_{\upsilon }v,$ the entropy is given by $S=\int\nolimits_{%
\upsilon }\vartheta ,$ when (\ref{conseq}) and (\ref{anisotrfl}) lead to
\begin{equation*}
\widehat{\mathcal{T}}\ \mathbf{v}^{\gamma }\ \widehat{\mathbf{D}}_{\gamma
}\vartheta :=\mathbf{v}^{\beta }\widehat{\mathbf{D}}_{\beta }(\rho v)+p%
\mathbf{v}^{\gamma }\widehat{\mathbf{D}}_{\gamma }v=v(\mathbf{v}_{\alpha }%
\widehat{\mathbf{J}}^{\alpha }-\widehat{\mathbf{D}}_{\beta }\mathbf{q}%
^{\beta }-\mathbf{q}^{\alpha }\mathbf{v}^{\gamma }\widehat{\mathbf{D}}%
_{\gamma }\mathbf{v}_{\alpha }-\sigma ^{\alpha \beta }\pi _{\alpha \beta }).
\end{equation*}%
The variables $\vartheta $ and $\widehat{\mathcal{T}},$ represent the
point-wise entropy and temperature. If we define $\widehat{\mathcal{T}}$
independently, we get $S=\int\nolimits_{\upsilon }\vartheta .$\footnote{%
For temperature, we use the symbol $\mathcal{T}$ instead of $T,$ or $t,$ in
order to avoid possible ambiguities with similar notations, respectively,
for tangent space, $TV,$ time like coordinate $t$ etc.}

We are seeking to construct a Ricci flow and gravitational analogue of the
fundamental laws of the thermodynamics in the form%
\begin{equation}
~^{g}\mathcal{T}d~^{g}S=d~^{g}U+~^{g}pdV,  \label{termmod}
\end{equation}%
where $~^{g}\mathcal{T},~^{g}S,~^{g}U$ and $~^{g}p$ are respectively the
effective temperature, entropy, internal energy, isotropic pressure and $V $
is the spatial volume. For the vacuum gravitational fields, we consider the
values
\begin{equation*}
\widehat{\mathbf{W}}=\frac{1}{4}(\widehat{\mathbf{E}}_{\alpha }^{\quad \beta
}\widehat{\mathbf{E}}_{\quad \beta }^{\alpha }+\widehat{\mathbf{H}}_{\alpha
}^{\quad \beta }\widehat{\mathbf{H}}_{\quad \beta }^{\alpha }),\quad
\widehat{\mathbf{J}}_{\alpha }=\frac{1}{2}[\widehat{\mathbf{E}},\widehat{%
\mathbf{H}}]_{\alpha },\quad \mathbf{v}^{\gamma }\widehat{\mathbf{D}}%
_{\gamma }\widehat{\mathbf{W}}+\widehat{\mathbf{D}}^{\alpha }\widehat{%
\mathbf{J}}_{\alpha }\simeq 0,
\end{equation*}%
where $\widehat{\mathbf{W}}$ is the 'super-energy density' and the
'super-Poynting vector' $\widehat{\mathbf{J}}_{\alpha }=-\mathbf{h}_{\alpha
}^{~\delta }\widehat{\mathbf{T}}_{\delta \beta \gamma \tau }\mathbf{v}%
^{\beta }\mathbf{v}^{\gamma }\mathbf{v}^{\tau }$ is determined by the
Bel-Robinson tensor (for holonomic configurations, see details in \cite%
{bel,maartens,krish,clifton}),%
\begin{equation*}
\widehat{\mathbf{T}}_{\alpha \beta \gamma \tau }:=\frac{1}{4}(\widehat{%
\mathbf{C}}_{\varepsilon \alpha \beta \varphi }\widehat{\mathbf{C}}_{\
\gamma \tau }^{\varepsilon \quad \varphi }+\ ^{\ast }\widehat{\mathbf{C}}%
_{\varepsilon \alpha \beta \varphi }\ ^{\ast }\widehat{\mathbf{C}}_{\ \gamma
\tau }^{\varepsilon \quad \varphi }),
\end{equation*}%
when the dual Wey tensor is $\ ^{\ast }\widehat{\mathbf{C}}_{\alpha \beta
\varphi \tau }:=\frac{1}{2}\eta _{\alpha \beta \varepsilon \delta }\widehat{%
\mathbf{C}}_{\ \varphi \tau }^{\varepsilon \delta }.$ There is a two--index
'square--root' \cite{sch}, $\mathfrak{t}_{\alpha \beta },$ constructed as a
solution of%
\begin{equation*}
\widehat{\mathbf{T}}_{\alpha \beta \gamma \tau }=\mathfrak{t}_{(\alpha \beta
}\mathfrak{t}_{\gamma \tau )}-\frac{1}{2}\mathfrak{t}_{\varepsilon (\alpha }%
\mathfrak{t}_{\beta }^{\ \varepsilon }\mathbf{g}_{\gamma \tau )}+\frac{1}{24}%
[\mathfrak{t}_{\varepsilon \mu }\mathfrak{t}^{\varepsilon \mu }+\frac{1}{2}(%
\mathfrak{t}_{\beta }^{\ \varepsilon })^{2}]\mathbf{g}_{(\alpha \beta }%
\mathbf{g}_{\gamma \tau )}.
\end{equation*}%
The solutions of these equations depend on the algebraic type (following
Petrov's classification) of spacetime and related classes of solutions of
the Einstein equations, see details in \cite{bonill,clifton}. There are
variants such as {\small
\begin{equation*}
\mathfrak{t}_{\alpha \beta }=\left\{
\begin{array}{ccc}
3\epsilon |\widehat{\Psi }_{2}|(\mathbf{m}_{(\alpha }\mathbf{\bar{m}}_{\beta
)}+\mathbf{l}_{\alpha }\mathbf{k}_{\beta }), &  &
\mbox{ Petrov type N, similar
to  pure radiation;} \\
\epsilon |\widehat{\Psi }_{4}|\mathbf{k}_{\alpha }\mathbf{k}_{\beta }, &  &
\mbox{
Petrov type D,    Coloumb-like gravitational
configurations;} \\
\mbox{ various forms }, &  &
\mbox{other Petrov types, factorized as D or N, or
more complicated},%
\end{array}%
\right.
\end{equation*}%
} where $\epsilon =\pm 1,$ $\widehat{\Psi }_{2}=\widehat{\mathbf{C}}%
_{\varepsilon \alpha \beta \varphi }\mathbf{k}^{\varepsilon }\mathbf{m}%
^{\alpha }\mathbf{\bar{m}}^{\beta }\mathbf{l}^{\varphi }$ and $\widehat{\Psi
}_{4}=\widehat{\mathbf{C}}_{\varepsilon \alpha \beta \varphi }\mathbf{\bar{m}%
}^{\varepsilon }\mathbf{l}^{\alpha }\mathbf{\bar{m}}^{\beta }\mathbf{l}%
^{\varphi }.$ The first two cases above contain very interesting examples \
corresponding to stationary black hole solutions and the cases of scalar
perturbations of FLRW geometries and their off--diagonal deformations \cite%
{afdm}.

Using respective $\mathfrak{t}_{\alpha \beta }$ and $|\widehat{\Psi }_{2}|=%
\sqrt{2\widehat{\mathbf{W}}/3},$ we can construct a thermodynamic model \ (%
\ref{termmod}) for Coulomb--like gravitational fields with effective
energy--momentum tensor of type (\ref{anisotrfl}),%
\begin{equation*}
\mathfrak{t}_{\alpha \beta }\simeq \varkappa \ ^{g}\mathbf{T}_{\alpha \beta
}=\hat{\alpha}\sqrt{2\widehat{\mathbf{W}}/3}[\mathbf{x}_{\alpha }\mathbf{x}%
_{\beta }+\mathbf{y}_{\alpha }\mathbf{y}_{\beta }+2(\mathbf{v}_{\alpha }%
\mathbf{v}_{\beta }-\mathbf{z}_{\alpha }\mathbf{z}_{\beta })],
\end{equation*}%
for a constant $\hat{\alpha}$ to be determined from certain experimental
data and/or other theoretic considerations. The corresponding effective
energy density and pressure are $\varkappa \ ^{g}\rho =2\hat{\alpha}\sqrt{2%
\widehat{\mathbf{W}}/3}\geq 0$ and $\ ^{g}p=0.\ $We can take $\
^{g}q_{\alpha }=0$ and write the effective fundamental thermodynamic
equation for $\ ^{g}S=\int\nolimits_{\upsilon }\ ^{g}\vartheta $ in the
presence of perfect matter field in the form%
\begin{equation*}
~^{g}\mathcal{T\ }\mathbf{v}^{\gamma }\widehat{\mathbf{D}}_{\gamma
}(~^{g}\vartheta )=\mathbf{v}^{\gamma }\widehat{\mathbf{D}}_{\gamma }(\
^{g}\rho v)=-v\sigma _{\alpha \beta }[\pi ^{\alpha \beta }+\varkappa (\rho
+p)\widehat{\mathbf{E}}^{\alpha \beta }/2\hat{\alpha}\sqrt{6\widehat{\mathbf{%
W}}}].
\end{equation*}

For wave--like gravitational fields $|\Psi _{4}|=2\sqrt{\widehat{\mathbf{W}}}%
.$ Working with plane wave geometries in Kundt's class of solutions (for
holonomic configurations, see details in \cite{clifton}), the effective
thermodynamic quantities%
\begin{equation*}
\varkappa \ ^{g}\rho =2\hat{\beta}\sqrt{\widehat{\mathbf{W}}},\ ^{g}p=\
^{g}\rho /3,\ \varkappa \ ^{g}q_{\alpha }=2\hat{\beta}\sqrt{\widehat{\mathbf{%
W}}}\mathbf{z}_{\alpha },\ 6\varkappa \ ^{g}\pi _{\alpha \beta }=-2\hat{\beta%
}\sqrt{\widehat{\mathbf{W}}}(\mathbf{x}_{\alpha }\mathbf{x}_{\beta }+\mathbf{%
y}_{\alpha }\mathbf{y}_{\beta }-2\mathbf{z}_{\alpha }\mathbf{z}_{\beta })
\end{equation*}%
can be taken as
\begin{equation*}
\mathfrak{t}_{\alpha \beta }\simeq \varkappa \ ^{g}\mathbf{T}_{\alpha \beta
}=2\hat{\beta}\sqrt{\widehat{\mathbf{W}}}\mathbf{k}_{\alpha }\mathbf{k}%
_{\beta }
\end{equation*}%
with a constant $\hat{\beta}$ (in general, $\hat{\beta}\neq \hat{\alpha}).$
In the presence of a perfect fluid, the resulting fundamental thermodynamic
equation for $\ ^{g}S=\int\nolimits_{\upsilon }\ ^{g}\vartheta $ is%
\begin{eqnarray*}
~^{g}\mathcal{T\ }\mathbf{v}^{\gamma }\widehat{\mathbf{D}}_{\gamma
}(~^{g}\vartheta ) &=&\mathbf{v}^{\gamma }\widehat{\mathbf{D}}_{\gamma }(\
^{g}\rho v)+\ ^{g}p\mathbf{v}^{\gamma }\widehat{\mathbf{D}}_{\gamma }v \\
&=&-v[\sigma _{\alpha \beta }\ ^{g}\pi ^{\alpha \beta }+\mathbf{g}^{\alpha
\beta }\widehat{\mathbf{D}}_{\beta }(\ ^{g}\mathbf{q}_{\alpha })+\ ^{g}%
\mathbf{q}_{\alpha }\mathbf{v}^{\gamma }\widehat{\mathbf{D}}_{\gamma }%
\mathbf{v}^{\alpha }]-\varkappa (\rho +p)v\sigma _{\alpha \beta }\widehat{%
\mathbf{E}}^{\alpha \beta }/4\hat{\beta}\sqrt{\widehat{\mathbf{W}}}.
\end{eqnarray*}

There are alternative definitions of the gravitational temperature which
depend on the type of solutions considered, for instance, in a black hole or
cosmological model. Usually, one postulates the expression%
\begin{equation}
~^{g}\mathcal{T}=8|\mathbf{k}^{\gamma }\mathbf{l}^{\beta }\widehat{\mathbf{D}%
}_{\gamma }\mathbf{v}_{\beta }|/\varkappa \ =4|\mathbf{z}^{\beta }\mathbf{v}%
^{\gamma }\widehat{\mathbf{D}}_{\gamma }\mathbf{v}_{\beta }+\widehat{H}%
+\sigma _{\alpha \beta }\mathbf{z}^{\alpha }\mathbf{z}^{\beta }|/\varkappa ,
\label{eftemp}
\end{equation}%
where $\widehat{H}=\widehat{\Theta }/3$ is the isotropic Hubble rate. Such a
formula reproduces in appropriate limits the formulas from quantum field
theory in curved spacetimes, black hole thermodynamics, de Sitter spaces, or
Unrugh temperature etc. It can be defined in MGTs with effective modelling
by off--diagonal deformations of Einstein spaces.

The smooth function $f$ on a $3$--dimensional closed hypersurface can be
considered as a function determining the natural log of partition function $%
Z=\int e^{-\beta E}d\omega (E),$ with density of states measure $\omega (E),$
\begin{equation}
\log Z=\int_{\Sigma _{t}}[-f+3/2](4\pi \tau )^{-3/2}e^{-f}d\check{v}.
\label{logpf}
\end{equation}%
To find thermodynamic values in explicit form we consider the log of
partition function $\log Z$ (\ref{logpf}) on a closed 3--d hypersurface $%
\Sigma _{t}$ of volume $V=\int\nolimits_{\upsilon }v,$ when the entropy is
given by $S=$ $\int\nolimits_{\upsilon }\vartheta $ and the constant $\tau
=\beta ^{-1}$ is treated as an effective temperature. In order to reproduce
via relativistic Ricci flows effective thermodynamics models for
gravitational fields with temperature (\ref{eftemp}), the corresponding
Perelman type entropy
\begin{equation}
S=(1-\beta \frac{\partial }{\partial \beta })\log Z=-\int\nolimits_{\upsilon
}[\tau (\mathbf{\breve{R}}+|\mathbf{\breve{D}}f|^{2})+f-3]d\check{v}
\label{peren}
\end{equation}%
can be considered as a functional $~^{g}S(S)$ when%
\begin{equation}
~~^{g}\mathcal{T}d~^{g}S=\tau dS.  \label{thermtr}
\end{equation}%
This way, we transform the thermodynamical variables for $%
~^{g}U(~^{g}S,V)=U(S,V)$ and when the fundamental laws of the thermodynamics
(\ref{termmod}) are re--written in the form%
\begin{equation*}
dU=\tau dS-~^{g}pdV,
\end{equation*}%
where $p=~^{g}p$ and the relation $\partial S/\partial \tau =\sigma
^{2}/\tau ^{3}$ follows from the definition of W--entropy.

In explicit form, the effective entropy of gravitational fields and
thermodynamic transforms should be defined differently, for instance, for
the Couloumb and/or wave like gravitational fields, when the respective
constants $\hat{\alpha}$ and $\hat{\beta}$ are chosen to obtain equivalent
models for $~^{g}S$ and/or $S.$ Prescribing any values for independent data $%
~^{g}S$ and $~^{g}\mathcal{T}$, we can construct $S$ and $\log Z$ using
respectively the formulas (\ref{thermtr}) and (\ref{peren}). We denote the
solution of the last equation in the \ form $\log ~^{g}Z.$ Inverting the
formula (\ref{logpf}), we can find a corresponding function $f=~^{g}f$ $\ $%
describing the geometric evolution of a configuration of gravitational
fields characterized by certain relativistic thermodynamics data $%
(~^{g}S,~^{g}\mathcal{T}).$

In \cite{perelman1}, the W--entropy was defined by analogy to statistical
thermodynamics but a microscopic description was not considered. In
statistical mechanics, the partition function $Z=\sum\nolimits_{i}e^{-\beta
E_{i}}$ for a thermodynamic system is $\Gamma \lbrack \varphi
_{i},E_{i},\beta ]$ in a canonical ensemble with $E_{i}$ being energies
associated with corresponding "microstates" $\varphi _{i}$ at inverse
temperature $\beta $ (we consider summation on all available "microstates").
For non-discrete microstates, one considers integrals over the spaces of
microstates $\Omega ,$ $Z=\int\nolimits_{\upsilon }e^{-\beta E(\omega
)}d\omega ,$ with associated energy functions $E:\Omega \rightarrow \mathbb{R%
}$ and $d\omega $ being the density of states measure on $\Omega .$ The
entropy $S$ (\ref{peren}) of $\Gamma $ was computed for equilibrium states
at temperature $\tau .$ In the relativistic case, we can consider such
constructions for an effective theory for evolution of certain 3--d metrics
embedded into 4--d spacetime for which a local gravitational equilibrium
exists for a timelike variable $t.$ As a thermodynamic system, the analogous
$\Gamma $ is characterized by a full set of "macrostates" describing its
large--scale properties and equivalently modelled by data $~^{g}S$ and $~^{g}%
\mathcal{T}$.

It should be noted that G. Perelman called $f$ as the dilaton field and used
also an $F$--functional, $F=-\tau \log Z,$ with \textquotedblleft first
variation\textquotedblright\ formula which "can be found in the literature
on string theory, where it describes the low energy effective action; ...",
see the end of section 1 in \cite{perelman1}. Recently, the analogy with
string theory in the Polyakov formulation \cite{polyakov} was exploited in
Ref. \cite{lin}, where a general scheme of defining partition functions
associated to relevant geometric flows was proposed. The C. Lin's approach
is based on the functional determinant of differential operators and has
well--defined microstates as members of a functional spaces. For each
microstate, it associates the "Dirichlet energy" which in turn is associated
to the underlying operator. Using double, 3+1 and 2+2, fibrations,
corresponding assumptions and data (on analogous macroscopic thermodynamical
values associated to solutions of the Einstein equations, additional smooth
functions etc) we can determine the operator for energy--determining or
macrostates.

\section{Final Remarks and Conclusions}

\label{s6} G. Perelman's proof of the Poincar\'{e} conjecture provided a
fundamental result on topological structure of the Universe. He elaborated a
general schematic procedure in geometric analysis and speculated on a number
of perspectives for applications in modern physics. It was developed a
statistical and thermodynamic analogy for geometric evolution scenarios
using Lyapunov type functionals. The approach was based on the concept of
W--entropy. It was supposed from the very beginning \cite{perelman1} that
geometric flows may have certain implications for black hole physics and
string theory but the original theory of Ricci evolution flows was
formulated in a non--relativistic form. So, Perelman's ideas could not be
developed in a framework of theories with geometric flow evolution of three
dimensional, 3-d, Riemannian metrics.

This article has a very exact goal: to understand the W--entropy in a wide
range of general relativistic geometric flows and analyze possible
connections to other formulations of gravitational thermodynamics describing
general nonhomogeneous cosmological solutions and black hole configurations.
To this end, we need to provide a satisfactory relativistic account of
analogous statistical thermodynamical models associated to geometric flows
of exact solutions in general relativity, GR, resulting in effective
polarizaton (running) of fundamental constant and generic off--diagonal
effects. This will remain a challenge with many interesting directions in
modified gravity theories, MGTs, and modern accelerating cosmology emerging
from the work presented here. An interesting problem is that, for instance,
certain scenarios in modern cosmology can be modeled equivalently by generic
off--diagonal interactions in GR and/or by analogous solutions in a MGT.
Nevertheless, such gravitational field configurations have very different
properties in a framework of geometric flow theory, its self--similar
configurations as (relativistic) Ricci solitons. The corresponding analogous
thermodynamics is determined, in general, by different type physical values.
This motivates our approach to study general relativistic models of Ricci
flow evolution of exact solutions in gravity theories.

In this paper, we have shown in explicit form how the W--entropy can be
generalized for relativistic geometric evolution of solutions of the
Einstein equations. This allows a statistical thermodynamic description of
gravitational interactions, their (fractional) diffusion and kinetic
processes \cite{vkin,vfracrf,velatdif}. The marriage of the analogous
Perelman's thermodynamics and GR is an issue of very active debates on
physical meaning of such constructions and new mathematics. A temperature
like evolution parameter can be considered also for relativistic
generalizations but we can not address the problem as a specific type heat
propagation in the the context of relativistic dynamics of a non-perfect
fluid like in various approaches to relativistic thermodynamics, nonlinear
diffusion and relativistic kinetic theory \cite%
{rttoday1,rttoday2,rttoday3,rttoday4,rttoday5,rttoday6}. In theories with
geometric flows, we consider evoluton of metrics and fundamental geometric
objects (in general, these can be certain generalized connections,
curvatures and torsions). Similarly to heat fluxes, we can consider
nonrelativistic gradient flows and consider an unbounded speed of effective
thermal disturbances (fluctuations of tensor fields). This is due to the
parabolic nature of partial differential equations, PDEs, describing the
heat transport. In the relativistic cases, we get not only an inconvenience
but indeed a fundamental problem. There were studied various issues related
to this problem in relativistic hydrodynamical and dissipative theories \cite%
{rtcrit1,rtcrit2,rtcrit3,rtcrit4,rtcrit5,moe} (for review, see \cite{lopez}).

Such constructions, in a macroscopic relativistic thermodynamic formulation
\cite{clifton}, seem to be important for the structure formation of
Universes, black hole thermodynamics and characterization of nonlinear waves
locally anisotropic and/inhomogeneous gravitational configurations. It is
possible to formulate an underlying operator formalism and microscopic
approach in the spirit of Polyakov's approach to string theory. This scheme
requests a method of constructing generic off--diagonal solutions depending
on all spacetime variables \cite{afdm,vepc} which drive in nonlinear
parametric form the relativistic Ricci flows evolution of 3--d hypersurface
metrics in a 4--d spacetime.

We conclude that anholonomic frame deformation method, AFDM, involved in our
constructions can be applied to a wide range in GR and MGTs \cite%
{mav,stav,capoz,odint,sami,esv,sv2016} and generalized geometric evolution models
\cite{vnhrf,vnoncjmp,vnoncjmprf}. Such constructions are more general and
different from those elaborated in \cite{sc,head} where Ricci flows were
studied in connection with the standard black hole thermodynamics and
possible extensions to quantum diffusion and informational entropy.

\vskip5pt

\textbf{Acknowledgments:\ }
SV  was  supported by a travel grant from  MG14 at Rome and reports certain research related to his basic activity at UAIC, a DAAD fellowship and  the Program IDEI, PN-II-ID-PCE-2011-3-0256. Possible  implications in modern cosmology and geometric mechanics of such geometric methods were discussed with N. Mavromatos, E. Saridakis and P. Stavrinos. He is grateful for DAAD hosting to  D. L\"{u}st and O. Lechtenfeld. The final version of the paper was completed with results of a research supported by QGR-Topanga, USA.


\end{document}